\documentclass[12pt]{article}

\usepackage[title]{appendix}

\usepackage{authblk}

\usepackage{geometry}
\usepackage{bm,amsmath,amssymb,
}
\usepackage{amsmath}
\usepackage{graphicx}
\usepackage{mathtools}
\usepackage{mathtools}
\usepackage{float}
\usepackage{empheq}
\usepackage{subcaption}
\usepackage{epstopdf}
\usepackage{geometry}
\usepackage{lipsum}
\usepackage{epstopdf}
\usepackage{amssymb}
\usepackage{caption}
\usepackage{picture}
\usepackage{mwe,tikz}
\usepackage{onimage}
\usepackage[percent]{overpic}
\usepackage{pgf}
\usepackage{psfrag}
\usepackage{pict2e}
\usepackage[linktoc=all]{hyperref}
\usepackage{cleveref}
\usepackage{color}
\usepackage{transparent}
\usepackage{textcomp}
\usepackage[numbers,sort&compress]{natbib}
\usetikzlibrary{shapes,arrows}

\newcommand{\uvc}[1]{\bm{\mathrm{\hat #1}}} 

\definecolor{Lblu}{RGB}{193,193,234}
\definecolor{Mblu}{RGB}{131,131, 221}
\definecolor{Dblu}{RGB}{0,0, 255}
\definecolor{Lgr}{RGB}{192,192,192}
\definecolor{Mgr}{RGB}{138,138,138}

\definecolor{OR}{cmyk}{0, 0.58, 0.91, 0} 
\definecolor{BR}{RGB}{170,0,0} 
\definecolor{GR}{cmyk}{0.185, 0, 0.474, 0.322} 
\definecolor{VIO}{RGB}{170,0,170} 
\definecolor{PI}{cmyk}{0.523, 0, 0.0739, 0.31} 

\hypersetup{
	colorlinks,
	citecolor=red,
	filecolor=red,
	linkcolor=black,
	urlcolor=blue
}

\usetikzlibrary{shapes}
\usetikzlibrary{plotmarks}
\usetikzlibrary{positioning}

\newcommand{\bX}{{\bf X}}
\newcommand{\bv}{{\bf v}}

\newcommand{\bR}{{\bf R}}
\newcommand{\tbv}{\tilde{{\bf v}}}
\newcommand{\tbq}{\tilde{{\bf q}}}

\newcommand{\pll}{{\parallel}}

\graphicspath{{./phi/}{./Rnu/}{./special/}{./abstract/}{./hetro/}{./1endB/}{./stress_transition_1/}{./stress_sph/}{./stress_transition_2/}{./3dcurves/}{./towing/}{./types/}{./pureax/}{./abstract2/}}

\begin{document}

\tikzset{%
	mynode/.style={circle,minimum width=.5ex, fill=none,draw}, 
	myfillnode/.style={circle,minimum width=.5ex, fill=black,draw}, 
}

\title{Catenaries in viscous fluid}
\author{Brato Chakrabarti\thanks{bchakrab@eng.ucsd.edu}}
\affil{Department of Biomedical Engineering and Mechanics,\\
Virginia Polytechnic Institute and State University, Blacksburg, VA 24061, U.S.A.\\}
\author{J A Hanna\thanks{hannaj@vt.edu}}
\affil{
Department of Biomedical Engineering and Mechanics,\\
Department of Physics,\\
Virginia Polytechnic Institute and State University, Blacksburg, VA 24061, U.S.A.}

\date{June 17, 2016}

\maketitle

\begin{abstract}
This work explores a simple model of a slender, flexible structure in a uniform flow, providing analytical solutions for the translating, axially flowing equilibria of strings subjected to a uniform body force and drag forces linear in the velocities.   
The classical catenaries are extended to a five-parameter family of curves.  A sixth parameter affects the tension in the curves.  
Generic configurations are planar, represented by a single first order equation for the tangential angle.
The effects of varying parameters on representative shapes, orbits in angle-curvature space, and stress distributions are shown.
As limiting cases, the solutions include configurations corresponding to ``lariat chains'' and the towing, reeling, and sedimentation of flexible cables in a highly viscous fluid.
Regions of parameter space corresponding to infinitely long, semi-infinite, and finite length curves are delineated.  Almost all curves subtend an angle less than $\pi$ radians, but curious special cases with doubled or infinite range occur on the borders between regions.  
Separate transitions in the tension behavior, and counterintuitive results regarding finite towing tensions for infinitely long cables, are presented.  Several physically inspired boundary value problems are solved and discussed.
\end{abstract}


\section{Introduction}

The catenary, or hanging chain, is one of the oldest problems in analytical mechanics.  
In the late 17th century, Hooke identified its shape with that of an inverted arch.  In keeping with scientific customs of his time, he hid this knowledge in a very long anagram, appended as filler to an unrelated work on helioscopes \cite{Heyman98, Espinasse56}.  Not long after, in response to a challenge by Jacob Bernoulli, the solution was found independently by Johann Bernoulli, Huygens, and Leibniz \cite{Antman05}.
In the mid-19th century, the catenaries were recognized to be solutions of a broader dynamical problem, namely that of determining the shape equilibria of a moving string subjected to a uniform body force.  This was perhaps first recognized by a 
 Tripos examiner in 1854, one of whose victims, Routh, incorporated the fact in a mechanics text a few years later \cite{Routh55}.  By that time, the discovery had also been made by Airy (Astronomer Royal) and Thomson (Lord Kelvin), 
 whose impetus to study this problem was the massive failure of the first attempt to lay transatlantic telegraph cable \cite{Airy1858, Thomson1857both}.  Their treatments of the effects of the fluid medium fell short, with Airy modeling drag as isotropic with respect to the cable geometry, and Kelvin neglecting it entirely.

As shapes, the catenaries are a family of curves with a single scaling parameter that can effectively describe the length of a string hung, or arching, between two points.  Additional axial flow along the tangents of the curve serves only to increase the tension in the string, leaving shape equilibria intact.  The tension increase has consequences for the stability of equilibria, as has been observed in translating belts \cite{PerkinsMote89} and in the ``chain fountain'' phenomenon \cite{Biggins14, Virga14}.  To the symmetry of the system under axial flow, we may add another under uniform translations of the curve, which correspond to Galilean boosts that affect neither the shape nor the tension.

In this paper, we extend these classical results to include the effects of linear, anisotropic drag forces, such as would arise if the translating, axially flowing catenary were immersed in a highly viscous fluid.  The presence of fluid breaks multiple symmetries and turns the one-parameter family of catenary curves into a five-parameter family.  The additional parameters beyond scaling are the magnitude and direction of the translational velocity, the magnitude of the axial velocity, and the anisotropy of the drag forces.
We find that, in contrast to symmetry breaking by gyroscopic means \cite{Guven13skirts, HannaPendar15}, some symmetry still remains, in that inertial forces continue to only affect the string's tension, not its shape.  This is because Coriolis terms are absent in our problem; centripetal contributions to inertia behave identically in both cases.

Slender bodies immersed in fluid may be encountered across a wide range of scales in the industrial and natural worlds.
High-altitude power devices 
and offshore structures rely on pendant tethers, mooring lines, risers, and umbilical cables loaded by currents of air and water.  Towed cable systems are widely used for retrieval and sensing under the ocean.  Some aquatic and terrestrial plants bend and twist to avoid damage or uprooting, and the flexibility of such structures suggests adaptive motifs for drag reduction or control \cite{Alben04, BaroisdeLangre13}.  
Biofilm streamers form at high \cite{Hall-Stoodley04} and low \cite{Rusconi11, Karimi15} Reynolds number.  Chemical and fluidic applications may employ microscopic wall-mounted tubes or rods in shear or other non-uniform flows \cite{Ahmed96, Pozrikidis10, Wexler13}.
Suspensions of fibers at low Reynolds number arise in classical and industrial sedimentation problems \cite{GuazzelliHinch11, MartonRobie69}.
Other examples include spinning yarn \cite{BatraFraser15}, bacterial flagella \cite{Powers10}, and the ballooning threads of migrating spiders \cite{Humphrey87}.  
These systems often involve nonlinear drag forces, nonlocal effects, nonuniform flows, and bending stiffness, which are complications we do not address in this paper.  Before tackling such problems, it behoves us to consider our simple system, and try to use our conclusions as a basis for generalization.

Our augmented catenaries are generically planar curves, and admit an analytical description as a first-order dynamical system for the tangential angle, with arc length as the independent variable.  This suggests an approach to organizing our results, by classifying the corresponding phase space in terms of the presence and location of fixed points and poles.
We illustrate the role of the several parameters on the solutions.
We also discuss the nature of the string tension, which, surprisingly, undergoes its own bifurcations in a manner noncoincident with those of the phase portraits.
Our discovery and exposition of the solutions is aided by our initial disregard for any specific physical boundary conditions or integral constraints, which are connected to the axes of our parameter space only through implicit integral relationships.  However, towards the end of the paper, we discuss several boundary value problems in which practical parameters like the positions of end points and the total length of the curve are specified.

The general solution we consider covers several interesting problems as limiting cases.  Aside from the classical catenary, we may use our results to examine the towing of a neutrally buoyant cable (or, equivalently, such a cable tethered in a uniform flow), the reeling in and out of a belt or ``lariat chain'' \cite{LariatChain} between two points, or the free sedimentation of a perfectly flexible filament.
One limit, that of a string snaking along its own tangents under the action of drag forces alone, is a singular axis of our parameter space, admitting only the very restricted set of straight-line solutions.  In contrast, the hypersurface in parameter space where drag anisotropy vanishes is a degenerate case, admitting arbitrary shapes as freely sedimenting solutions.

In Section \ref{general}, we pose our problem and provide the general form of its solutions.  In Section \ref{solutions}, we present examples of generic solutions and organize them according to their singularity structure in phase space.  In Section \ref{bvp}, we present a few solutions of physically realistic boundary value problems.  The literature on several related problems is briefly discussed in Section \ref{discussion}.  The Appendices include brief details on the symmetry properties of the phase space, and an alternate derivation and representation of a subset of our solutions that includes the classical catenaries.

Throughout, we will use the following terms interchangeably: ``tension'' and ``stress'', ``string'' and ``cable''.

\section{Formulation and general solution}\label{general}

We first consider catenary curves in three dimensions in Section \ref{fullproblem}, before casting and solving the problem in two-dimensional form in Section \ref{planarproblem}.  Additional three-dimensional aspects of the solution are reserved for Section \ref{towing}.

\subsection{The three-dimensional problem}\label{fullproblem}

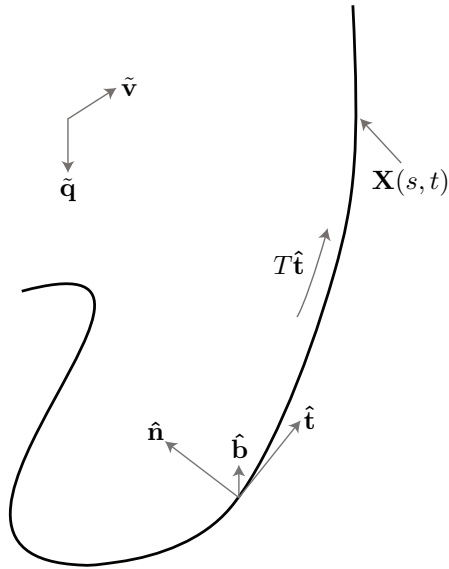
\begin{figure}[H]
	\centering
	\begin{tikzonimage}[scale=0.52]{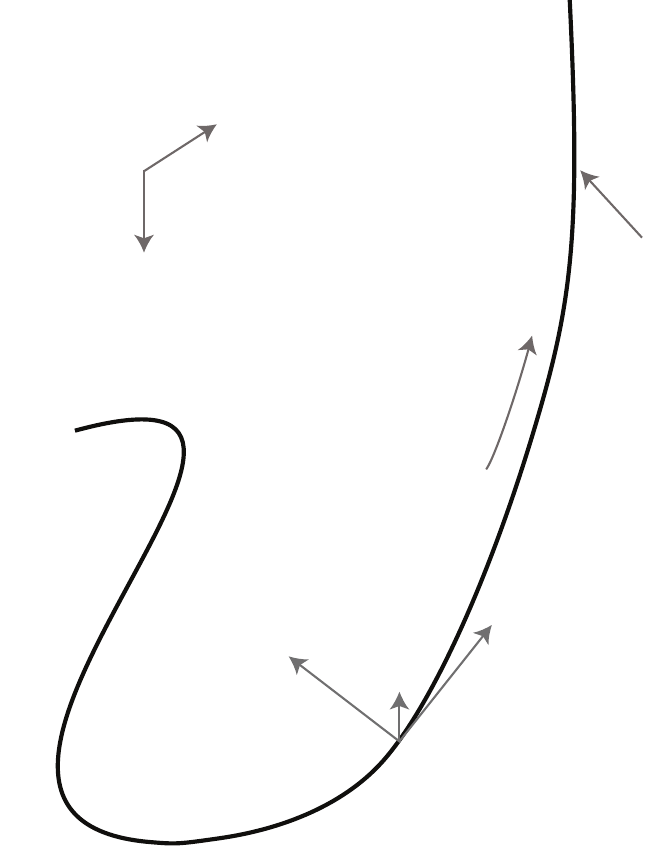} 
		\draw  (0.22,0.677) node[]{\footnotesize{$\tbq$}};
		\draw  (0.36,0.857) node[]{\footnotesize{$\tbv$}};
		\draw  (0.723,0.55) node[]{\footnotesize{$T \uvc{t}$}};
		\draw  (1.0,0.69) node[]{\footnotesize{$\bX(s,t)$}};
		\draw  (0.77,0.28) node[]{\footnotesize{$\uvc{t}$}};
		\draw  (0.42,0.257) node[]{\footnotesize{$\uvc{n}$}};
		\draw  (0.611,0.23) node[]{\footnotesize{$\uvc{b}$}};
	\end{tikzonimage}
	\caption{A catenary $\bX$ is an inextensible curve acted on by a uniform body force $\tbq$.  The body carries the orthonormal frame $(\uvc{t},\uvc{n},\uvc{b})$, and is translating with uniform velocity $\tbv$, with a superimposed uniform material flow $T\uvc{t}$ along its tangents.}
	\label{curve3d}
\end{figure}

Consider a curve $\bX(s,t)$ parameterized by arc length $s$ and time $t$, carrying an adapted orthonormal frame $\left(\uvc{t}(s,t),\uvc{n}(s,t),\uvc{b}(s,t)\right)$ whose spatial rotations are described by two extrinsic curvatures $\kappa(s,t)$ and $\tau(s,t)$.  Hence,
\begin{equation}\label{framerotation}
	\partial_s \left(\begin{array}{c}
			\bX\\
			\uvc{t}\\
			\uvc{n}\\
			\uvc{b}
			\end{array}\right) = \left(\begin{array}{cccc}
							0 & 1 & 0 & 0\\
							0 & 0 & \kappa & 0\\
							0 & -\kappa & 0 & \tau \\
							0 & 0 & -\tau & 0
							\end{array}\right)
	\left(\begin{array}{c}
	\bX\\
	\uvc{t}\\
	\uvc{n}\\
	\uvc{b}
	\end{array}\right) \, .
\end{equation}
We seek solutions in the form of shapes translating with uniform velocity $\tbv$, with a uniform axial flow of material given by a tangential velocity $T\partial_s\bX = T\uvc{t}$.  These velocities lead to linear, anisotropic drag forces with coefficients $\nu_\pll$ for drag along the tangent and $\nu_\perp$ for drag in the directions of either of the two normals $\uvc{n}$ or $\uvc{b}$.  A uniform body force $\tbq$ acts on the string.  These elements are shown schematically in Figure \ref{curve3d}.

Material time derivatives, taken at fixed $s$, provide the velocity and acceleration for the assumed form of motion, 
\begin{align}
	\partial_t \bX &= \tbv + T \uvc{t} \, , \\
	\partial_t^2 \bX &= T^2 \partial_s\uvc{t} = T^2\kappa\uvc{n}\, ,
\end{align}
allowing us to write a force balance on the string,
\begin{equation}\label{force1}
mT^2\partial_s\uvc{t} +  \nu_\pll \left(T+\tbv\cdot\uvc{t}\right)\uvc{t} + \nu_\perp \tbv\cdot\left(\uvc{n}\uvc{n} + \uvc{b} \uvc{b} \right)  =  \partial_s\left(\sigma\uvc{t}\right) +  \tbq \, .
\end{equation}
The first term represents inertial forces (per unit volume) arising from centripetal accelerations of material moving along the rigidly translating shape.  The mass coefficient $m$ represents both the material of the string as well as added mass effects from the surrounding fluid, while the body force $\tbq$ could be linked to the density of the body, an electric field acting on a charged body, or any other uniform effects.  The drag coefficients $\nu_\pll$ and $\nu_\perp$ are taken to be constant over the range of velocities experienced by the body\footnote{In the Stokes flow limit, where the Reynolds number vanishes, the drag coefficients can be 
derived analytically from the leading-order singular terms in a boundary-integral representation of the fluid velocity.  They are such that $2\nu_\pll = \nu_\perp$, and are logarithmically singular in the filament radius (normalized by some larger cutoff length).  The derivation is similar to those found in \cite{Higdon79, Lighthill96, TornbergShelley04}, and breaks down if the curvatures get too large.}.  Finally, $\sigma$ is the tension in the string, which acts as a multiplier enforcing the inextensibility constraint $\partial_s\bX\cdot\partial_s\bX = 1$.  

Equation \eqref{force1} has several limiting cases of interest.  When the drag terms vanish, we recover the classical catenaries.  If, in addition, the body force vanishes, any arbitrary configuration $\bX$ with uniform tension $\sigma = mT^2$ is a solution; these are the Routh ``lariats'' that snake along their own tangents \cite{Routh55, HealeyPapadopoulos90}.  Arbitrary configurations $\bX$ are also solutions when drag forces are present, but isotropic ($\nu_\pll = \nu_\perp \equiv \nu$); these sediment as rigid, tensionless bodies ($T=\sigma=0$) at a velocity such that $\nu \tbv = \tbq$.

As is true for many string problems, generic solutions of the equation of motion \eqref{force1} are planar, meaning that $\tau=0$ and $\uvc{b}$ is a constant vector.  This may be seen by considering the projections onto the frame vectors,
\begin{align}
\nu_\pll (T + \tbv\cdot \uvc{t}) &= \partial_s \sigma + \tbq\cdot \uvc{t} \label{tproj} \, , \\ 
m T^2 \kappa + \nu_\perp \tbv \cdot \uvc{n} &= \sigma \kappa + \tbq \cdot \uvc{n} \label{nproj} \, , \\ 
\nu_\perp \tbv \cdot \uvc{b} &= \tbq \cdot \uvc{b} \label{bproj} \, ,
\end{align}
taking repeated $s$-derivatives of \eqref{bproj}, remembering that $\tbv$ and $\tbq$ are constant vectors, and using the relations \eqref{framerotation}.  One finds that the only fully three-dimensional solutions are the limiting cases already mentioned.  Additionally, equations (\ref{tproj}-\ref{nproj}) may be used to show that the only lariats ($\tbv = \tbq = 0$, $T \ne 0$) in the presence of tangential drag forces are straight lines ($\kappa  =0$).  

Before discussing generic planar solutions, let us pause and consider these simplest of equilibria, straight lines, known to arise when a cable has one free end.  For straight lines, the $\kappa$ terms in \eqref{nproj} are zero, and the distinction between any vectors orthogonal to the tangent disappears.  
An equation like \eqref{bproj} holds for any such vector, which implies that the straight cable lives in the plane spanned by any choice of nonvanishing, noncollinear velocity $\tbv$ and body force $\tbq$.  
Imagine towing a cable with one end on a reel and the other end free.  Given $\tbv$ and $\tbq$, the angle of attack of the cable is determined, and \eqref{tproj} relates the tension $\sigma$ and the payout speed $T$.  The tension is a linear function of $s$.  If one end is free, the constant obtained by integrating \eqref{tproj} is used to set the tension to zero there.  The payout speed does not affect the angle of attack, but is linearly related to the tension at the tow point, and there is a particular speed at which an untensioned cable ($\sigma=0$) will be pulled off of the reel.  
For collinear $\tbv$ and $\tbq$, parallel straight lines can move at any velocity, while for perpendicular straight lines there is a single velocity satisfying $\nu_\perp\tbv = \tbq$.  Either case admits any value of axial flow $T$, although in the parallel case the distinction between $\tbv$ and $T\uvc{t}$ is ambiguous.
Similar calculations can be performed for nonlinear drag forces, when 
 the geometry is linear like this.

\subsection{The planar problem}\label{planarproblem}

\begin{figure}[H]
	\centering
	\begin{tikzonimage}[scale=0.6]{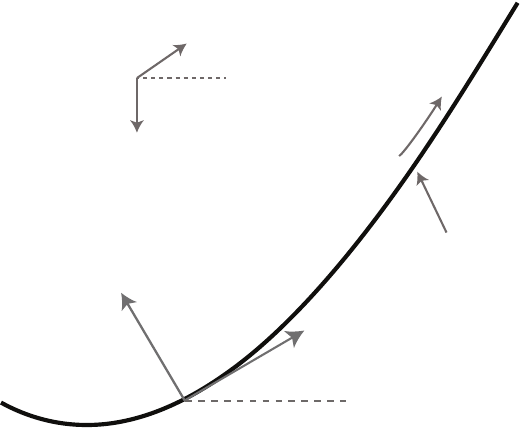} 
		\draw (0.733,0.726) node[]{\footnotesize $T \uvc{t}$ };
		\draw (0.85,0.4) node[]{\footnotesize $\bX (s,t)$};
		\draw (0.605,0.235) node[]{\footnotesize $\uvc{t}$};
		\draw (0.22,0.35) node[]{\footnotesize $\uvc{n}$};
		\draw (0.51,0.116) node[]{\footnotesize $\theta$};
		\draw (0.25,0.65) node[]{\footnotesize $-q \uvc{z}$};		
		\draw (0.31,0.9) node[]{\footnotesize $\bv$};
		\draw (0.37,0.856) node[]{\scriptsize $\phi$};
	\end{tikzonimage}
\caption{A planar catenary $\bX$ with orthonormal frame $(\uvc{t},\uvc{n})$.  The translational velocity and body force have been projected onto the $\uvc{t}$-$\uvc{n}$ plane to obtain $\bv$ and $-q\uvc{z}$, respectively.  The flow $T\uvc{t}$ is unaffected.  The parameter $\phi$ is the fixed angle between the projected velocity $\bv$ and the $\uvc{x}$ direction (not labeled, but indicated by dotted lines) perpendicular to $\uvc{z}$.  The body $\bX$ can be described by the local angle $\theta$ between its tangent $\uvc{t}$ and $\uvc{x}$.}
\label{curve2d}
\end{figure}

We may use the planarity of generic solutions to simplify their description.  Note that while the curves $\bX$ lie in a plane, this plane does not, in general, contain the body force or the direction of translation.  Specification of the plane requires use of the binormal force balance \eqref{bproj} and an additional condition that will typically be provided by a boundary condition.  We reserve discussion of this process until Section \ref{towing}, where we discuss a boundary value problem for towing a cable with specified endpoints.  For now, we note that the specification of the binormal in accordance with these requirements leaves us to consider the balance of forces in the plane of the curve.  
This situation is illustrated in Figure \ref{curve2d}.  The quantities $\bv$ and $-q\uvc{z}$ 
are the projections of the translational velocity $\tbv$ and body force $\tbq$ onto the plane of the curve.  A planar curve is specifiable in terms of a single function $\theta(s)$, the tangential or Whewell angle, which we measure with respect to what we denote as the $\uvc{x}$ direction, perpendicular to the $\uvc{z}$ direction.  We also specify a parameter $\phi$, the fixed angle between $\uvc{x}$ and the projected velocity $\bv$.  These elements are shown schematically in Figure \ref{curve2d}.

Let us rescale quantities using an arbitrary reference length scale $L$ and a velocity scale \\ \mbox{$\Gamma \equiv \sqrt{\bv\cdot\bv + T^2 + \left(q/\nu_\perp \right)^2}\,$}, so that $s \rightarrow s/L$, $|\bv| \rightarrow |\bv|/\Gamma$, $T \rightarrow T/\Gamma$, $q \rightarrow q/(\nu_\perp\Gamma)$, and $m \rightarrow m\Gamma/(\nu_\perp L)$.  These rescalings lead to the constraint $T^2 + \bv\cdot\bv + q^2 = 1$.  They will need to be revisited in Section \ref{reeling}, when we discuss a limiting case with vanishing $\bv$, for which $\nu_\perp$ is meaningless.  

The planar, dimensionless form of \eqref{force1} is
\begin{equation} \label{dimlesseq}
mT^2\partial_s\theta\uvc{n} + R_{\nu}\left(T+\bv\cdot\uvc{t}\right)\uvc{t} + \bv\cdot\uvc{n}\uvc{n} = \partial_s\left(\sigma\uvc{t}\right) - q\uvc{z} \, ,
\end{equation}
where a new parameter, the drag anisotropy ratio $R_\nu \equiv \nu_\pll/\nu_\perp$, appears.
Using the relations
\begin{equation}\label{relations}
	\left(\begin{array}{c}
		\uvc{t}\\
		\uvc{n}
		\end{array}\right) = \left(\begin{array}{cc}
						\cos \theta & \sin \theta \\ 
						-\sin \theta & \cos \theta
						\end{array}\right)
							\left(\begin{array}{c}
								\uvc{x}\\
								\uvc{z}
								\end{array}\right) \, ,
\end{equation}
we obtain the projections of \eqref{dimlesseq} onto the two osculating frame vectors $\uvc{t}$ and $\uvc{n}$,
\begin{align}
\partial_s\sigma &= R_\nu\left(T + \bv\cdot\uvc{x}\cos\theta + \bv\cdot\uvc{z}\sin\theta\right) + q\sin\theta \label{dimlesstproj} \, , \\ 
\sigma\partial_s\theta &= mT^2\partial_s\theta + \left(-\bv\cdot\uvc{x}\sin\theta + \bv\cdot\uvc{z}\cos\theta\right) + q\cos\theta \label{dimlessnproj} \, .
\end{align}
The projections $\bv\cdot\uvc{x}$ and $\bv\cdot\uvc{z}$ appearing in these equations are constants related to the magnitude $|\bv|$ and direction $\phi$ of the translational velocity.  The stress $\sigma$ can be eliminated to obtain a single second order equation for $\theta$,
\begin{equation}
\partial_s\left[ \frac{ \left(-\bv\cdot\uvc{x}\sin\theta + \bv\cdot\uvc{z}\cos\theta\right) + q\cos\theta }{ \partial_s\theta } \right] =  R_\nu\left(T + \bv\cdot\uvc{x}\cos\theta + \bv\cdot\uvc{z}\sin\theta\right) + q\sin\theta \, .
\end{equation}
Multiplying by the inverse of the bracketed quantity on the left, we obtain
\begin{equation} \label{totalder}
\partial_s\ln\left| \frac{ \left(-\bv\cdot\uvc{x}\sin\theta + \bv\cdot\uvc{z}\cos\theta\right) + q\cos\theta }{ \partial_s\theta } \right| = \left[ \frac{ R_\nu \left(T + \bv\cdot\uvc{x}\cos\theta + \bv\cdot\uvc{z}\sin\theta\right) + q\sin\theta } { \left(-\bv\cdot\uvc{x}\sin\theta + \bv\cdot\uvc{z}\cos\theta\right) + q\cos\theta } \right] \partial_s\theta \, .
\end{equation}
Both sides of this equation are total $s$-derivatives.  Before proceeding, let us introduce a second rescaling, whose purpose is to simplify the expressions that will result from integrating the equation.  We define $\Delta \equiv \sqrt{(\bv \cdot \uvc{x})^2 + (\bv \cdot \uvc{z} + q)^2}\,$, $\bar{U} \equiv \bv \cdot \uvc{x}/\Delta$, $\bar{W} \equiv \bv \cdot \uvc{z}/\Delta$, $\bar{T} \equiv R_\nu T/\Delta$, $\bar{Q} \equiv q/\Delta$, which leads to the constraint $\bar{U}^2 + \left(\bar{W}+\bar{Q}\right)^2 = 1$.  We may now write the integrated equation \eqref{totalder} as 
\begin{align}
\ln\left| \frac{ g(\theta) }{ \partial_s\theta } \right| &= \tilde{C} - \bar{T}\ln \left| f(\theta) \right| -\left(k-1\right)\ln\left| g(\theta) \right| - \left(1 - R_\nu\right)\bar{U}\bar{Q}\theta \, , \nonumber \\
\mathrm{where}\quad  f(\theta) &= \frac{ 1- \bar{U} - \left(\bar{W} + \bar{Q}\right)\tan\frac{\theta}{2} }{ 1+  \bar{U} + \left(\bar{W} + \bar{Q}\right)\tan\frac{\theta}{2} }\, , \\
g(\theta) &= -\bar{U}\sin\theta + \left(\bar{W} + \bar{Q} \right)\cos\theta \, , \\
k &= 1+R_\nu \bar{U}^2 + \left(R_\nu \bar{W} + \bar{Q}\right)\left(\bar{W} + \bar{Q} \right) \, , \label{keq}
\end{align}
and a term proportional to $\ln\Delta$ has been absorbed into the constant of integration $\tilde{C}$.  Rearranging, we obtain a single first order equation for $\theta$,
\begin{equation}\label{thetaeq}
	\partial_s\theta = Ce^{\left(1-R_\nu\right)\bar{U}\bar{Q} \theta} \left| f(\theta) \right|^{\bar{T}} \left| g(\theta) \right|^k \, .
\end{equation}
Once $\theta$ is determined by quadrature, $\sigma$ follows directly from \eqref{dimlessnproj} as
\begin{equation}\label{sigmaeq}
	\sigma = mT^2 + \frac{\Delta g(\theta)}{\partial_s\theta} \, ,
\end{equation}
and the Cartesian position vector $\bX$ follows by integration of the defining relation between tangent vector and tangential angle, $\partial_s\bX = \uvc{x}\cos\theta + \uvc{z}\sin\theta$.

The shape equation \eqref{thetaeq} is a five-parameter family of dynamical systems, written in terms of six constrained 
parameters.  A sixth independent parameter, the mass coefficient $m$, appears only in the stress equation \eqref{sigmaeq} and does not affect the geometry of the solutions.  We note, however, that the density of a massive cable acted on by gravity will indeed affect its geometry, through the body force parameter $\bar{Q}$.  Specifying the horizontal and vertical velocity parameters $\bar{U}$ and $\bar{W}$ is akin to specifying the translational velocity magnitude $|\bv|$ and direction $\phi$, and in the remainder of this paper we will use whichever description is simplest.  The other quantities appearing are the axial flow parameter $\bar{T}$, the drag anisotropy $R_\nu$, and a scaling constant $C$ which is inherited from the limiting case of the classical catenaries.  Due to the symmetry of the problem, we need only consider positive values of $\bar{U}$ and $\bar{Q}$.  We will also restrict $R_\nu$ to the physically realistic range between zero (purely normal drag) and unity (isotropic drag).

The right hand side of equation \eqref{thetaeq} is the product of an aperiodic exponential, a $2\pi$-periodic function $\left| f(\theta) \right|$, and a $\pi$-periodic function $\left| g(\theta) \right|$.  The $\pi$-periodicity of $\left| g(\theta) \right|$ is punctuated by zeroes at values $\theta_0$ for which $\bar{U}\tan\theta_0 = \bar{W} + \bar{Q}\,$.  Excepting special cases, which will be discussed in Section \ref{organization}, individual orbits generated by \eqref{thetaeq} will encounter either fixed points or poles at these locations, and cannot subtend an angle greater than $\pi$.  While it is not obvious by inspection, $f(\theta)$ has as its distinguished points alternating zeroes and poles that coincide with the zeroes of $g(\theta)$.  A generic phase portrait, drawn in the $\theta$-$\partial_s\theta$ plane by varying the constant of integration $C$, will consist of a series of lobes separated by fixed points and/or poles.  It turns out that a single lobe, subtending an angle of $\pi$, suffices to describe all the possible generic shape equilibria.  This is because a combination of lobe shift, sign change in axial velocity, and rescaling brings orbits in phase space into coincidence,
\begin{align}\label{scaleax}
	\partial_s \theta \left(\theta_0 - \pi + \theta, -\bar{T}\right) &= e^{-(1-R_\nu)\bar{U}\bar{Q}\pi} \left(\frac{1+\bar{U}}{1-\bar{U}} \right)^{\bar{T}}  \partial_s \theta \left(\theta_0 + \theta , \bar{T}\right) \, , \\
	\mathrm{where}\quad \theta_0 &= \arctan\tfrac{\bar{W}+\bar{Q}}{\bar{U}}\; \in \left[-\tfrac{\pi}{2}, \tfrac{\pi}{2}\right] \, ,
\end{align}
and reflection about the $\theta$ axis in phase space, achieved through a sign change in the scaling coefficient $C \rightarrow -C$, serves to identify configurations $\bX$ in physical space.
Details may be found in Appendix \ref{zeroesandpoles}.

While the full form of equation \eqref{thetaeq} appears complicated, there are many situations in which it can simplify.  For example, if $\bar{U}$ vanishes, so that the body's translational velocity is collinear with the body force, the exponential term reduces to unity, and $g(\theta)$ simplifies to $\pm\cos\theta$.  Similarly, if the axial flow $\bar{T}$ vanishes, so that the body translates rigidly, the complicated dependencies of $f(\theta)$ become irrelevant.  Curves for which both of these conditions hold can be found using an alternate method of solution.  This is presented in Appendix \ref{collinear} along with representations of the classical catenaries ($k=2$) in $\theta$-$\partial_s\theta$ and $\kappa$-$\partial_s\kappa$ spaces.

\section{Solutions}\label{solutions}

Before organizing our findings in Section \ref{organization}, we present several examples in Section \ref{examples} that represent the spectrum of typical solutions.

\subsection{Examples of generic shapes}\label{examples}

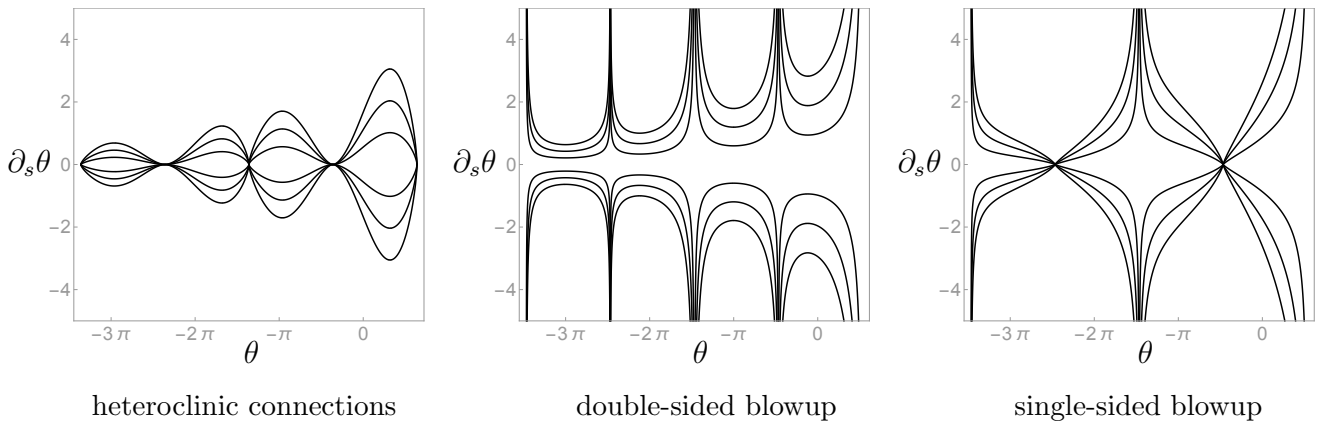
\begin{figure}[H]	
	
	\centering
	\subcaptionbox*{\hspace*{3mm} heteroclinic connections}[.31\linewidth][c]{%
		\hspace{-6mm}		\begin{tikzonimage}[scale=0.35]{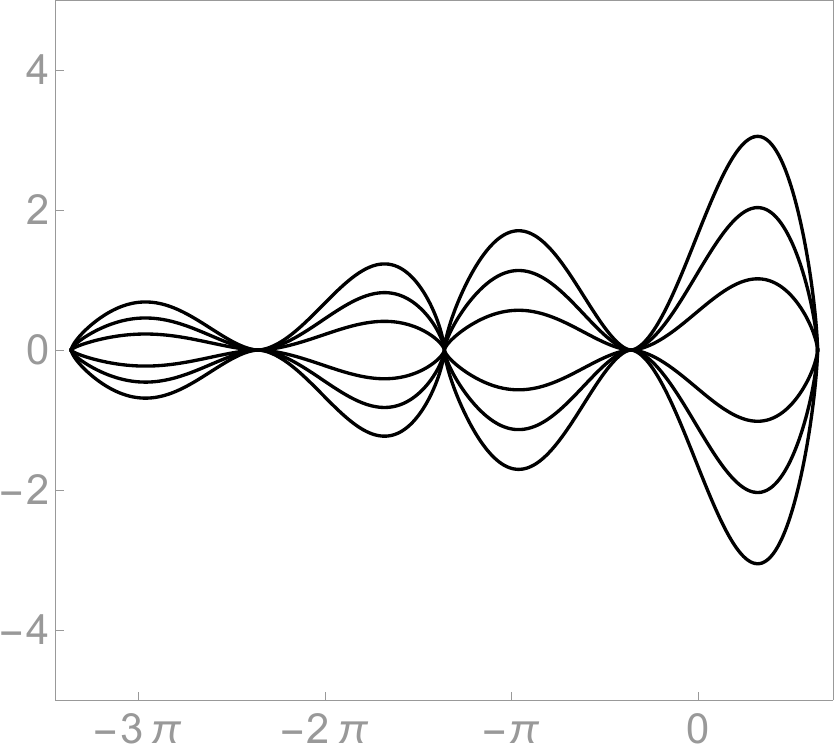} 
			\draw  (-0.05,0.54) node[]{$\partial_s \theta$};
			\draw  (0.54,-0.02) node[]{$\theta$};
		\end{tikzonimage}}\quad	
		\subcaptionbox*{\hspace*{1.1cm} double-sided blowup}[.31\linewidth][c]{%
			\hspace*{-3.2mm}			\begin{tikzonimage}[scale=0.35] {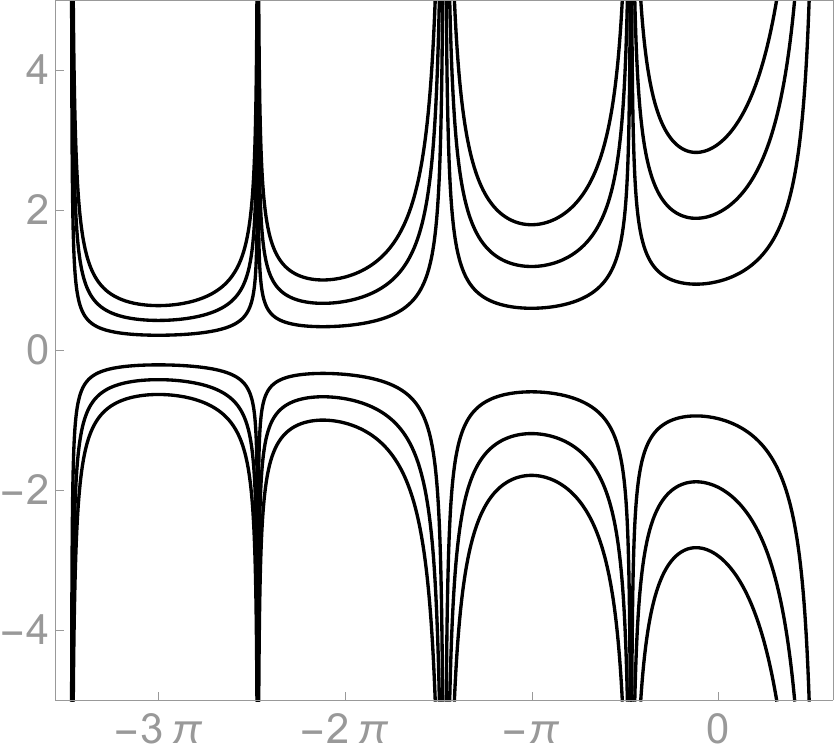} 
				\draw  (-0.05,0.54) node[]{$\partial_s \theta$};
				\draw  (0.55,-0.02) node[]{$\theta$};
			\end{tikzonimage}}\quad	
			\subcaptionbox*{\hspace*{1.1cm} single-sided blowup}[.31\linewidth][c]{%
				\hspace*{-1.5mm}		\begin{tikzonimage}[scale=0.35]{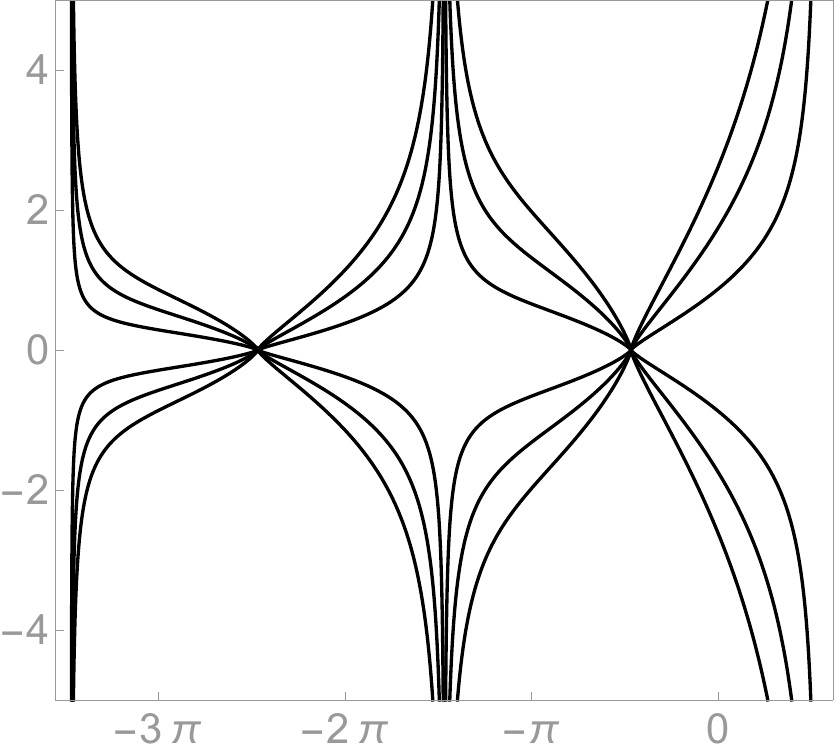} 
					\draw  (-0.05,0.54) node[]{$\partial_s \theta$};
					\draw  (0.55,-0.02) node[]{$\theta$};				
				\end{tikzonimage}}\quad

				\caption{The three generic types of phase portraits generated by equation \eqref{thetaeq}, using $C = \pm (1,2,3)$ and the following other parameters.  Heteroclinic connections: $\phi=-75^\circ$, $|\bv| = 0.8$, $T=0.5$, $q = 0.33$, $R_\nu = 0.5$.  Double-sided blowup: $\phi=-88^\circ$, $|\bv| = 0.8$, $T=0.3$, $q = 0.52$, $R_\nu = 0.1$.  Single-sided blowup: $\phi=-88^\circ$, $|\bv| = 0.76$, $T=0.4$, $q = 0.51$, $R_\nu = 0.4$.  See Figure \ref{curve2d} and the adjacent text for explanations of the parameters.}
				\label{types}	
				
\end{figure}

Almost all phase portraits generated by equation \eqref{thetaeq} are of three types, with either fixed points, poles, or alternating fixed points and poles occurring at intervals of $\pi$.  A fixed point solution is a straight line, with fixed angle $\theta_0$.  All other orbits are qualitatively the same for each type of phase portrait.  They are, respectively, heteroclinic connections, double-sided and single-sided blowup solutions, where the quantity that blows up is the curvature $\partial_s\theta$ at finite angle $\theta$ and arc length $s$.   Figure \ref{types} shows four lobes from each type of phase portrait.  The effect of the term exponential in $\theta$ is apparent.

In this section, we present several examples of the three qualitatively different types of solutions.
When displaying and comparing curves generated by different parameters, the scaling constant $C$ is not particularly interesting.  However, simply keeping $C$ fixed does not always result in illustrative comparisons.  
We wish to compare several curves of equal length, while assuring display of a result corresponding to as much of the orbit as is possible.  We achieve this by scaling solution curves such that a length of unity is associated with subtension of an angle slightly less than $\pi$.  This requires $C$ to vary along with the other parameters.  The condition we impose leads to a pair of curves corresponding to positive and negative values of $C$; one of these curves will be purely in tension ($\sigma \ge 0$), and this is the curve we choose.  Its partner, which may be partly or entirely in compression, and thus unstable, has a configuration obtained by rotation of the tensile shape by $\pi$ radians.
This representation is an efficient means of displaying families of curves, but often does not correspond to the variation of conditions arising naturally in a realistic boundary value problem.  If this is not kept in mind, some of the curve comparisons may seem counterintuitive.  

Because orbits approach fixed points asymptotically at infinite $s$, we cannot integrate the full $\pi$ radians and obtain finite curves for display.  Likewise, integrating up to a curvature singularity is numerically impractical.  Instead, we choose a small cutoff angle $\epsilon$ so that a typical shape will be represented as a curve of length unity subtending an angle of $\pi - 2\epsilon$ from $\theta_0 - \pi + \epsilon$ to $\theta_0 - \epsilon$.
The choice of cutoff affects the magnitudes of the curvature and stress, so care must be taken in comparing solutions with different values of $\epsilon$.  All of the curves in this section (\ref{examples}) were generated with a cutoff $\epsilon = 0.05$.
The normalization is achieved by using the fact that the length of the curve is
\begin{align}
	s &= \int^\theta \!\!\! d\tilde{\theta}\, \frac{1}{\partial_s\tilde{\theta}} \, ,\\
	&= \frac{1}{C}\int^\theta \!\!\! d\tilde{\theta}\, \frac{C}{\partial_s\tilde{\theta}} \, ,
\end{align}
so that a length of unity is achieved by setting
\begin{align}
	C = \int_{\theta_0 - \pi + \epsilon}^{\theta_0 - \epsilon} \!\!\! d\tilde{\theta}\, \frac{C}{\partial_s\tilde{\theta}} \, ,
\end{align}
first substituting the expression obtained from \eqref{thetaeq} for the integrand on the right hand side.  With this value of $C$, we integrate the system described in Section \ref{planarproblem} from $s=0$ to $s=1$ or some other desired length.  

Each curve $\bX$ is shown alongside its phase space orbit and its stress distribution.  The orbits in the figures will in general overlap, as they are extracted from different phase portraits.  We set the mass coefficient $m=1$ for all solutions in this paper, as it has no effect on the shapes, and simply shifts the stresses by a constant $mT^2$.  A dotted line on the end of a curve indicates that the curve could be continued out to infinite length, while an open circle indicates that the curve terminates.  Of course, one can always cut a solution subtending an angle less than $\pi$ from any of these mother curves.

\begin{figure}[H]
	\hspace*{0.85cm}
	\scalebox{1}{%
		\begin{overpic}[scale=0.5]{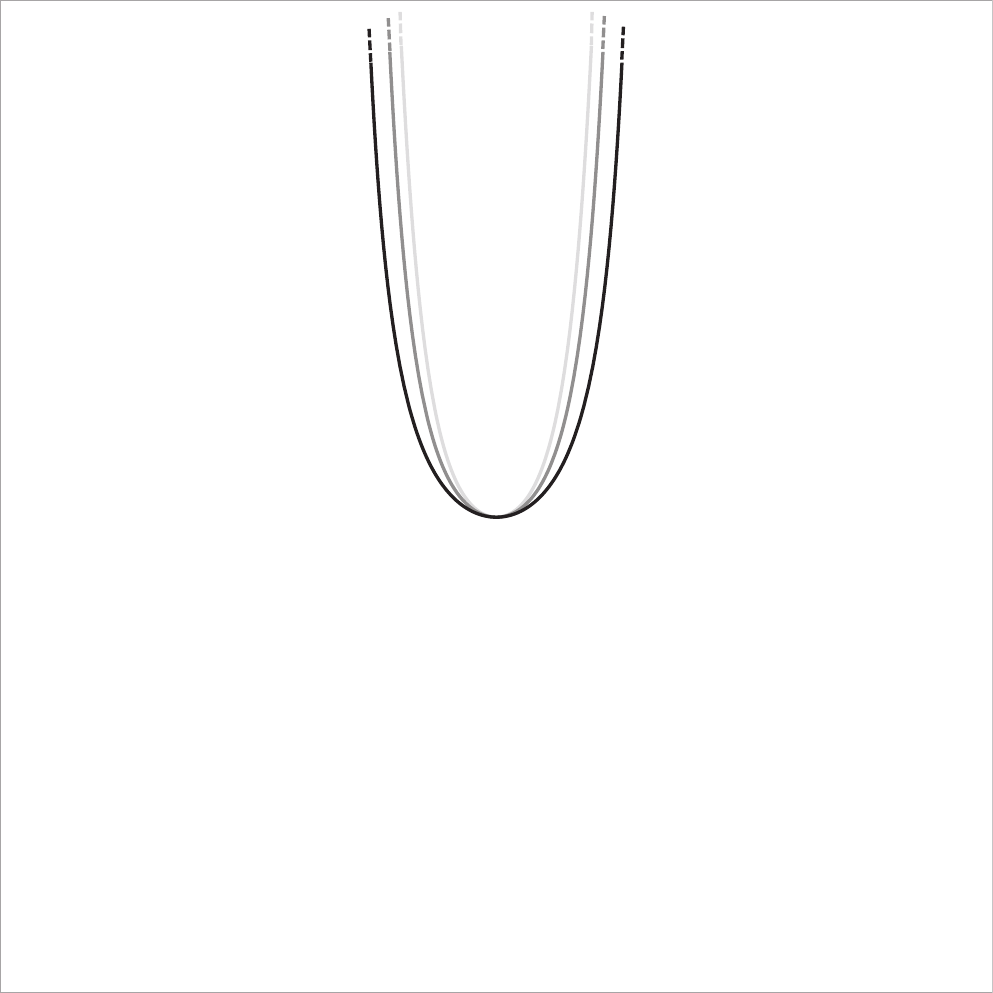}
			\put(120,-10){\includegraphics[scale=0.27]{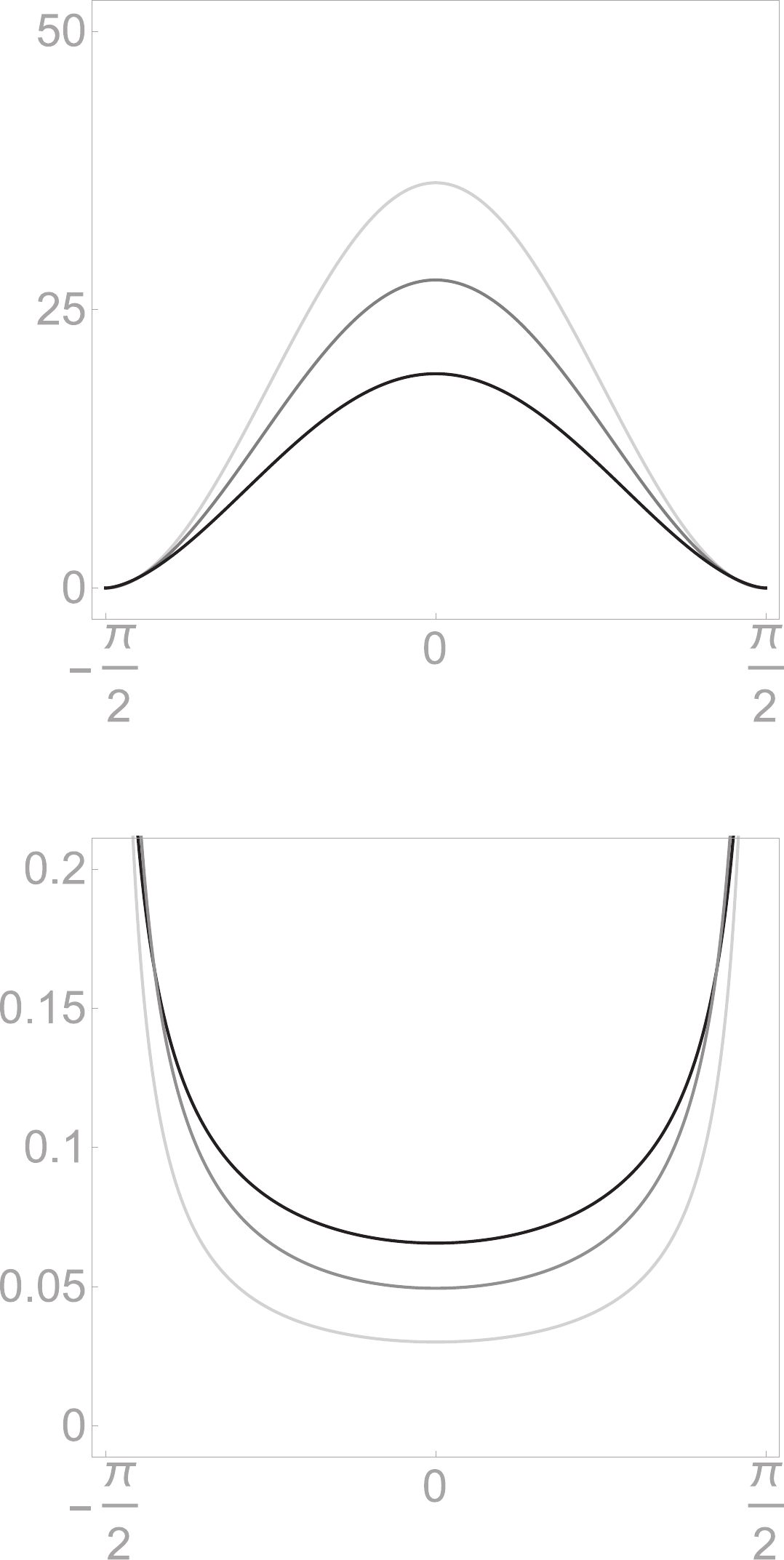}}
			\put(152,-13.8){$\theta$}
			\put(152,49){$\theta$}
			\put(114,81){$\partial_s \theta$}
			\put(117,19.2){$\sigma$}
			\put(6.5,70){\color{black}\vector(0,1){15}}
			\put(3,86){$\uvc{z}$}
			\put(7,86){$\uvc{v}$}
		\end{overpic}}
		\vspace*{1cm}
		\caption{Configurations (left), orbits (upper right), and stresses (lower right) for catenaries rigidly translating with velocity $\bv$ under the action of a body force $-q\uvc{z}$.  Darker curves indicate higher velocities.  Parameters are $\phi = 90^\circ$, $|\bv| = $ (\textcolor{Lgr}{0.1}, \textcolor{Mgr}{0.5}, 0.95), $T=0$, $q =$ (\textcolor{Lgr}{0.99}, \textcolor{Mgr}{0.87}, 0.31), $R_\nu = 0.5$, $m=1$, $\epsilon=0.05$.  Infinite length curves generated by heteroclinic orbits.
		}
		\label{curve90}
	\end{figure}

\begin{figure}[H]
	\hspace*{0.85cm}
	\scalebox{1}{%
		\begin{overpic}[scale=0.5]{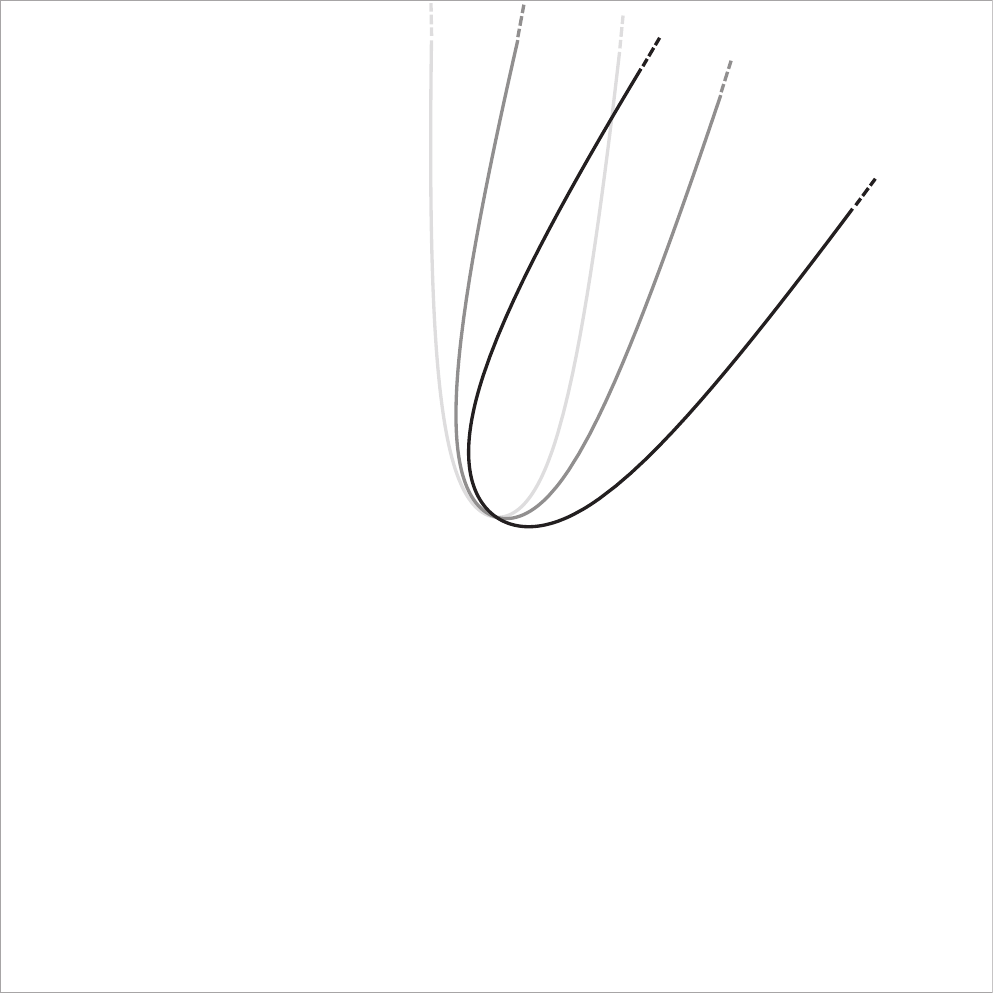}
			\put(120,-10){\includegraphics[scale=0.27]{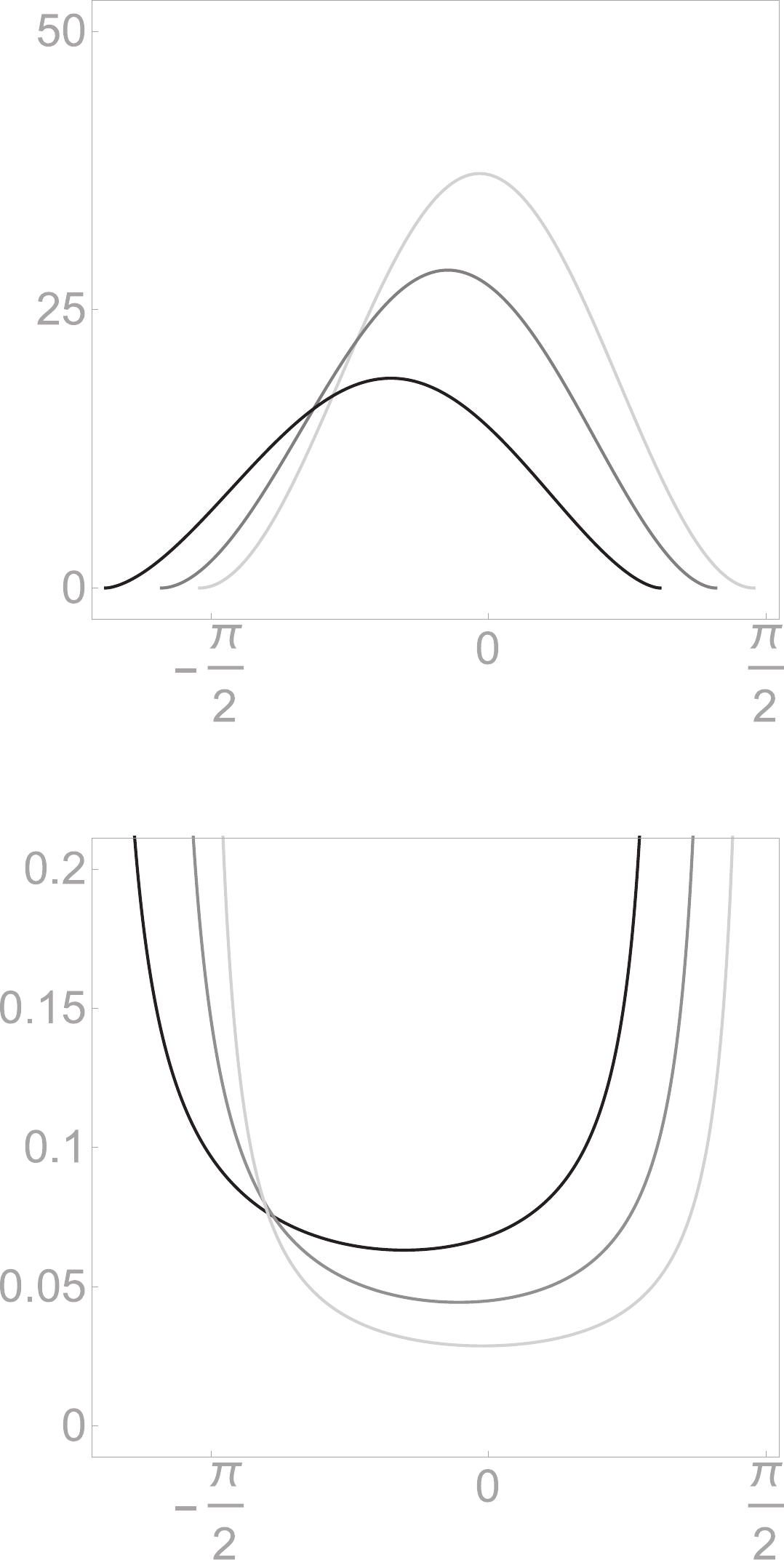}}
			\put(152,-13.8){$\theta$}
			\put(152,49){$\theta$}
			\put(114,81){$\partial_s \theta$}
			\put(117,19.2){$\sigma$}
			\put(6.5,70){\color{black}\vector(0,1){15}}
			\put(5.5,86){$\uvc{z}$}
			\put(6.5,70){\color{black}\vector(1,1){11}}
			\put(17.5,81){$\uvc{v}$}
		\end{overpic}}
		\vspace*{1cm}
		\caption{As in prior figure, with $\phi = 45^\circ$.}
		\label{curve45}
	\end{figure}

\begin{figure}[H]
		\hspace*{0.85cm}
		\scalebox{1}{%
			\begin{overpic}[scale=0.5]{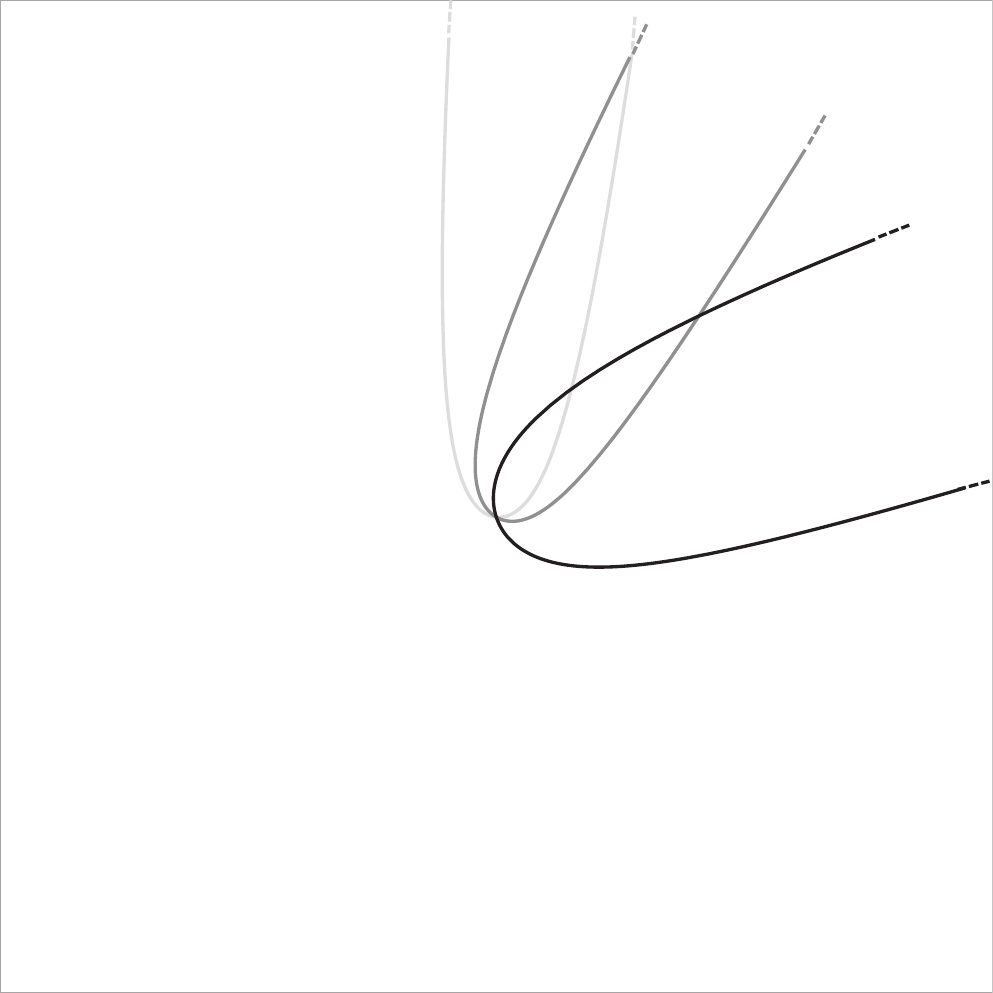}
				\put(120,-10){\includegraphics[scale=0.27]{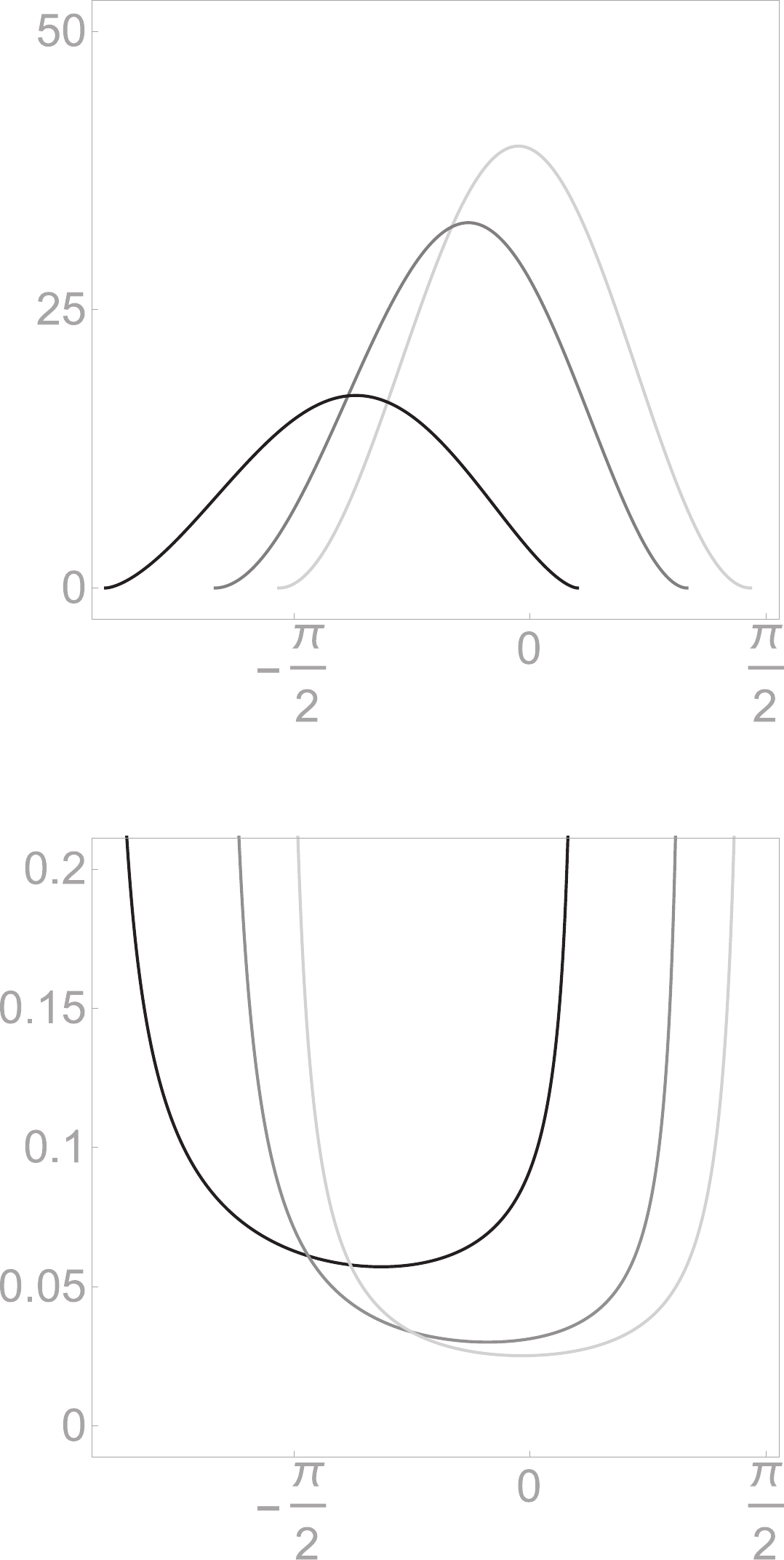}}
				\put(152,-13.8){$\theta$}
				\put(152,49){$\theta$}
				\put(114,81){$\partial_s \theta$}
				\put(117,19.2){$\sigma$}
				\put(6.5,70){\color{black}\vector(0,1){15}}
				\put(5.5,86){$\uvc{z}$}
				\put(6.5,70){\color{black}\vector(1,0){15}}
				\put(22.5,69){$\uvc{v}$}
			\end{overpic}}
			\vspace*{1cm}
			\caption{As in prior figure, with $\phi = 0^\circ$.} 
			\label{curve0}
		\end{figure}

\begin{figure}[H]
			\hspace*{0.85cm}
			\scalebox{1}{%
				\begin{overpic}[scale=0.5]{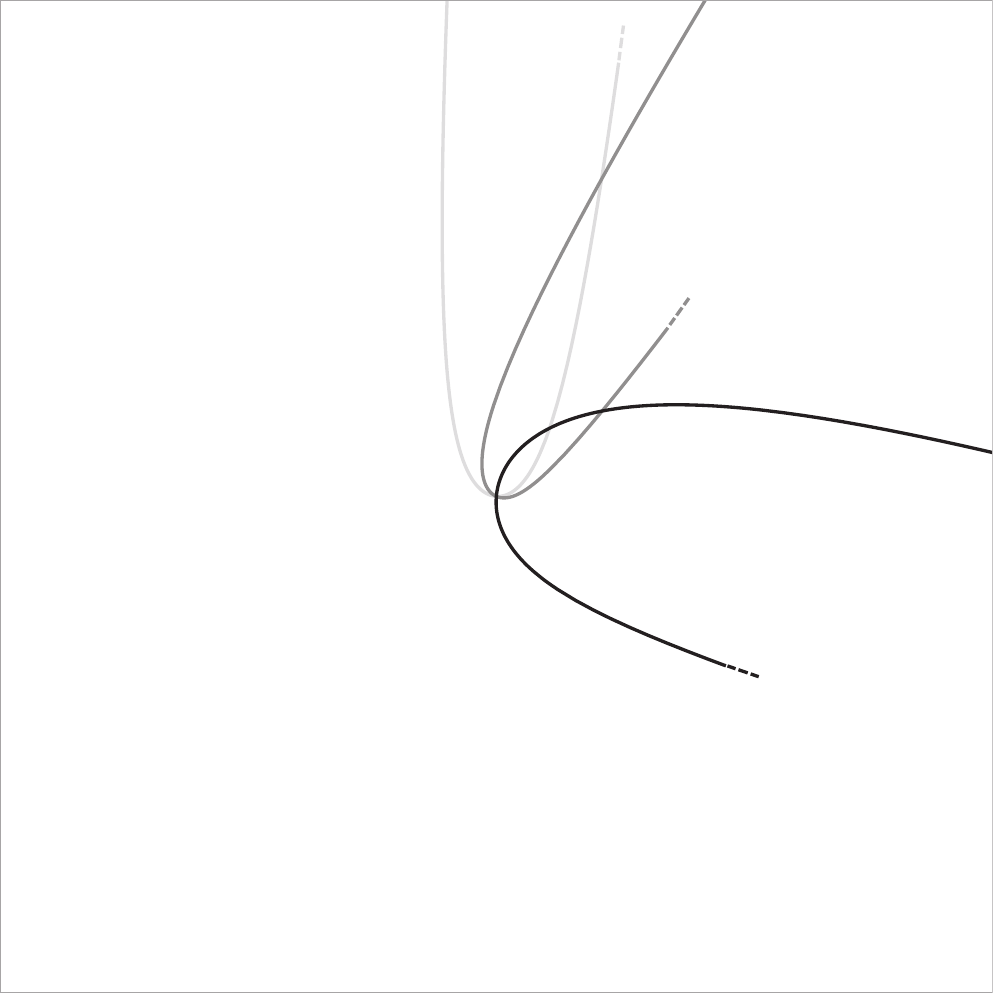}
					\put(120,-10){\includegraphics[scale=0.27]{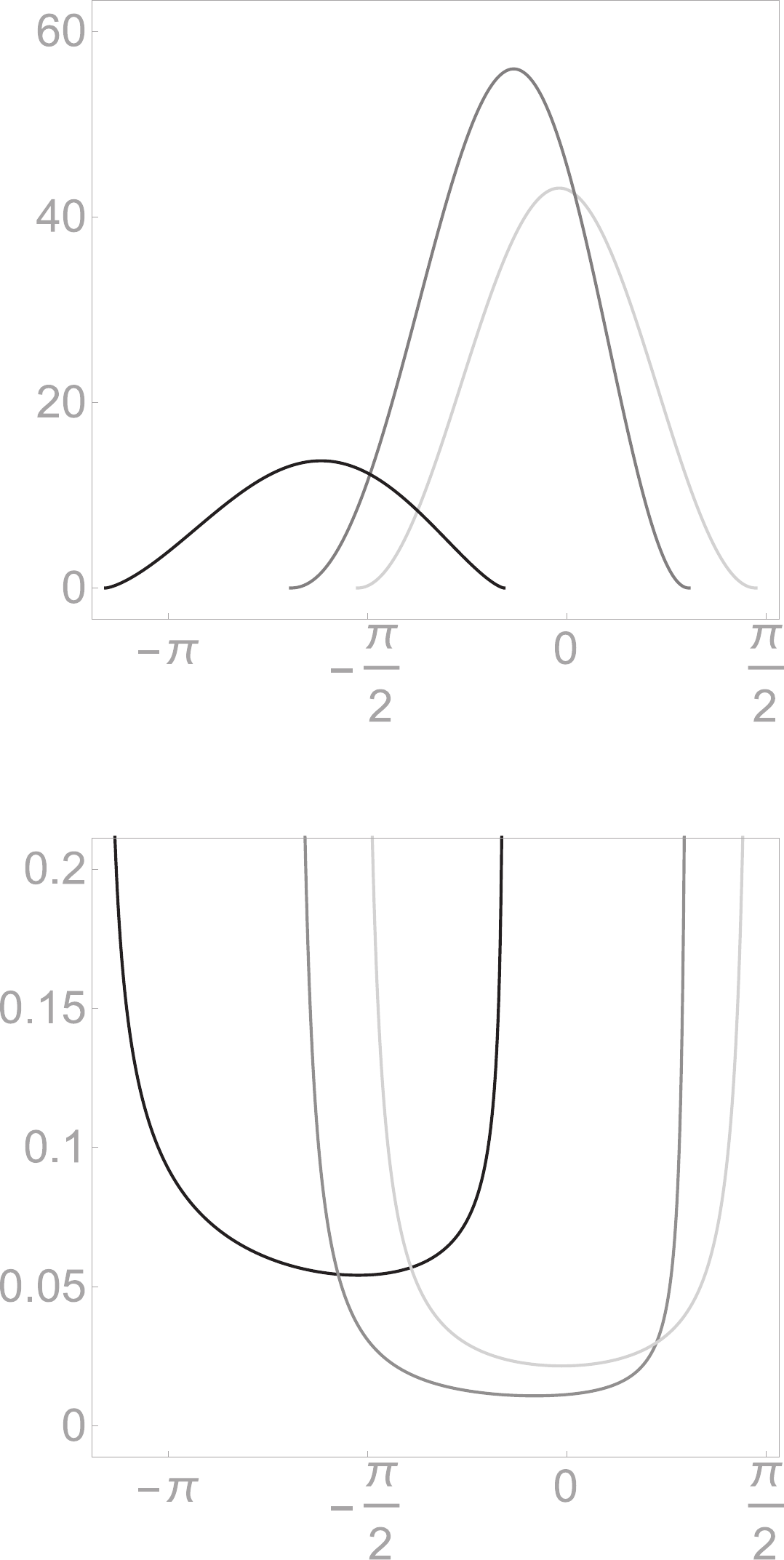}}
					\put(152,-13.8){$\theta$}
					\put(152,49){$\theta$}
					\put(114,81){$\partial_s \theta$}
					\put(117,19.2){$\sigma$}
					\put(6.5,70){\color{black}\vector(0,1){15}}
					\put(5.5,86){$\uvc{z}$}
					\put(6.5,70){\color{black}\vector(1,-1){11}}
					\put(17.5,55){$\uvc{v}$}
				\end{overpic}}
				\vspace*{1cm}
				\caption{As in prior figure, with $\phi = -45^\circ$.}
				\label{curvem45}
			\end{figure}

\begin{figure}[H]
	\hspace*{0.85cm}
		\scalebox{1}{%
			\begin{overpic}[scale=0.5]{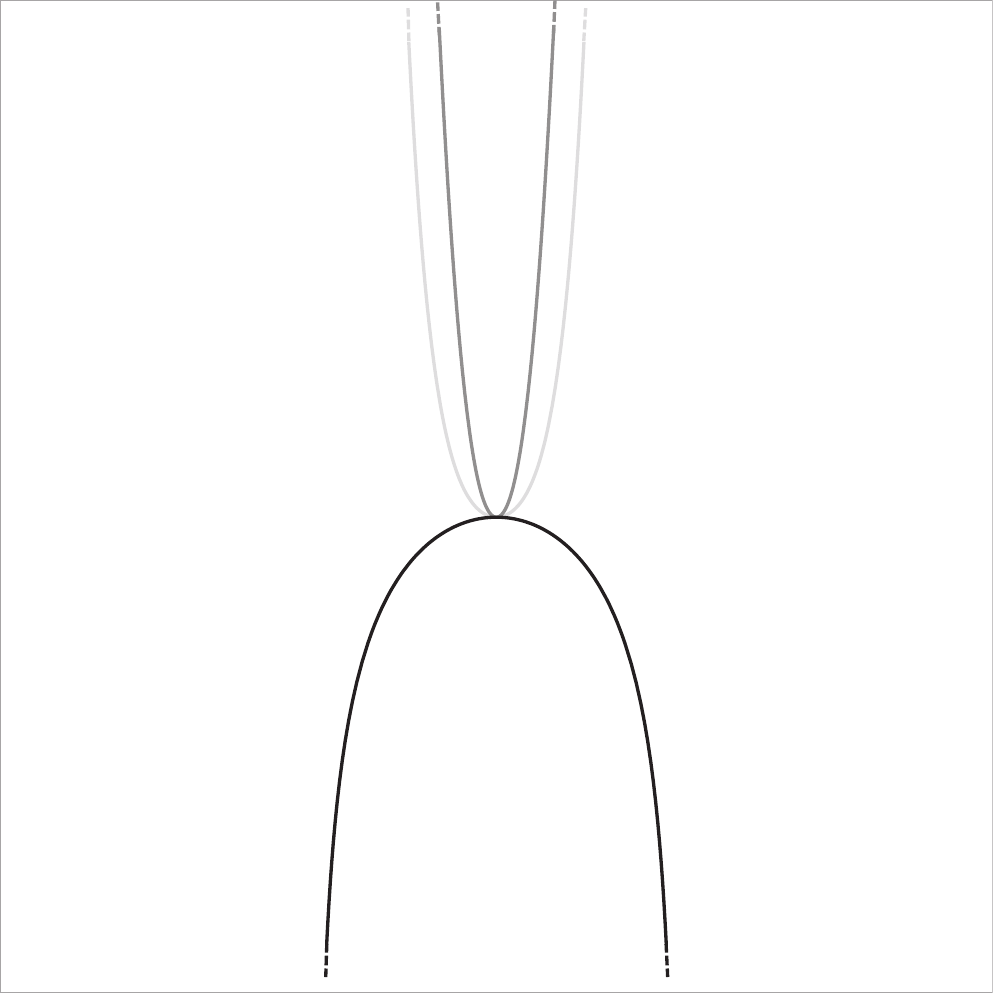}
				\put(120,-10){\includegraphics[scale=0.27]{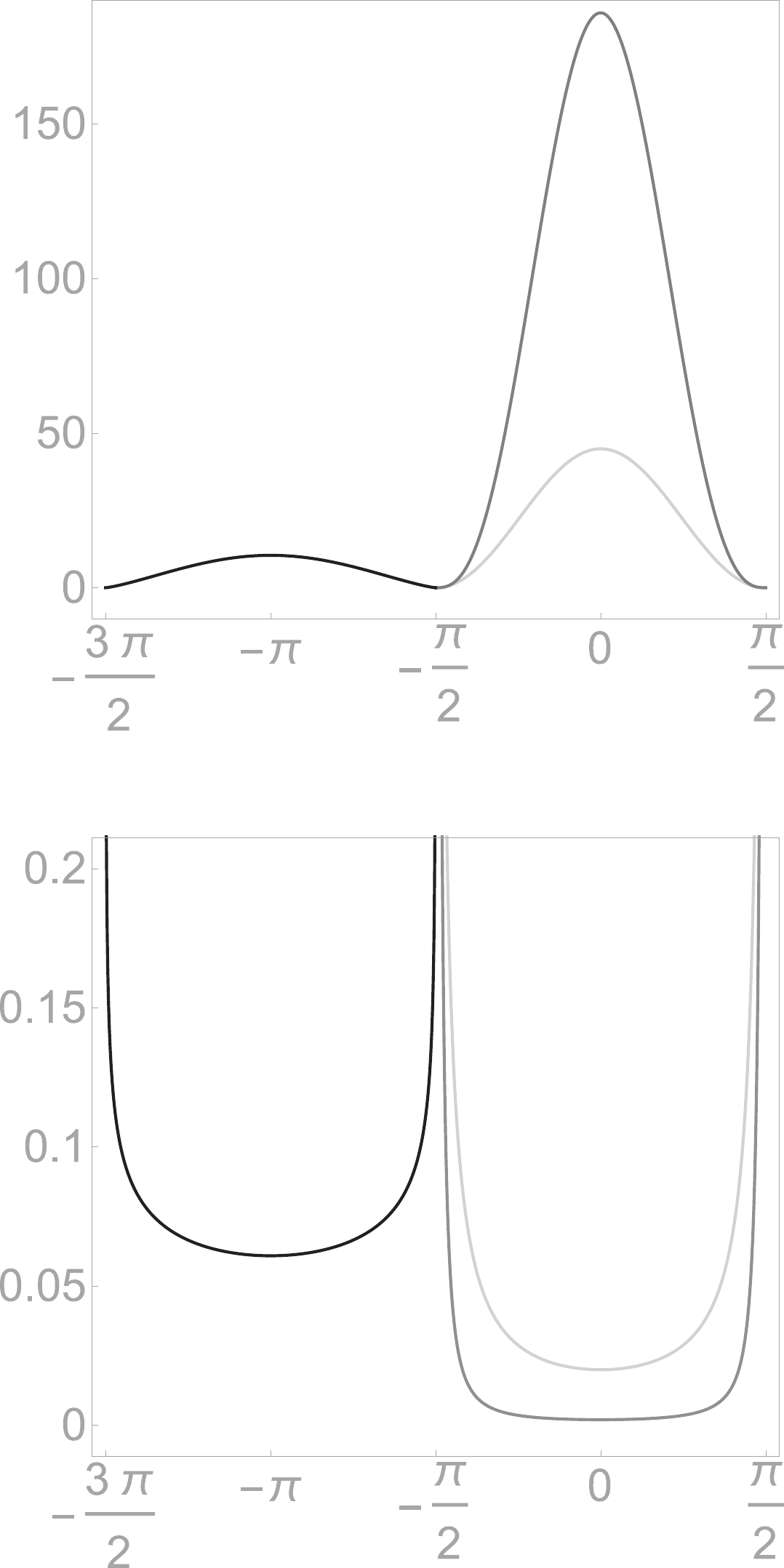}}
				\put(152,-13.8){$\theta$}
				\put(152,49){$\theta$}
				\put(114,81){$\partial_s \theta$}
				\put(117,19.2){$\sigma$}
				\put(6.5,70){\color{black}\vector(0,1){15}}
				\put(5.5,86){$\uvc{z}$}
				\put(6.5,70){\color{black}\vector(0,-1){15}}
				\put(5.3,50.5){$\uvc{v}$}
			\end{overpic}}
			\vspace*{1cm}
			\caption{As in prior figure, with $\phi = -90^\circ$.}
			\label{curvem90}
		\end{figure}

Figures \ref{curve90}-\ref{curvem90} show catenaries rigidly moving at three different speeds in five different directions, with drag anisotropy taking the Stokes flow value $R_\nu = 0.5$.  These solutions are all heteroclinic connections corresponding to infinitely long configurations.  Not surprisingly, the alignment of the curves is set by competition between the body force $-q\uvc{z}$ and translational velocity $|\bv|\uvc{v}$.  
There are also subtle differences in the configurations generated by a uniform body force or anisotropic drag forces, best seen in Figure \ref{curve90} where the two vectors are aligned.
The orbits in Figure \ref{curvem45} are significantly asymmetric, as is quite apparent from inspection of the configurations generated with a symmetric cutoff $\epsilon$ from an infinitely long mother curve.  When looking at these figures, it is important to recall that the solutions are generated by constraining the curves to be of unit length while subtending an angle just shy of $\pi$ radians.  This is not equivalent to fixing the location of the curve ends and varying the velocity.  The latter more realistic boundary value problem will be addressed in Section \ref{bvp}.  The tension in all of these curves increases indefinitely as the curves go off to infinity in either direction.  It is intuitively plausible that holding or towing an infinitely long curve should require infinite force, but in Section \ref{organization} we will show that this is not always the case.

\begin{figure}[H]
	\hspace*{0.85cm}
	\scalebox{1}{%
		\begin{overpic}[scale=0.5]{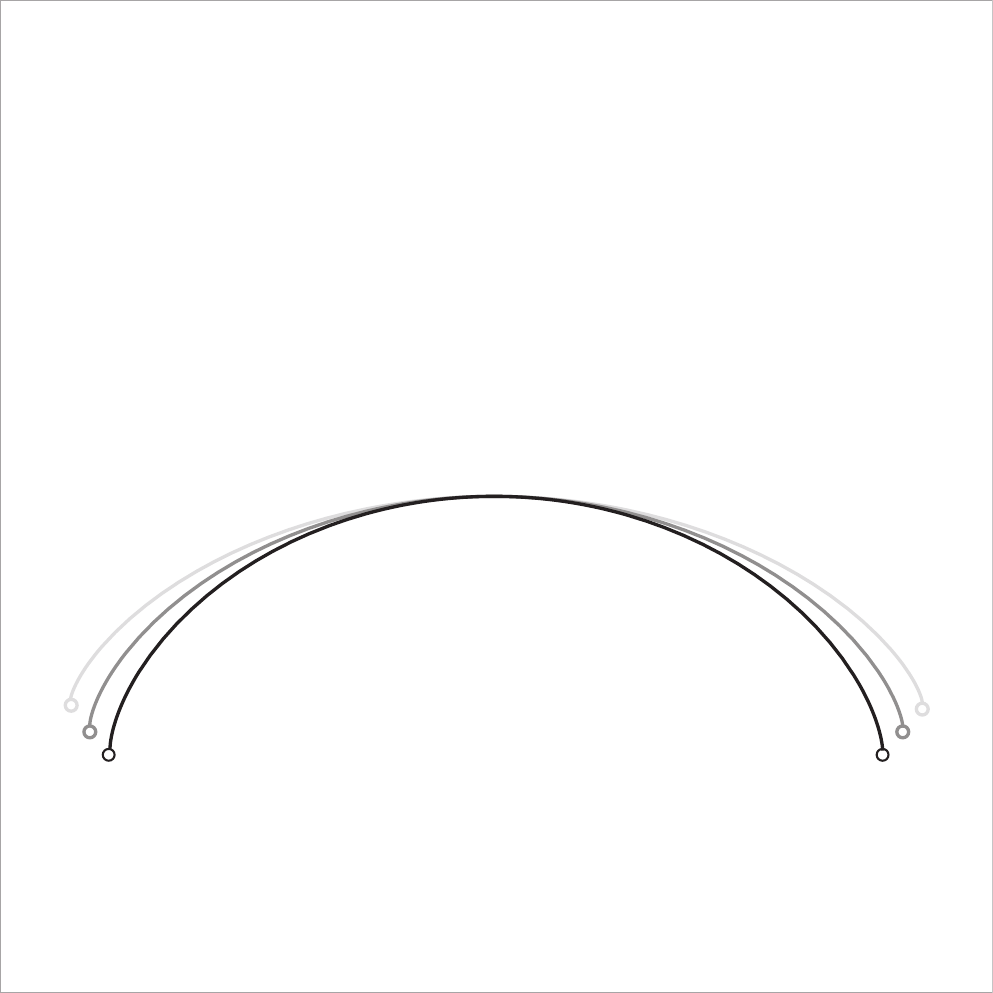}
			\put(120,-10){\includegraphics[scale=0.27]{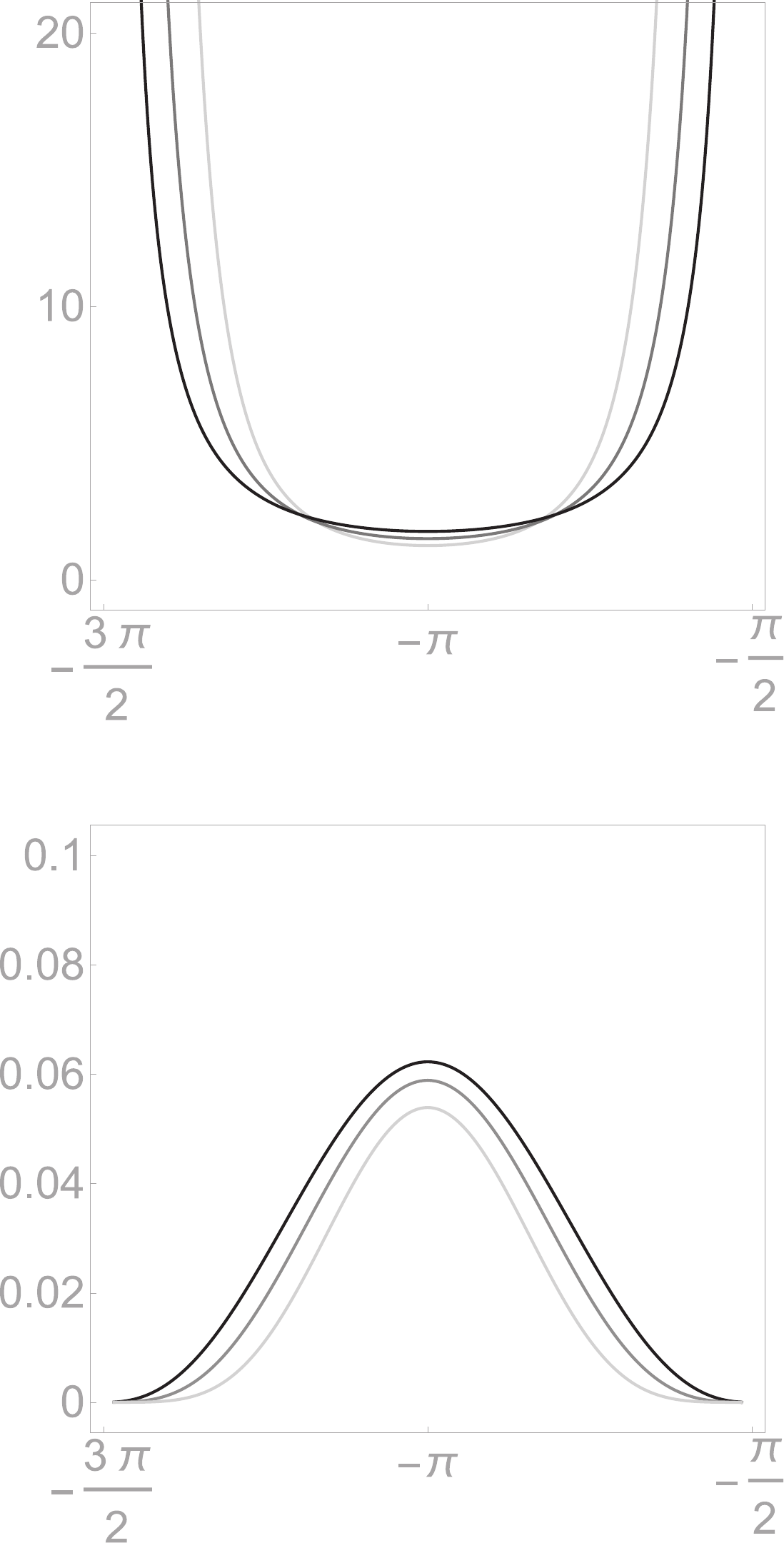}}
			\put(152,-13.8){$\theta$}
			\put(152,49){$\theta$}
			\put(114,81){$\partial_s \theta$}
			\put(117,19.2){$\sigma$}
			\put(6.5,70){\color{black}\vector(0,1){15}}
			\put(5.5,86){$\uvc{z}$}
			\put(6.5,70){\color{black}\vector(0,-1){15}}
			\put(5.3,50.5){$\uvc{v}$}
		\end{overpic}}
		\vspace*{1cm}
		\caption{Finite length ``parachute'' catenaries.  As in prior figures, with parameters $\phi = -90^\circ$, $|\bv| = $ (\textcolor{Lgr}{0.74}, \textcolor{Mgr}{0.75}, 0.76), $T=0$, $q =$ (\textcolor{Lgr}{0.67}, \textcolor{Mgr}{0.66}, 0.64), $R_\nu = 0.5$, $m=1$, $\epsilon=0.05$.}
		\label{sedimentm90}
	\end{figure}

In Figure \ref{curvem90}, the curves flip from being concave up to concave down.  When $\bv\cdot\uvc{z} = \bar{W} = -q$, our scaling is pathological and the solutions are straight lines perpendicular to $\bv$ and $\uvc{z}$.  However, this is not the full story.  The transition to concave down curves is not smooth.  For a small window of parameters such that the velocity and body force are nearly balanced, we encounter double-sided blowup solutions corresponding to finite-length curves, as in Figure \ref{sedimentm90}.  The tension profile now peaks in the middle and, for these rigidly moving curves, goes to zero at the ends.  Thus, these parachute-like objects are solutions to a free sedimentation problem for perfectly flexible filaments.	

\begin{figure}[H]
	\hspace*{0.85cm}
	\scalebox{1}{%
		\begin{overpic}[scale=0.5]{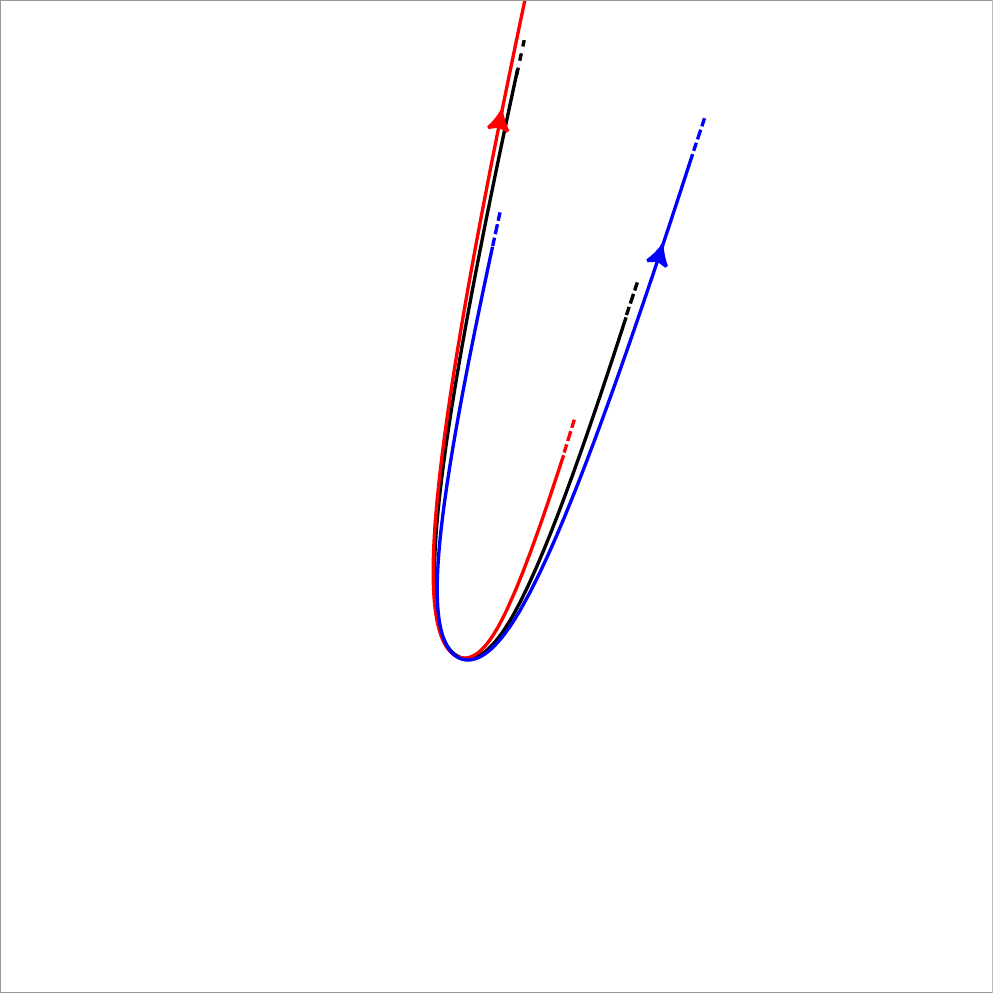}
			\put(120,-10){\includegraphics[scale=0.27]{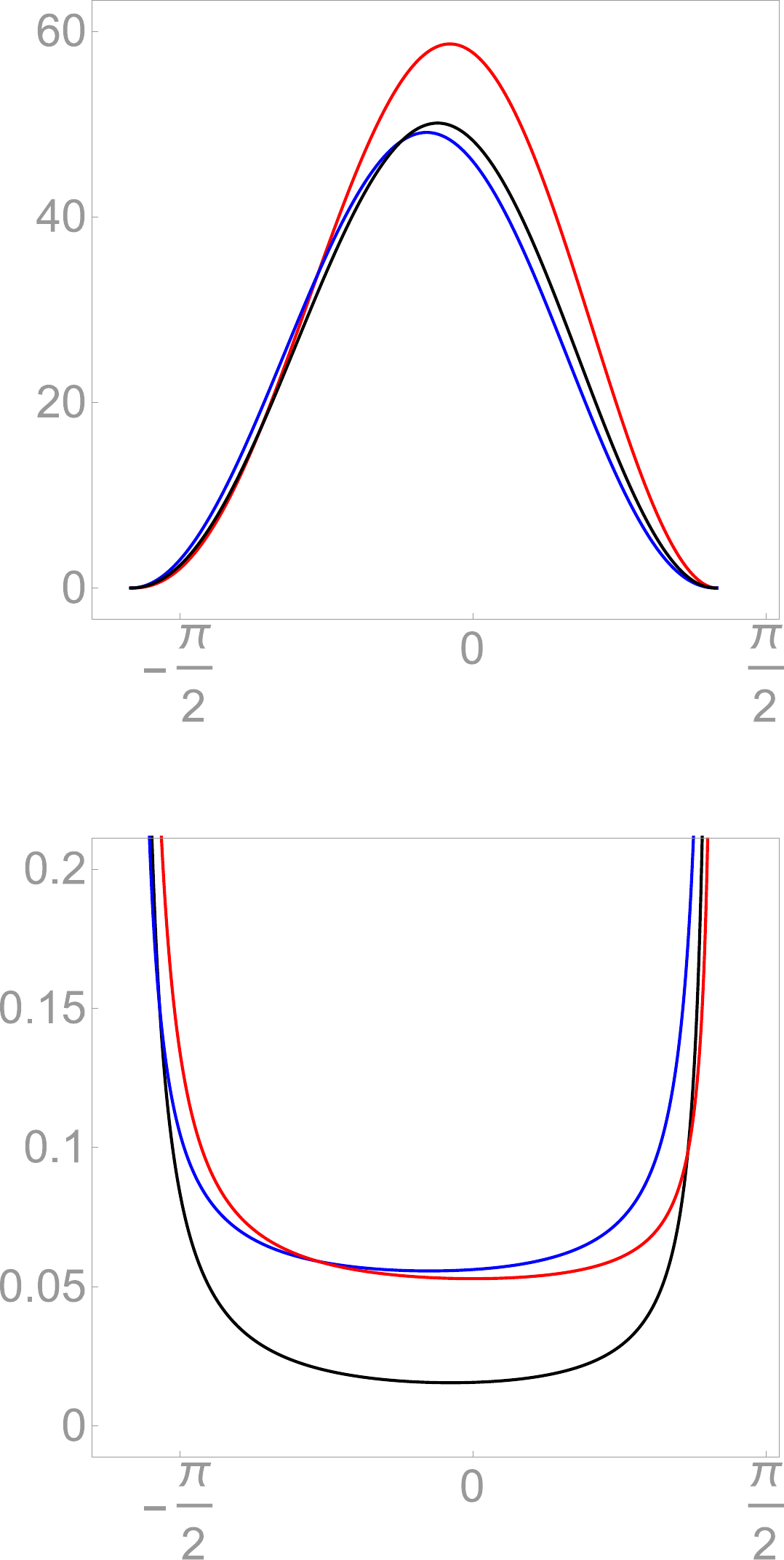}}
			\put(152,-13.8){$\theta$}
			\put(152,49){$\theta$}
			\put(114,81){$\partial_s \theta$}
			\put(117,19.2){$\sigma$}
			\put(6.5,70){\color{black}\vector(0,1){15}}
			\put(3.4,86){ $\uvc{z}$}
			\put(6.5,70){\color{black}\vector(1,-1){11}}
			\put(17.5,55){$\uvc{v}$}
		\end{overpic}}
		\vspace*{1cm}
		\caption{Effect of axial flow $T$ at fixed velocity to body force ratio $|\bv| / {q} = 0.3$.  As in prior figures, with parameters $\phi = -45^\circ$, $|\bv| = $ (\textcolor{red}{0.28}, 0.29, \textcolor{blue}{0.28}), $T=$(\textcolor{red}{-0.2}, 0, \textcolor{blue}{0.2}), $q =$ (\textcolor{red}{0.94}, 0.96, \textcolor{blue}{0.94}), $R_\nu = 0.5$, $m=1$, $\epsilon=0.05$.  Infinite length curves.}
		\label{curvem45T}
	\end{figure}

\begin{figure}[H]
	\hspace*{0.85cm}
	\scalebox{1}{%
		\begin{overpic}[scale=0.5]{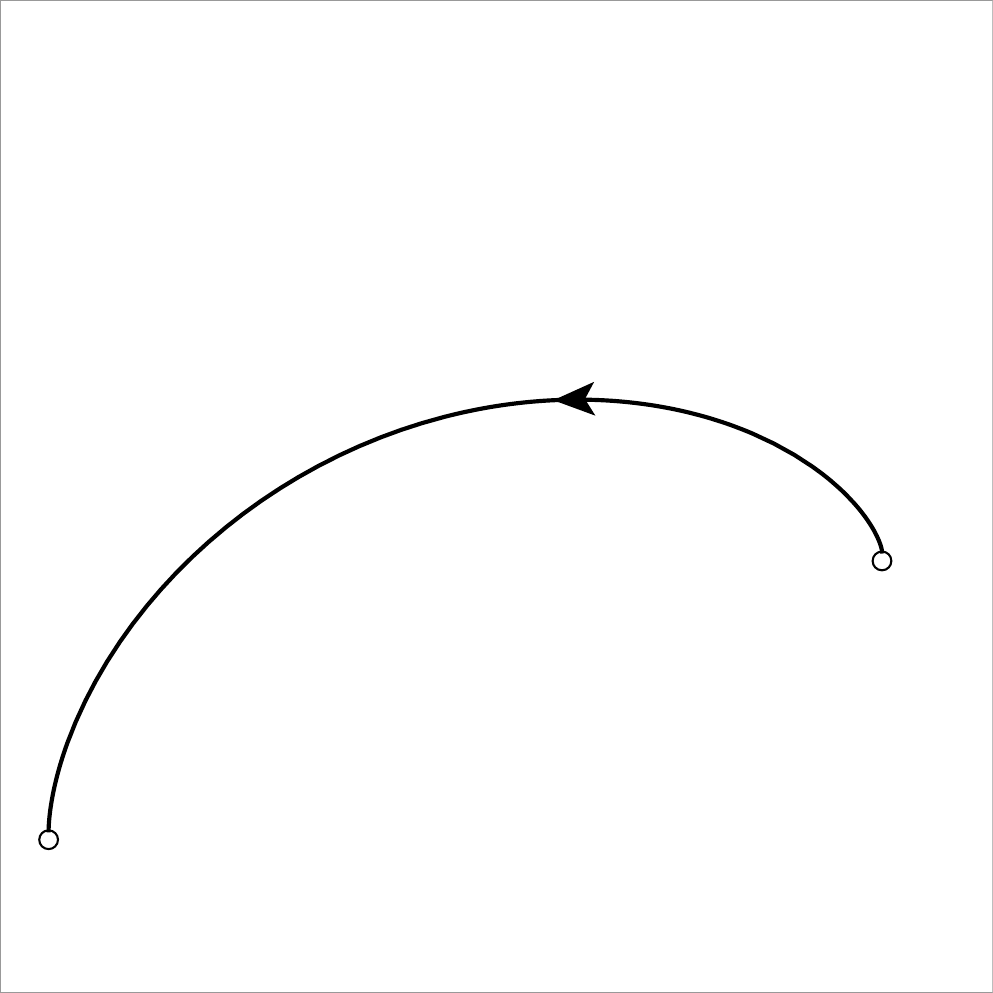}
			\put(120,-10){\includegraphics[scale=0.27]{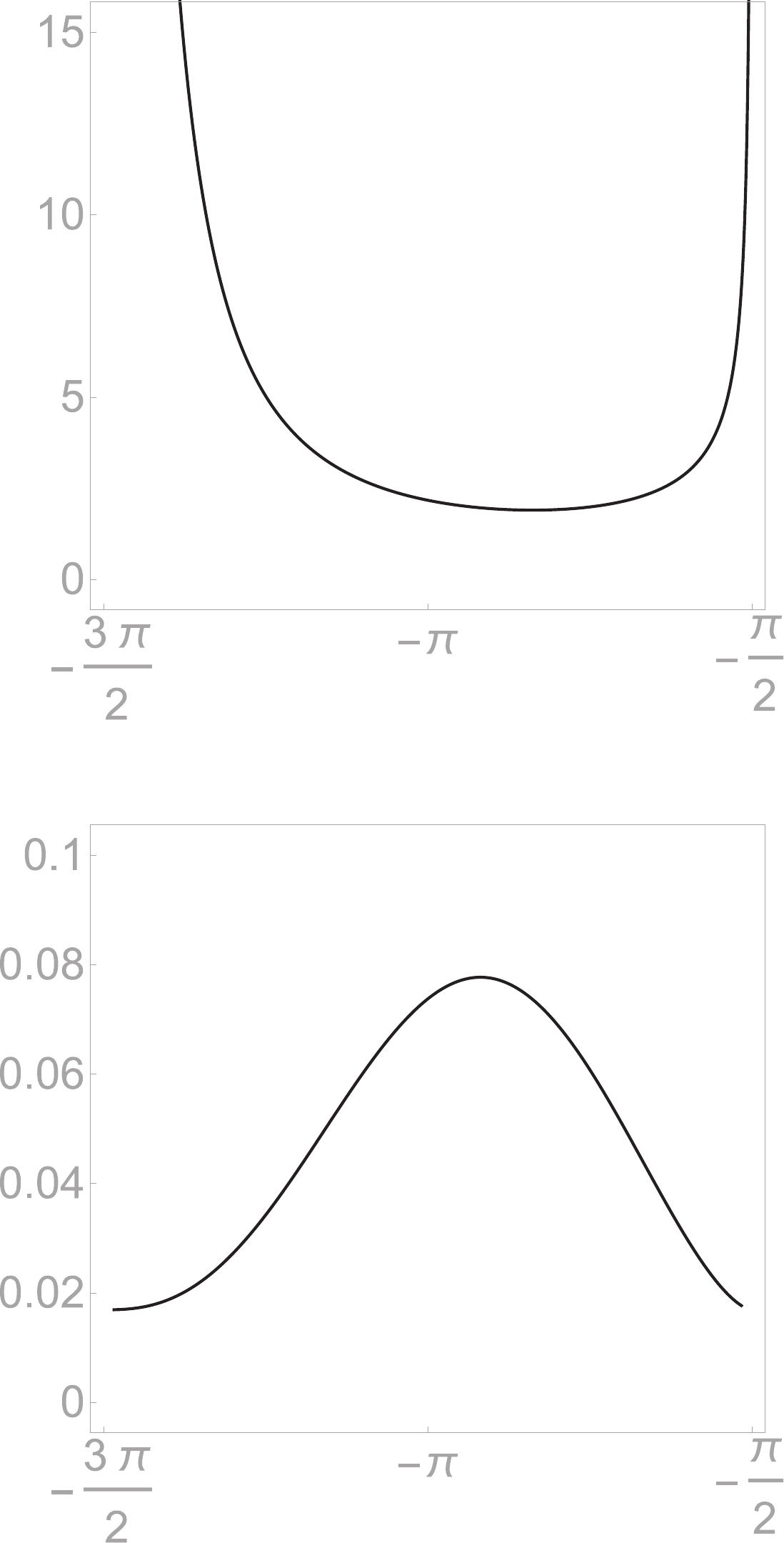}}
			\put(152,-13.8){$\theta$}
			\put(152,49){$\theta$}
			\put(114,81){$\partial_s \theta$}
			\put(117,19.2){$\sigma$}
			\put(6.5,70){\color{black}\vector(0,1){15}}
			\put(5.5,86){$\uvc{z}$}
			\put(6.5,70){\color{black}\vector(0,-1){15}}
			\put(5.3,50.5){$\uvc{v}$}
		\end{overpic}}
		\vspace*{1cm}
		\caption{An axially flowing finite length catenary.  As in prior figures, with parameters $\phi = -90^\circ$, $|\bv| = 0.76$, $T=0.13$, $q =0.64$, $R_\nu = 0.5$, $m=1$, $\epsilon=0.05$.}
		\label{sedimentm90T}
	\end{figure}

Figure \ref{curvem45T} shows the effect of a small amount of axial flow on an infinite length curve, with the ratio of drag to body forces maintained constant.  Arrows indicate the direction of flow.  The biggest contribution to the variation in tension with axial flow $T$ is a centripetal shift of $mT^2$.  Figure \ref{sedimentm90T} shows a finite length curve with weak axial flow, to be compared to the darkest curve in Figure \ref{sedimentm90} (while noting different overall scales).  Here one should imagine two reels traveling downward and exchanging a length of cable between them.  The tension at the ends is now $mT^2$ rather than zero.
				
\begin{figure}[H]
	\hspace*{0.85cm}
	\scalebox{1}{%
		\begin{overpic}[scale=0.5]{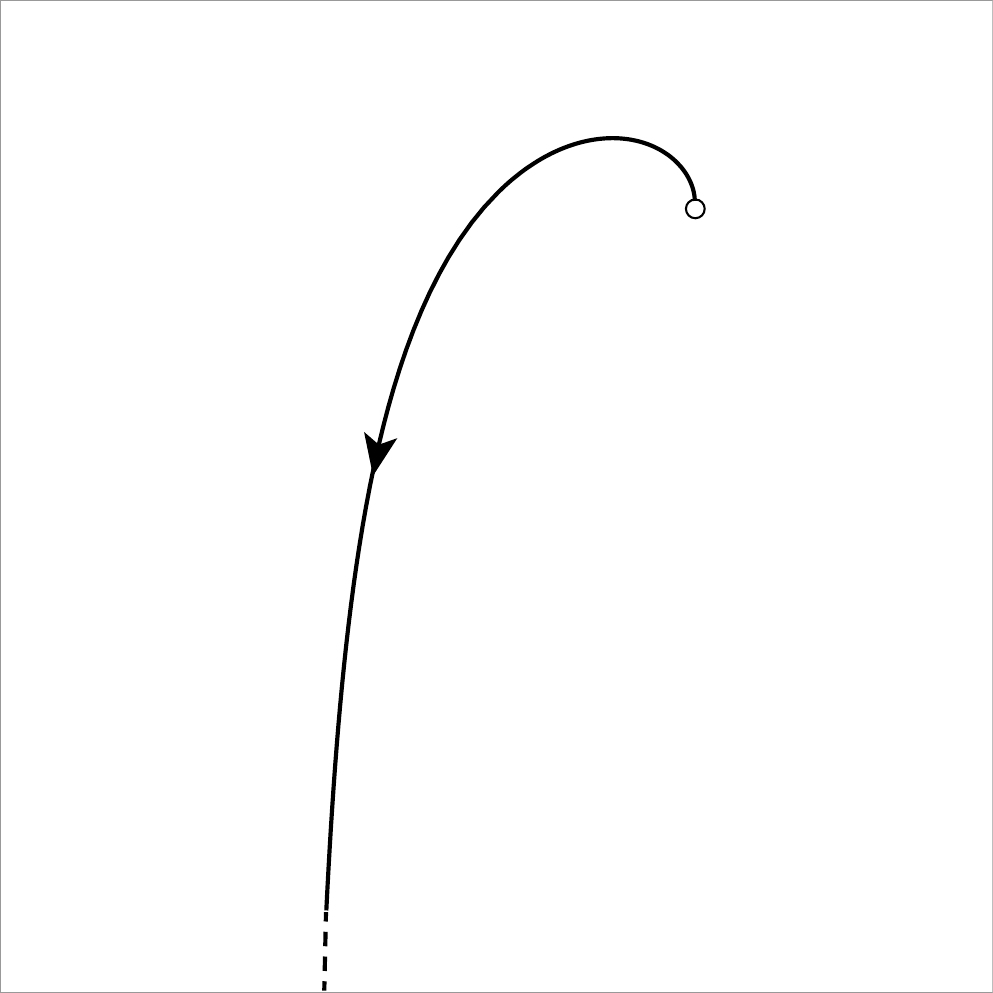}
			\put(120,-10){\includegraphics[scale=0.27]{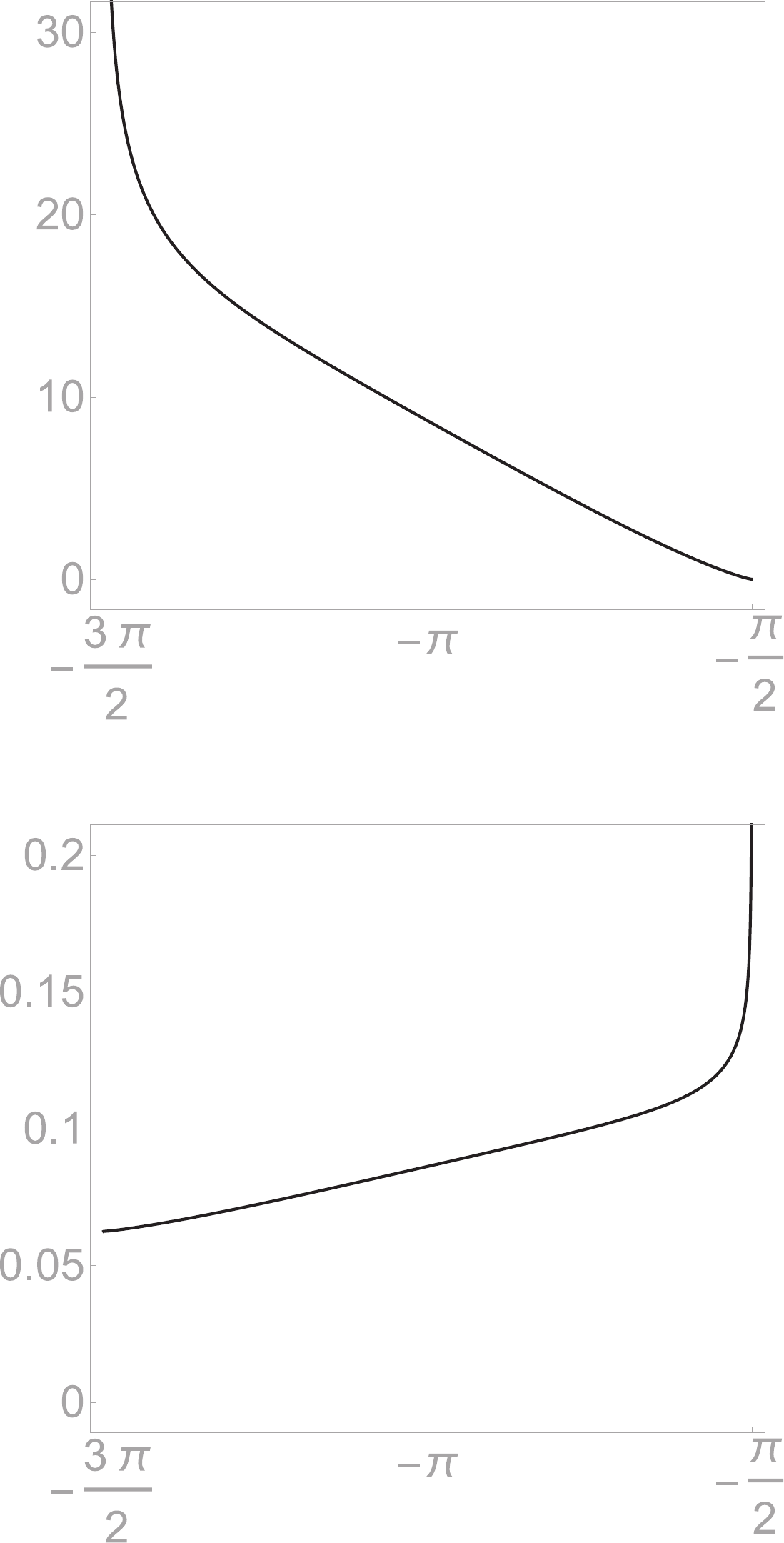}}
			\put(152,-13.8){$\theta$}
			\put(152,49){$\theta$}
			\put(114,81){$\partial_s \theta$}
			\put(117,19.2){$\sigma$}
			\put(6.5,70){\color{black}\vector(0,1){15}}
			\put(5.5,86){$\uvc{z}$}
			\put(6.5,70){\color{black}\vector(0,-1){15}}
			\put(5.3,50.5){$\uvc{v}$}						
		\end{overpic}}
		\vspace*{1cm}
		\caption{An axially flowing semi-infinite length catenary.  As in prior figures, with parameters $\phi = -90^\circ$, $|\bv| = 0.78$, $T=0.25$, $q =0.57$, $R_\nu = 0.6$, $m=1$, $\epsilon=0.05$.}
		\label{semim90T}
	\end{figure}
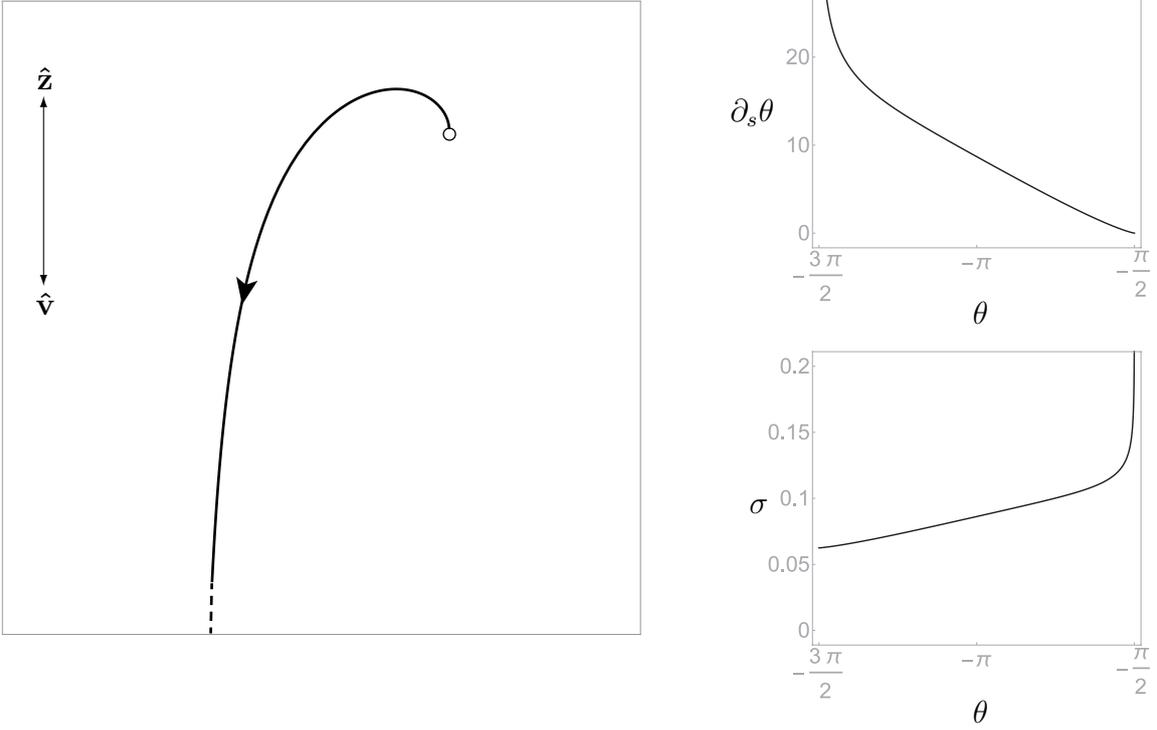

With sufficiently large axial flow, the solutions become of single-sided blowup type, corresponding to semi-infinite length curves.  Figure \ref{semim90T} shows an example, with drag anisotropy $R_\nu = 0.6$.  Here the tension is $mT^2$ on the finite end of the body, and goes to infinity on the infinite side.  As will be discussed in Section \ref{organization}, this is not always how the tension behaves in such curves.  

\vspace{0.5cm}
	
	\vspace{0.5cm}

\vspace{0.5cm}
		
\vspace{0.5cm}

\vspace{0.5cm}

\vspace{0.5cm}

\vspace{0.5cm}

\vspace{0.5cm}

\subsection{Organization of solutions}\label{organization}

The transitions between these qualitatively different configurations are controlled by the signs and relative magnitudes of the exponents of $f(\theta)$ and $g(\theta)$ in equation \eqref{thetaeq}.  These exponents are the twice-rescaled axial flow $\bar{T}$ and the quantity $k$ defined in \eqref{keq}.  Inequalities that govern the shapes are
\begin{align}
	k+\bar{T} > 0 \quad \mathrm{and} \quad k-\bar{T} > 0& \quad \quad \quad \textrm{heteroclinic connections} \, , \\
	k+\bar{T} < 0 \quad \mathrm{and} \quad k-\bar{T} < 0& \quad \quad \quad \textrm{double-sided blowup} \, , \\
	\mathrm{else}& \quad \quad \quad \textrm{single-sided blowup} \, .
\end{align}
The corresponding equalities are special cases that will be discussed below.

There are also transitions in the behavior of the tension at the zeroes or poles.  The extra factor of $g(\theta)$ in equation \eqref{sigmaeq} means that these do not coincide with the shape transitions.  Instead, we have
\begin{align}
	1-k+\bar{T} > 0 \quad \mathrm{and} \quad 1-k-\bar{T} > 0& \quad \quad \quad \sigma \to mT^2 \quad \textrm{on both sides} \, , \\
	1-k+\bar{T} < 0 \quad \mathrm{and} \quad 1-k-\bar{T} < 0& \quad \quad \quad \sigma \to \infty \quad\quad  \textrm{on both sides} \, , \\
	\mathrm{else}& \quad \quad \quad \sigma \to mT^2 \quad \textrm{on trailing side and } \\ 
	& \quad \quad \quad \sigma \to \infty \quad\quad \textrm{on leading side} \, . \nonumber
\end{align}
Again, equalities are special cases that will be discussed below.

\begin{figure}[H]
		\centering
		\scalebox{1.1}{
			\begin{tikzonimage}[scale=0.5]{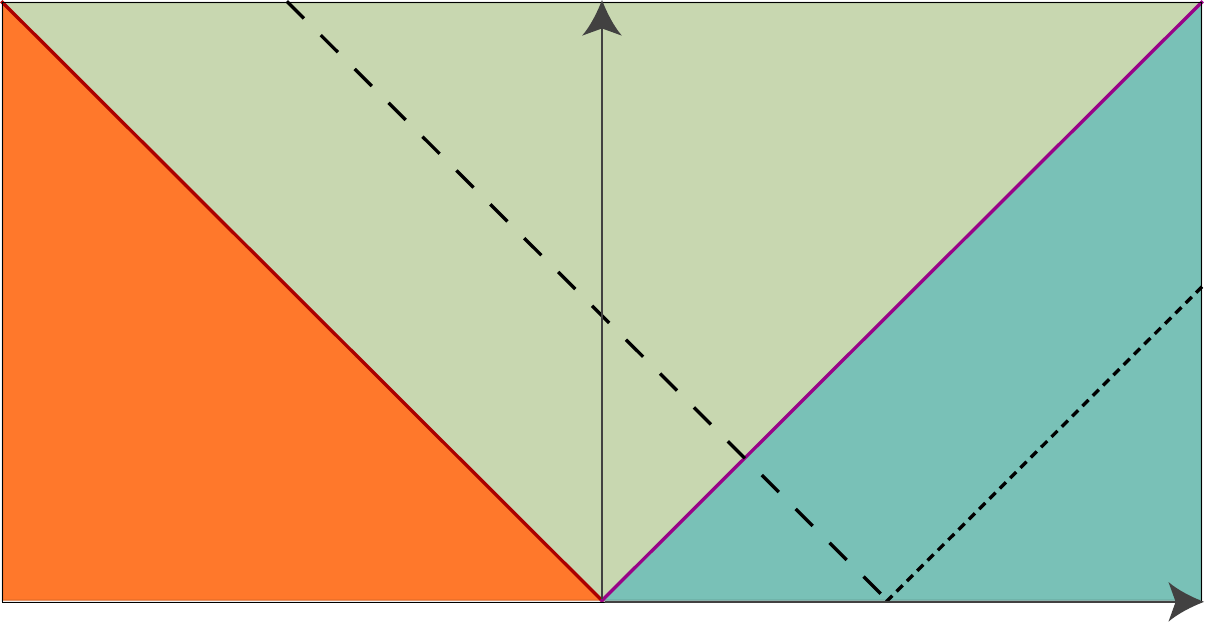}  
				\draw (1.025,0.04,0) node[] {$k$};
				\draw (0.505,1.05,0) node[] {$\bar{T}$};
				\draw (0.32,0.35,0) node[] {\scriptsize{$k + \bar{T}\,\,\,\,\,\, = 0$}};
				\draw (0.3,0.81,0) node[] {\scriptsize{$1- k - \bar{T}\,\,\,\,\,\,= 0$}};
				\draw (0.88,0.81,0) node[] {\scriptsize{$k - \bar{T}\,\,\,\,\,\,= 0$}};
				\draw (0.86,0.35,0) node[] {\scriptsize{$1-k + \bar{T}\,\,\,\,\,\,= 0$}};
			\end{tikzonimage}}
			\caption{A planar map of solution types, spanned by two exponents $k$ and $\bar{T}$ that appear in the shape equation \eqref{thetaeq}.  The teal, green, and orange regions correspond to \textcolor{PI}{heteroclinic connections}, \textcolor{GR}{single-sided blow up}, and \textcolor{OR}{double-sided blow up} solutions, respectively.  The \textcolor{VIO}{purple} border between teal and green, and the \textcolor{BR}{maroon} border between orange and green, correspond to special cases of the shape equation.  The long- and short-dashed lines correspond to special cases of the stress equation \eqref{sigmaeq}.  With respect to the dashed lines, the asymptotic end stresses are infinity in the region on the right, $mT^2$ in the region on the left, and one of each in the region in between.}
			\label{planarmap}
		\end{figure}

Figure \ref{planarmap} is a simple planar map, spanned by the two exponents, that shows where different types of solutions are to be found.  Teal, green, and orange regions denote heteroclinic, single-sided, and double-sided shapes, respectively.  The purple and maroon borders are special cases for the shapes.  The long- and short-dashed lines are special cases for the stress. Upon approach to a zero or pole, the stress approaches infinity in the rightmost region of the diagram and a finite value in the leftmost region.  Note that there is a green strip of semi-infinite curves with finite tensions, and a teal strip of infinite curves with tension approaching a finite value on one or both sides.  These results seem strange.  We would naively expect the dashed lines to coincide with the colored borders, so that the tension behavior echoed the shape of the curves.

\begin{figure}[H]
	\centering
		\scalebox{1.1}{
	\begin{tikzonimage}[scale=0.45]{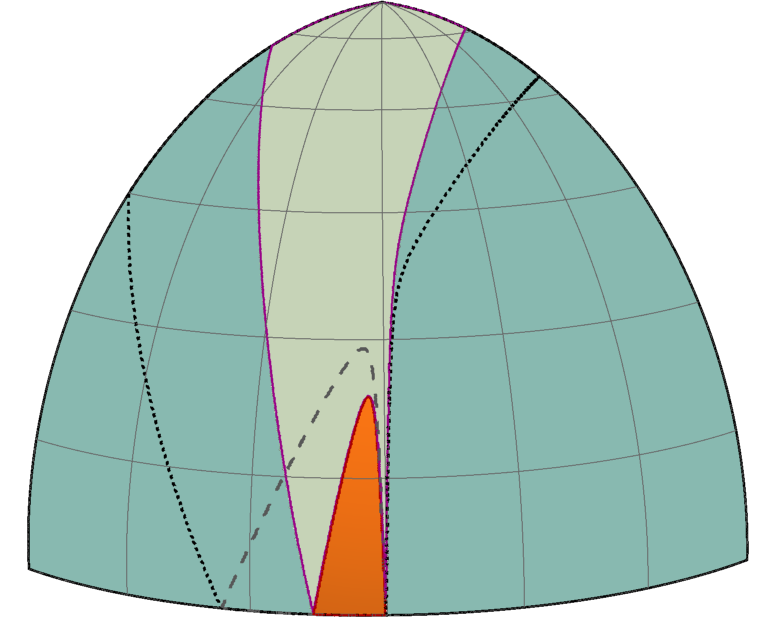} 
		\draw[->,>=stealth] (0.94,0.8,0) -- (0.94,0.93,0) node[above] {$\uvc{z}$};
		\draw[->,>=stealth] (0.94,0.8,0) -- (0.945,0.675,0); \draw (0.945,0.64) node[]  {$\uvc{v}$};
		\draw  (-0.03,0.1) node[]{ \large{$|\bv|$}};
		\draw  (1.02,0.1) node[]{\large{q}};
		\draw (0.5,1.027) node[]{\large{T}};
	\end{tikzonimage}}
		\caption{A spherical map of solution types spanned by the magnitudes of the translational velocity $| \bv |$, axial flow $T$, and body force $q$.  Fixed parameters are $\phi = -88^\circ$ and $R_\nu = 0.5$.  As in Figure \ref{planarmap}, colored regions correspond to \textcolor{PI}{heteroclinic connections}, \textcolor{GR}{single-sided blow up}, and \textcolor{OR}{double-sided blow up} solutions.  With respect to the dashed lines, the asymptotic end stresses are infinity in the regions adjacent to the $| \bv |$ and $q$ corners, $mT^2$ in the region enclosed by the long-dashed line and the equator, and one of each in the region adjacent to the $T$ pole.}
	\label{sphericalmap}
\end{figure}

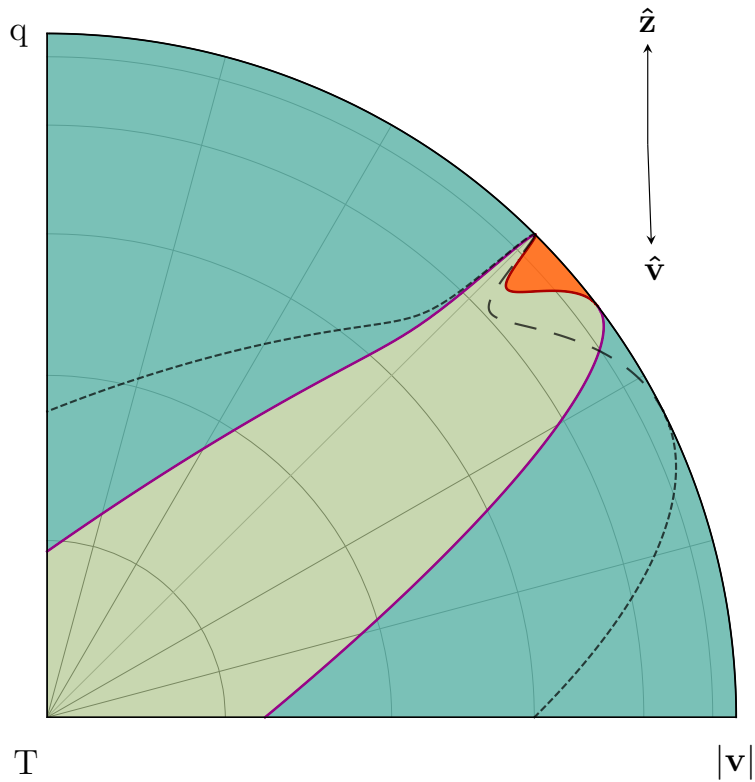
\begin{figure}[H]
	\centering
	\scalebox{1.1}{
		\begin{tikzonimage}[scale=0.43]{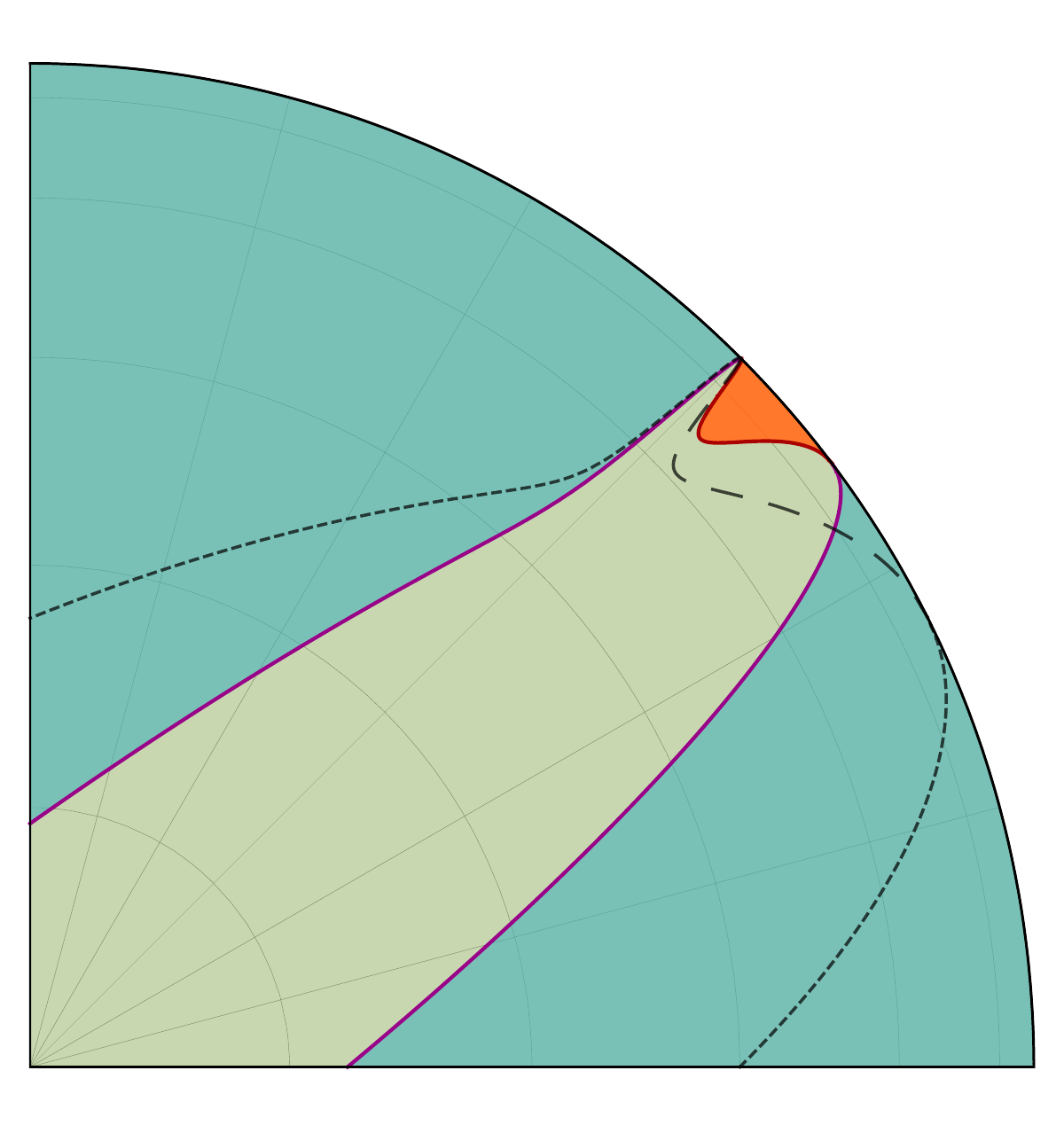} 
			\draw[->,>=stealth] (0.85,0.8,0) -- (0.85,0.93,0) node[above] {$\uvc{z}$};
			\draw[->,>=stealth] (0.85,0.8,0) -- (0.855,0.67,0); \draw (0.86,0.64) node[]  {$\uvc{v}$};
			
			\draw  (0.97,0) node[]{ $|\bv|$};
			\draw  (-0.01,0.94) node[]{q};
			\draw (0.0,0) node[]{T};
			
		\end{tikzonimage}}
		\caption{Orthographic projection of the octant of the $\phi = -88^\circ$, $R_\nu = 0.5$ sphere from Figure \ref{sphericalmap} onto the $|\bv|$-$q$ plane.}
		\label{ortho}
	\end{figure}

A more physically intuitive map may be drawn using the once-rescaled quantities.  Solutions for shapes and stresses depend on six independent parameters, but the scaling parameter $C$ and mass coefficient $m$ do not affect our qualitative classification.  Fixing the angle of translation $\phi$ and the drag anisotropy $R_\nu$ leaves us with two independent parameters.  These may be represented as coordinates on a 2-sphere in the space spanned by the magnitudes of the axial flow, translational velocity, and body force, $T$, $| \bv |$, and $q$, related by $T^2+ \bv\cdot\bv + q^2 =  1$.  Due to the symmetry of the problem, we need only consider one octant of the sphere.  Figure \ref{sphericalmap} shows an example for $\phi = -88^\circ$ and $R_\nu = 0.5$, with colors and dashing retaining the same meaning as in Figure \ref{planarmap}.  Solutions on the poleward $T$ side of the dashed lines have tension infinite on their leading ends and finite on their trailing ends.  Those in the teal heteroclinic regions that include the $| \bv |$ and $q$ corners have infinite tension on both ends, while those in the region that entirely encloses the orange double-sided blowup region have finite tension on both ends.
Moving along a lattitude line keeps the axial flow fixed while varying the relative strengths of the translational drag forces and body force.  Moving along a meridian keeps the ratio of drag to body forces fixed while varying the axial flow.  For ease of illustration and viewing, we use the orthographic projection of the octant onto the $|\bv|$-$q$ plane, a circular sector shown in Figure \ref{ortho}.  Physical chemists or the metallurgically inclined may note that this is somewhat reminiscent of a ternary phase diagram.  The corners of the sector represent the pure classical catenary ($q$), a neutrally buoyant towed cable ($|\bv|$), and a lariat snaking along itself ($T$).  However, the latter limit, corresponding to the pole of the sphere, has no meaning here because the rescaling by $\Delta$ is equivalent to dividing by zero.  Similarly, when the angle of translation is $-90^\circ$, the rescaling breaks down on the central meridian where $q$ and $|\bv|$ are equal.

\begin{figure}[H]
	\hspace*{7.3cm}
	\hspace*{0.7cm}\begin{overpic}[scale=0.32]{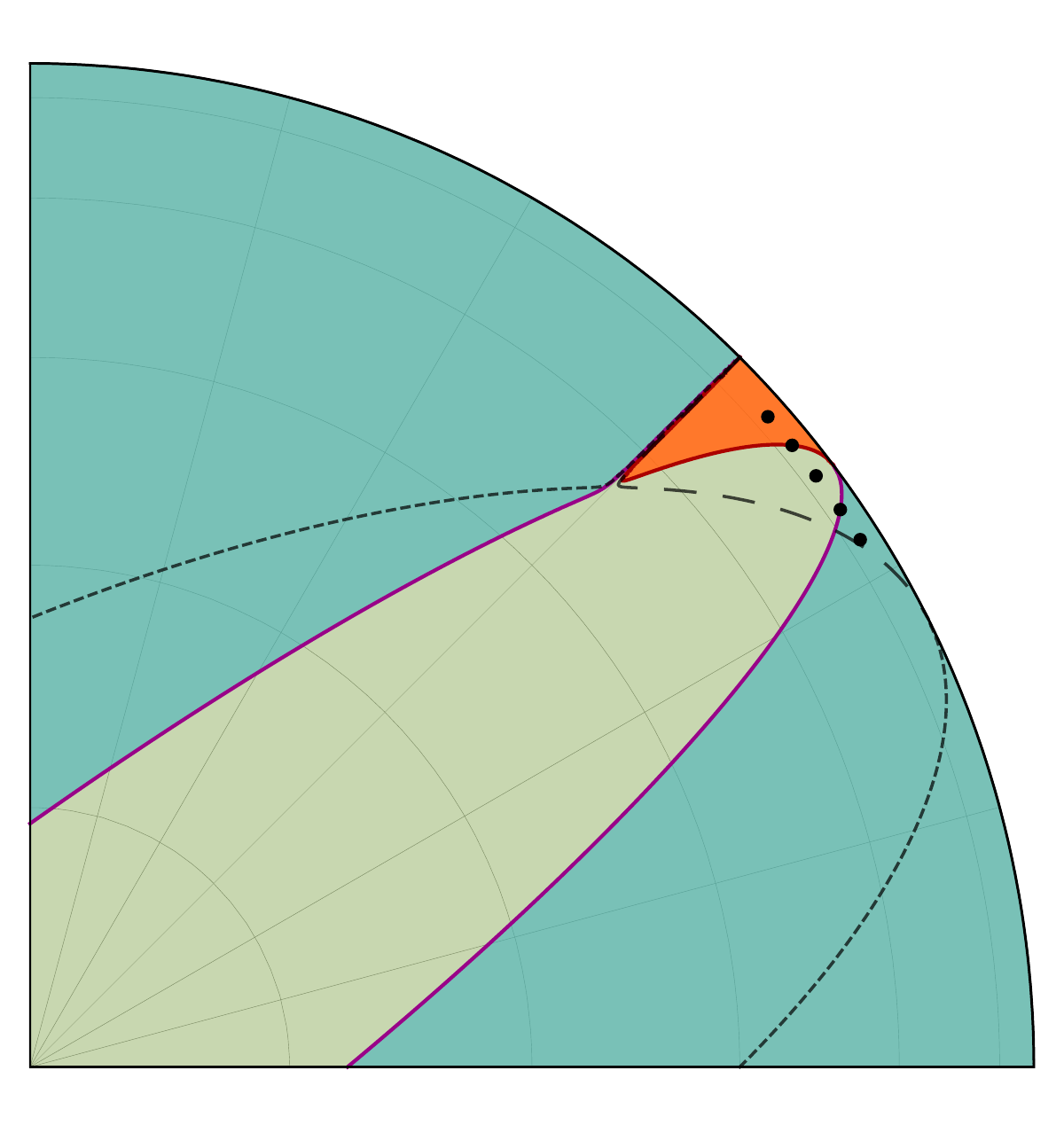}
		\put(103,4){\includegraphics[scale=0.2]{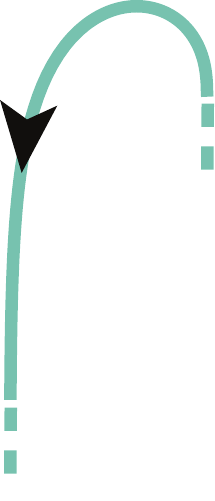}}
		\put(100,25){\includegraphics[scale=0.2]{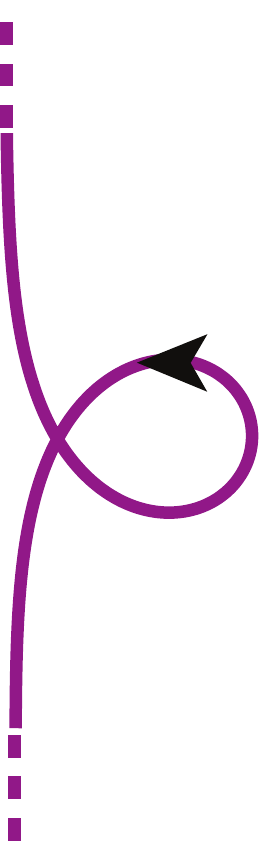}}
		\put(93,48){\includegraphics[scale=0.2]{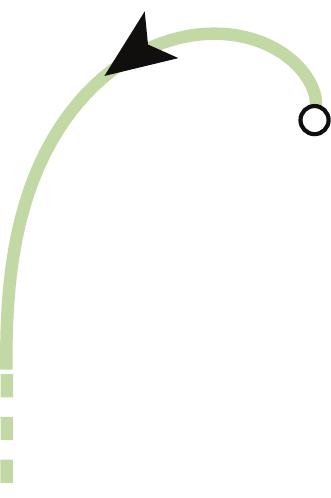}}
		\put(71,58){\includegraphics[scale=0.2]{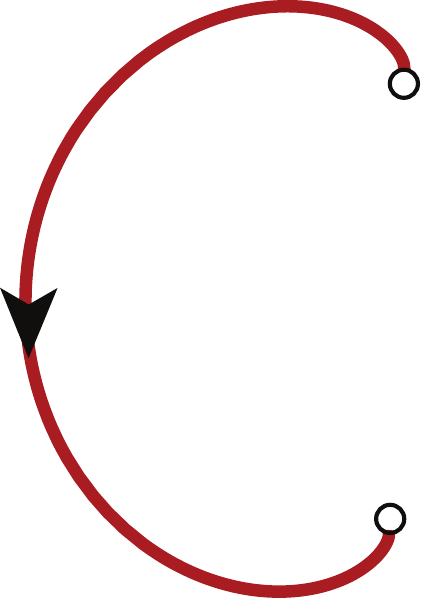}}
		\put(55,81){\includegraphics[scale=0.2]{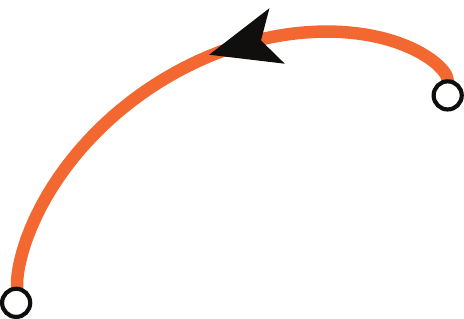}}
		\put(-107,5){\includegraphics[scale=0.34]{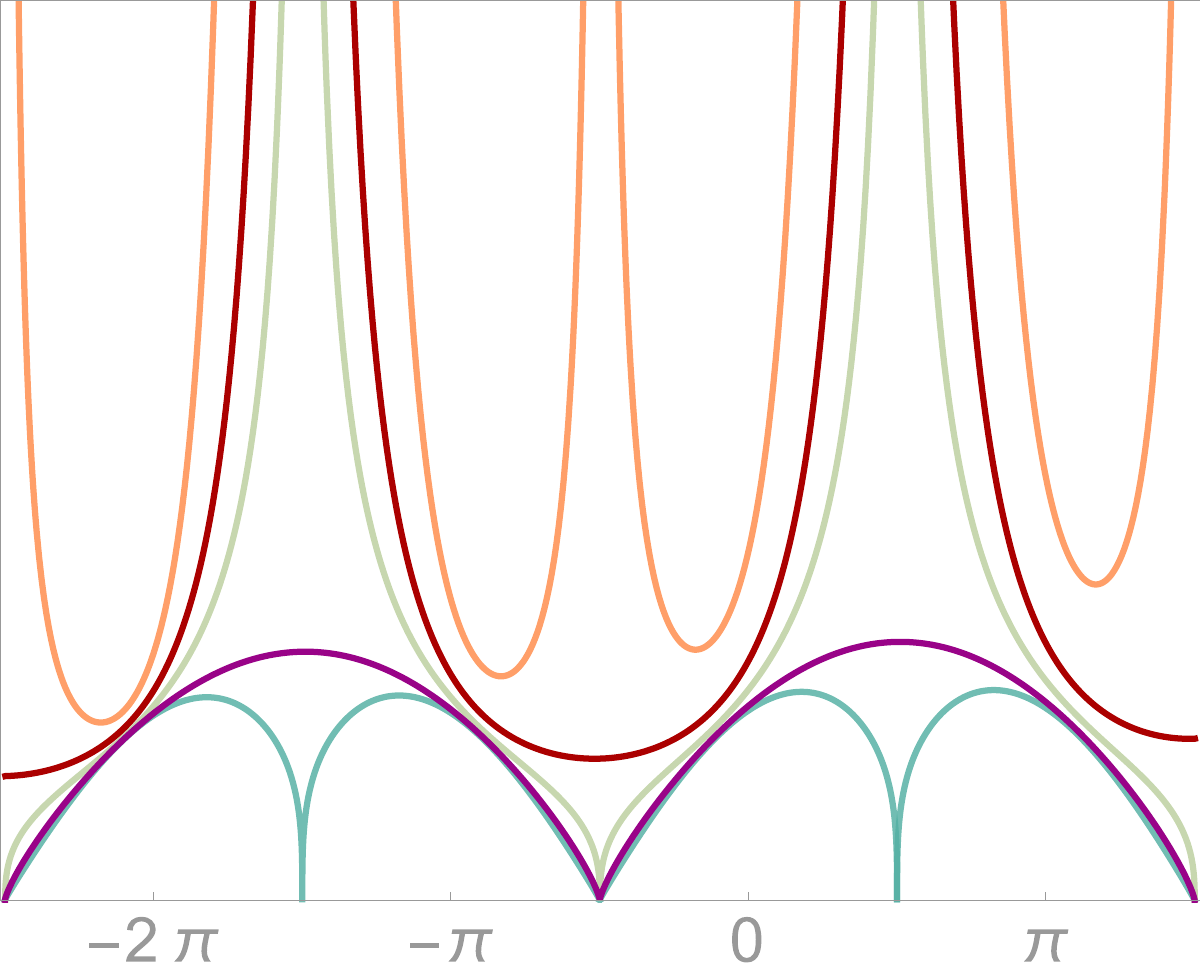}}
		\put(124,72){\color{black}\vector(0,1){12}}
		\put(122.5,85){$\uvc{z}$}
		\put(124,72){\color{black}\vector(0,-1){12}}
		\put(122.2,55){$\uvc{v}$}
		\put(-59,3){$\theta$}
		\put(-118,48){$\partial_s\theta$}
		\put(-1.5,93){$q$}
		\put(88,-0.5){$|\bv|$}
		\put(-1,-0.5){$T$}
	\end{overpic}
	\caption{Five axially flowing catenaries taken from the dotted locations along the $T = 0.2$ latitude line  in the projection (center) of a $\phi = -89.9^\circ$, $R_\nu = 0.5$ spherical map octant.  The configurations (right), and several periods of orbits (left), correspond to a generic $\pi$-subtending \textcolor{PI}{heteroclinic connection}, a $2\pi$-subtending heteroclinic \textcolor{VIO}{special case}, a generic $\pi$-subtending \textcolor{GR}{single-sided blow up} solution, a $2\pi$-subtending double-sided blow up \textcolor{BR}{special case}, and a generic $\pi$-subtending \textcolor{OR}{double-sided blow up} solution.  Configurations are drawn by adjusting $C$ so that a length of unity corresponds with a subtended angle of approximately $\pi$, while orbits are drawn using the scaling $C = 4 (1-R_\nu) \bar{U} \bar{Q} \pi$ to eliminate overlap in the range shown.  The other parameters are $|\bv| = $ (\textcolor{PI}{0.83}, \textcolor{VIO}{0.81}, \textcolor{GR}{0.78}, \textcolor{BR}{0.76}, \textcolor{OR}{0.74}),  $q =$ (\textcolor{PI}{0.52}, \textcolor{VIO}{0.55}, \textcolor{GR}{0.59}, \textcolor{BR}{0.62}, \textcolor{OR}{0.64}).	 
}
	\label{pseudophase1}
\end{figure}

In Figure \ref{pseudophase1}, we show five curves taken from the $T=0.2$ latitude line on the sphere for nearly downward translation $\phi = -89.9^\circ$ and Stokes drag anisotropy $R_\nu = 0.5$, and corresponding orbits in a $4\pi$-wide slice of phase space.  The teal, green, and orange curves correspond to the heteroclinic, single-sided, and double-sided solutions already discussed.  In between these are purple and maroon curves that subtend an angle of $2\pi$ rather than $\pi$.  They are special cases corresponding to the boundaries between the generic regions.  When $k = \pm \bar{T}$, $f(\theta)$ and $g(\theta)$ combine to form $2\pi$-periodic functions.  New orbits and configurations are formed from the fusion of two heteroclinic or two double-sided blowup solutions.  In the figure, we have scaled configurations as before, with a length of one unit per angle of $\pi$ subtended, but used the scaling $C = 4 (1-R_\nu) \bar{U} \bar{Q} \pi$ to draw the orbits.  This scaling prevents overlap of the orbits in the range shown, but is insufficient to prevent overlap at more negative values of $\theta$.  The result is a set of curves that seem to fit together within the range, and are suggestive of a phase portrait.  It is unknown whether the phase space, or some slice of it, can be foliated by orbits generated by all the intervening points along our circumferential cut through parameter space.  If so, it suggests that there is an additional symmetry hidden in our system.

\begin{figure}[H]
	\hspace{2.5cm}
	\scalebox{1}{%
		\begin{overpic}[scale=0.4]{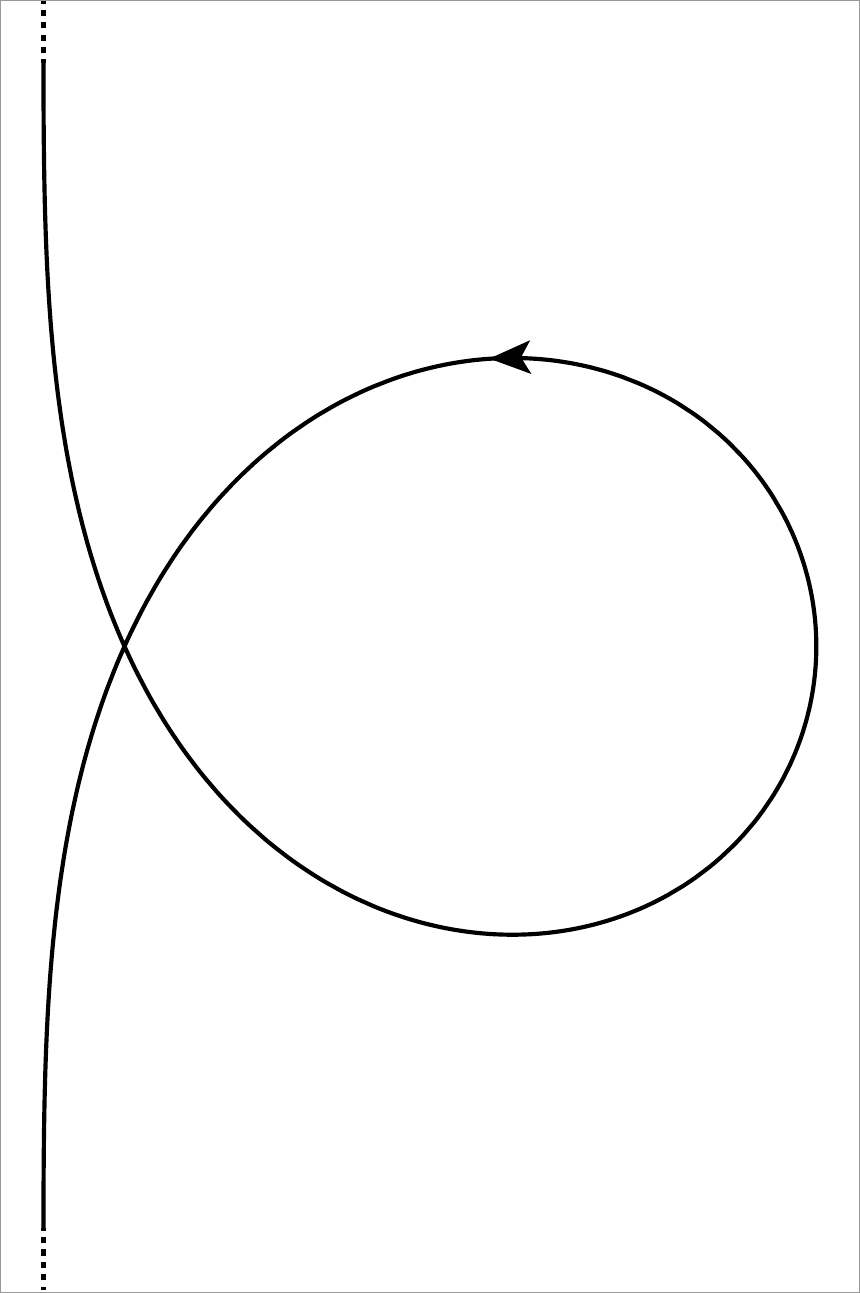}
			\put(-9,86){$\uvc{z}$}
			\put(-8,70){\color{black}\vector(0,-1){15}}
			\put(-9.3,51){$\uvc{v}$}
			\put(-8,70){\color{black}\vector(0,1){15}}
			\hspace*{1.3cm}
			\put(67,-8){\includegraphics[scale=0.27]{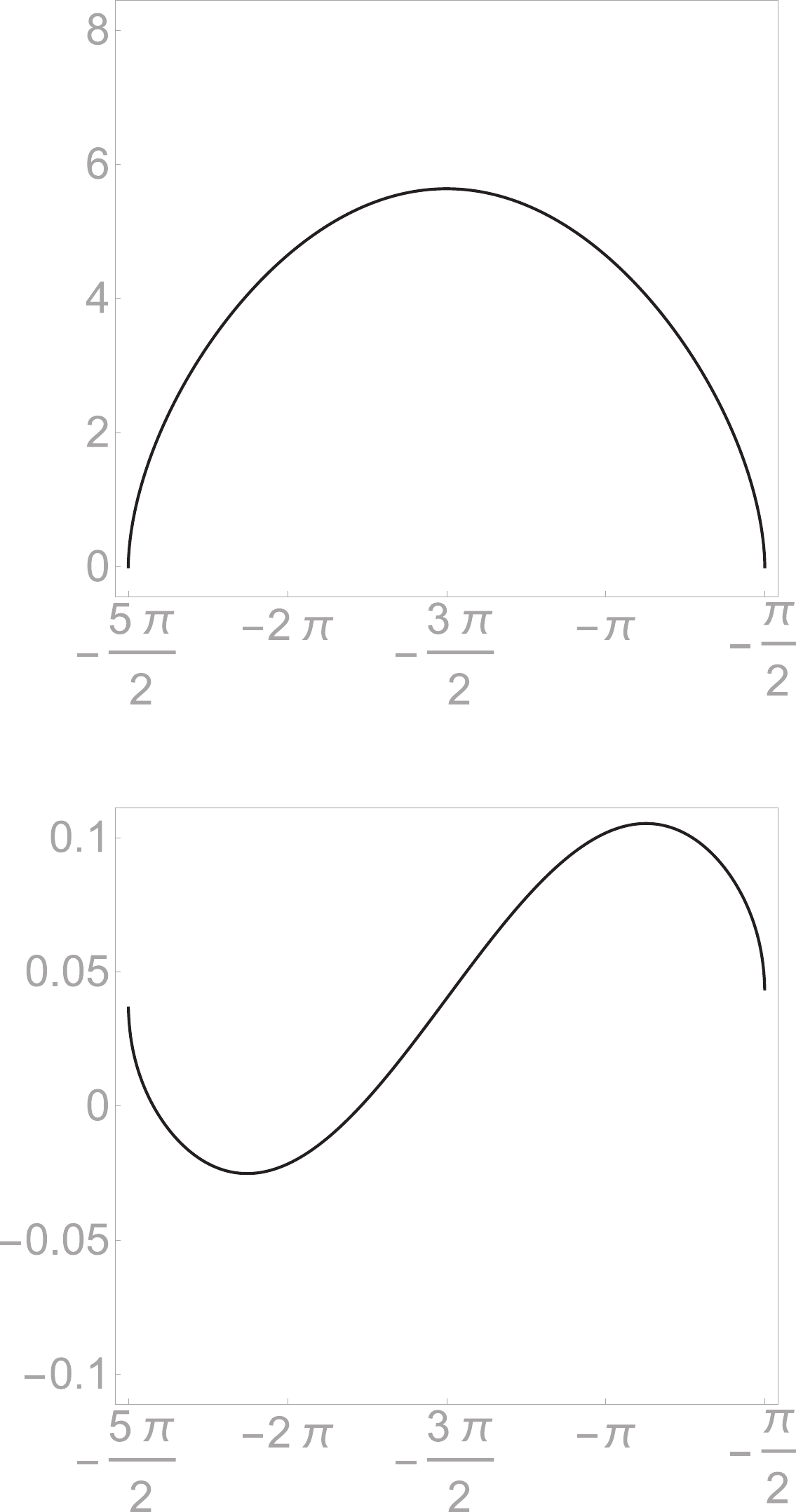}}
			\put(100,-12){$\theta$}
			\put(100,48){$\theta$}
			\put(64,81){$\partial_s \theta$}
			\put(67,21){$\sigma$}
		\end{overpic}}
		\vspace*{1cm}
		\caption{An axially flowing, self-intersecting 
		loop de loop catenary.  Details as in Figures \ref{curve90}-\ref{semim90T}, with parameters $\phi = -90^\circ$, $|\bv| = 0.82$, $T=0.2$, $q =0.53$, $R_\nu = 0.4$, $m=1$, $\epsilon=0.001$.  Infinite length curve subtending $2\pi$ due to the special condition $k = \bar{T}$.
}
		\label{loop}
	\end{figure}

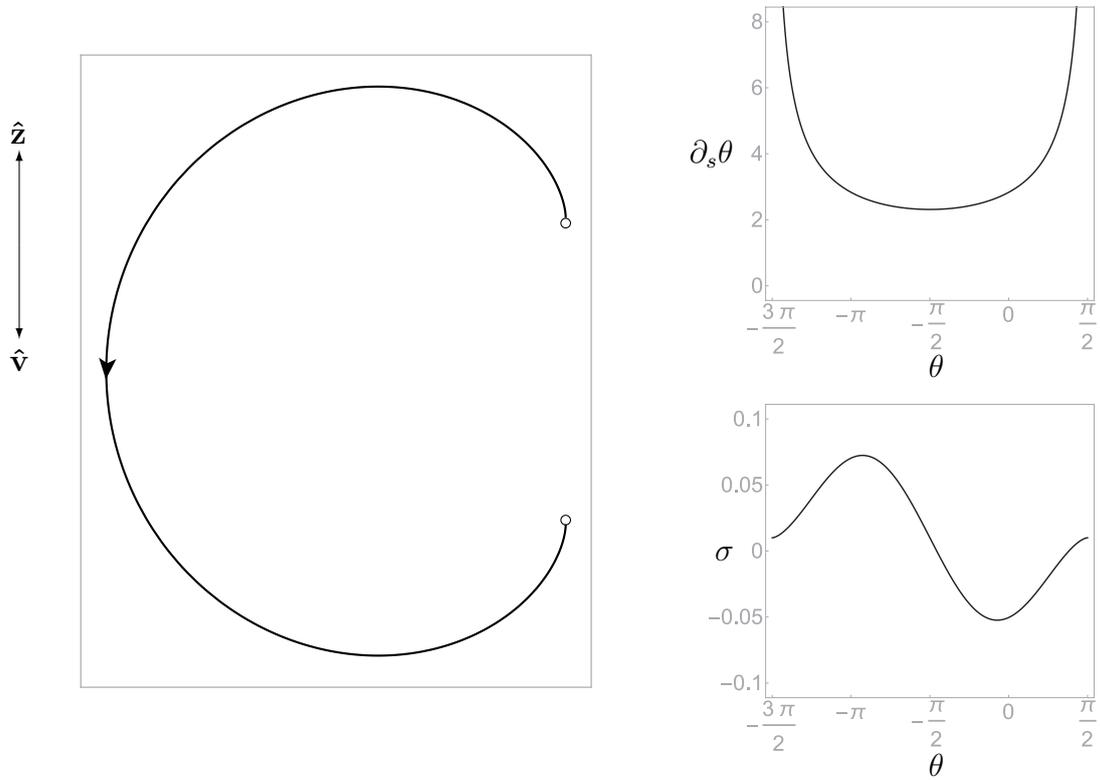
\begin{figure}[H]
	\hspace{2cm}
	\scalebox{1}{%
		\begin{overpic}[scale=0.4]{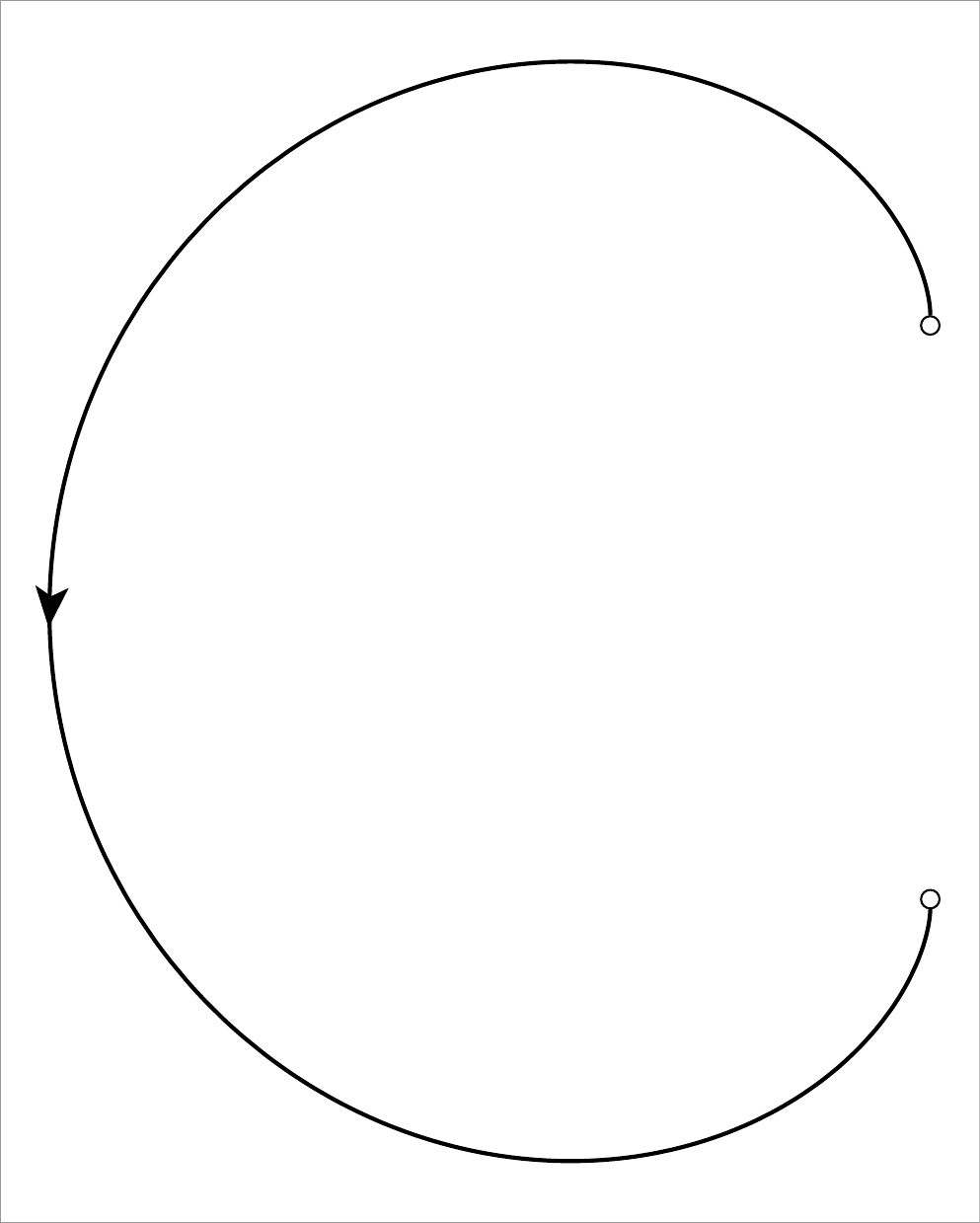}
			\put(-9.6,70){\color{black}\vector(0,1){15}}
			\put(-11,86){$\uvc{z}$}
			\put(-9.6,70){\color{black}\vector(0,-1){15}}
			\put(-11.1,50){$\uvc{v}$}
			\hspace*{1.5cm}
			\vspace*{6mm}
			\put(79,-10){\includegraphics[scale=0.27]{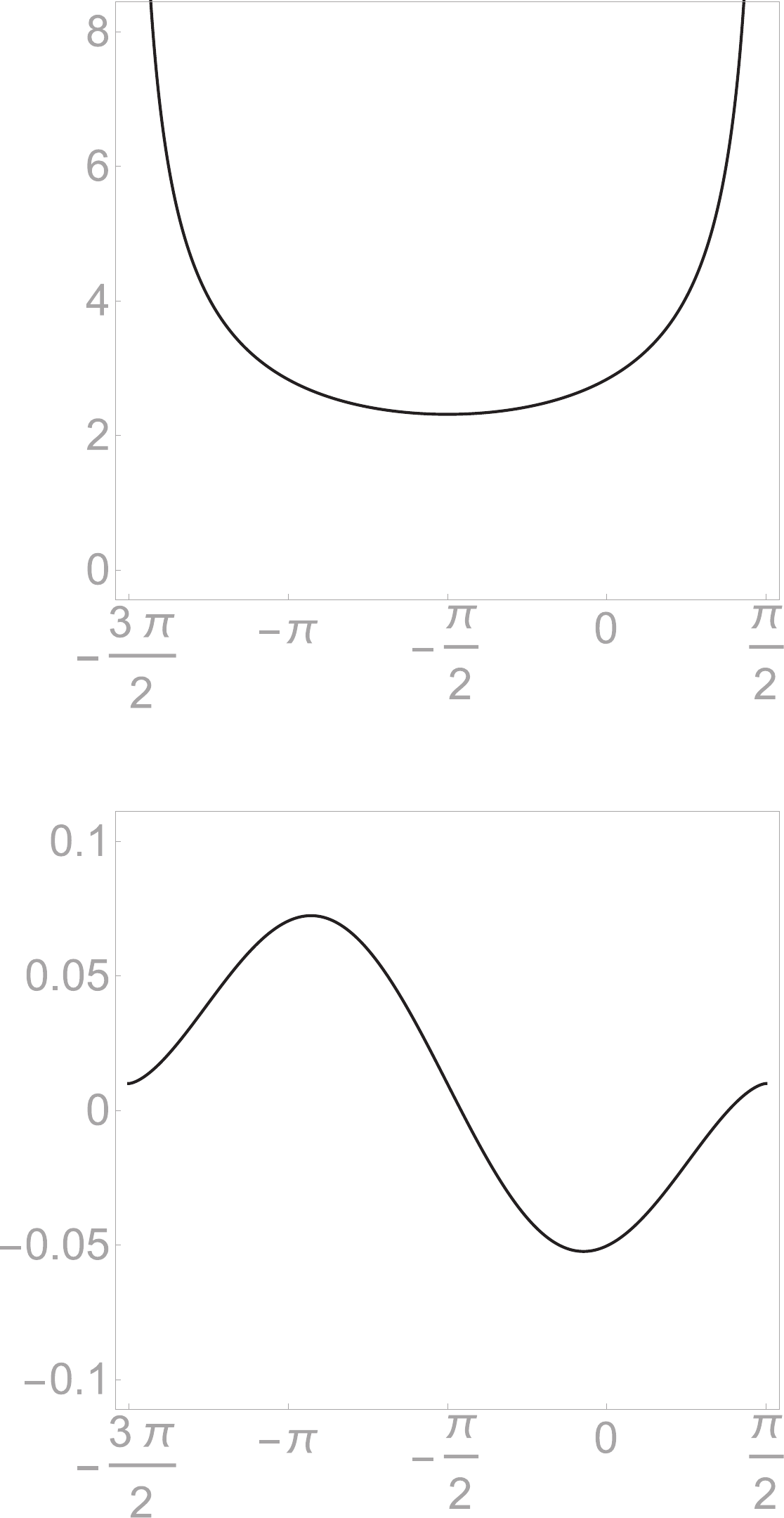}}	
			\put(113.5,-14){$\theta$}
			\put(113.5,49){$\theta$}
			\put(76,83){$\partial_s \theta$}
			\put(80,20){$\sigma$}
		\end{overpic}}
		\vspace*{1.3cm}
		\caption{An axially flowing C-clamp catenary.  As in prior figure, with parameters $\phi = -90^\circ$, $|\bv| = 0.78$, $T=0.1$, $q =0.61$, $R_\nu = 0.5$, $m=1$, $\epsilon=0.001$.  Finite length curve subtending $2\pi$ due to the special condition $k = -\bar{T}$. }
		\label{clamp}
	\end{figure}

Figures \ref{loop} and \ref{clamp} show examples of $2\pi$-subtending special cases for perfectly downward translation $\phi = -90^\circ$.    These two curves are symmetric about their midpoint, being formed from a $\pi$-subtending curve and its mirror image.  Their tensions are symmetric about the value $mT^2$.  Because the tension takes on negative values in portions of the concave-upward portions of the curves, these solutions are unstable.  Note that Figure \ref{loop} was generated with $R_\nu = 0.4$, and that both curves were generated with a smaller value of the cutoff, $\epsilon = 0.001$, than in Section \ref{examples}.  These curves have been scaled to have a length of two units.

\vspace*{1cm}

\begin{figure}[H]
	\hspace*{7.3cm}
	\hspace*{0.7cm}\begin{overpic}[scale=0.32]{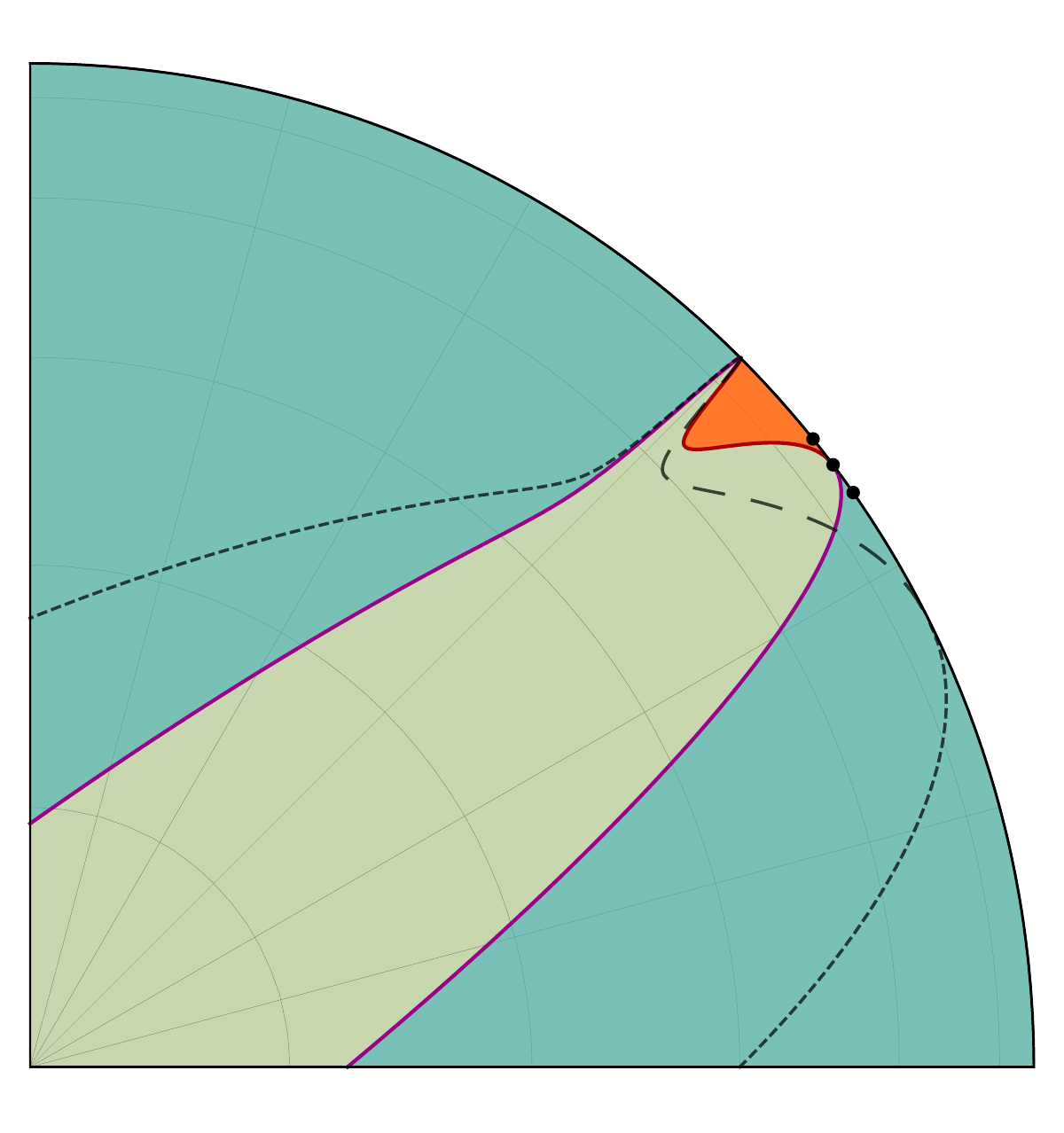}
		\put(89,41){\includegraphics[scale=0.2]{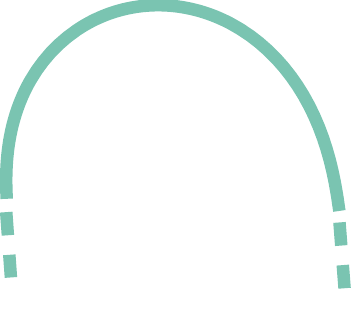}}
		\put(73,61){\includegraphics[scale=0.2]{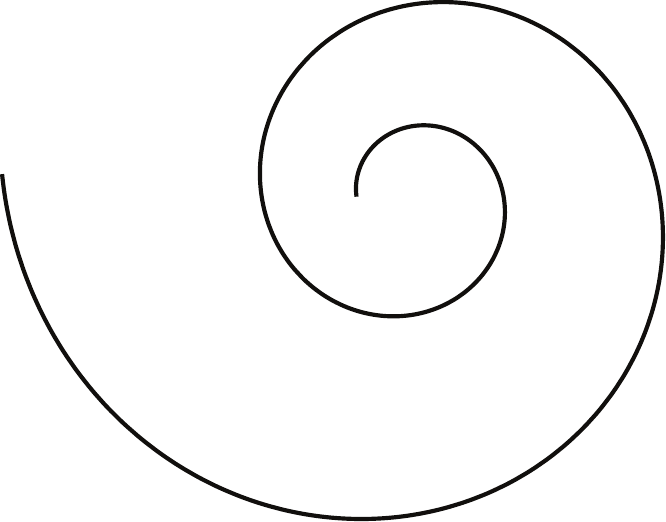}}
		\put(58,83){\includegraphics[scale=0.2]{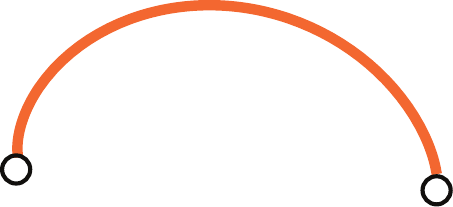}}
		\put(-107,5){\includegraphics[scale=0.34]{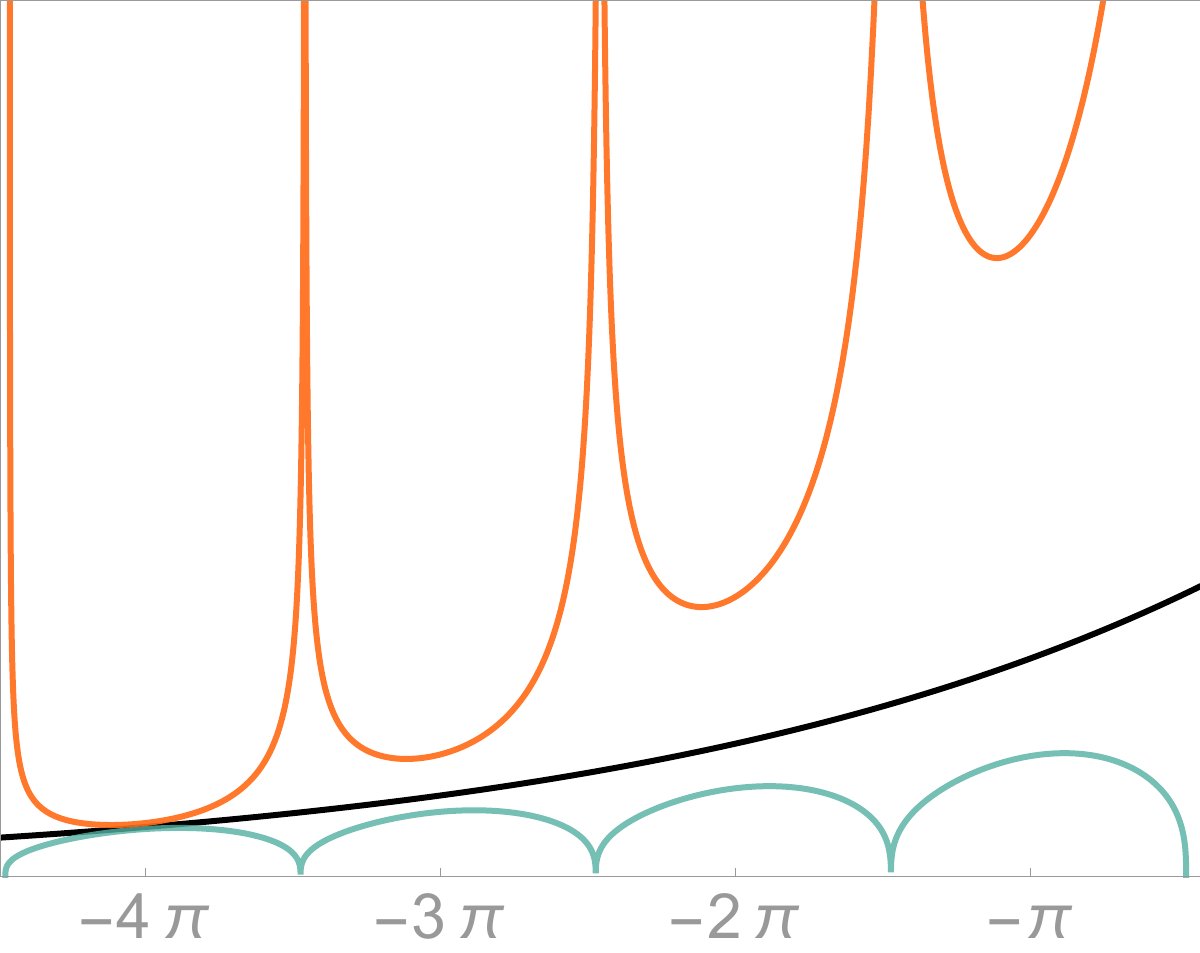}}
		\put(124,72){\color{black}\vector(0,1){12}}
		\put(122.5,85){$\uvc{z}$}
		\put(124,72){\color{black}\vector(0.1,-1){1}}
		\put(123.5,56){$\uvc{v}$}
		\put(-1.5,93){$q$}
		\put(88,-0.5){$|\bv|$}
		\put(-1,-0.5){$T$}		
		\put(-59,3){$\theta$}
		\put(-118,48){$\partial_s\theta$}
	\end{overpic}
	\caption{Three rigidly translating catenaries taken from the dotted locations along the $T = 0$ equator in the projection (center) of a $\phi = -88.5^\circ$, $R_\nu = 0.5$ spherical map octant.  The configurations (right), and several periods of orbits (left), correspond to a generic $\pi$-subtending \textcolor{PI}{heteroclinic connection}, a logarithmic spiral {\bf{special case}}, and a generic $\pi$-subtending \textcolor{OR}{double-sided blow up} solution.  Configurations are drawn by adjusting $C$ so that a length of unity corresponds with a subtended angle of approximately $\pi$, while orbits are drawn using the scaling $C = 4 (1-R_\nu) \bar{U} \bar{Q} \pi$ to eliminate overlap in the range shown.  The other parameters are $|\bv| = $ (\textcolor{PI}{0.82}, 0.8, \textcolor{OR}{0.78}), $q =$ (\textcolor{PI}{0.57}, 0.6, \textcolor{OR}{0.63}).}
	\label{pseudophase2}
\end{figure}

\begin{figure}[H]
	\centering
	\hspace*{9cm}
	\scalebox{1}{%
		\begin{overpic}[scale=0.27]{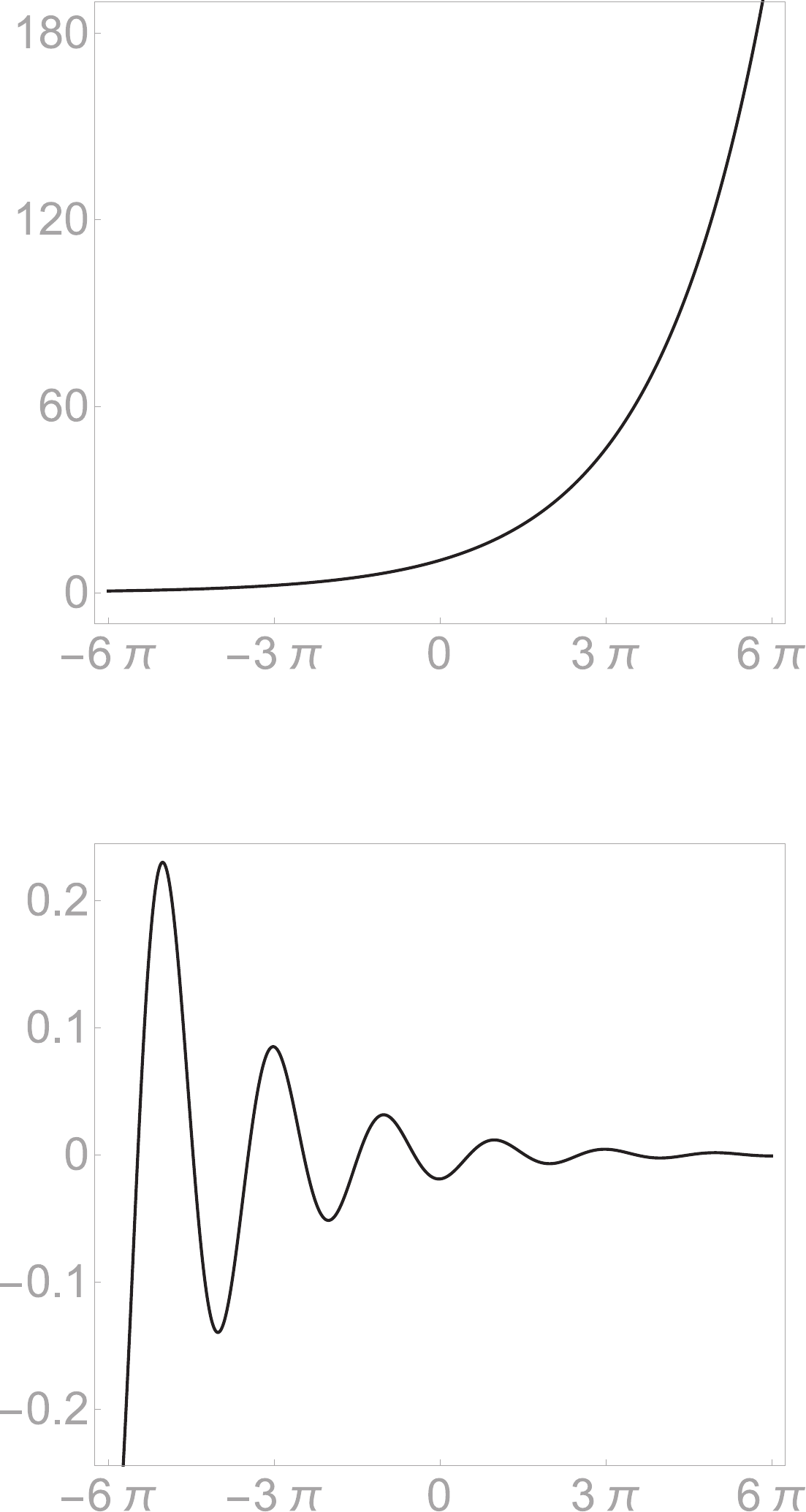}
			\put(28.5,-4.5){$\theta$}
			\put(28.5,51){$\theta$}
			\put(-4,77){$\partial_s \theta$}
			\put(0.3,23){$\sigma$}
			\hspace*{-1cm}
			\put(-79,7){\includegraphics[scale=0.4]{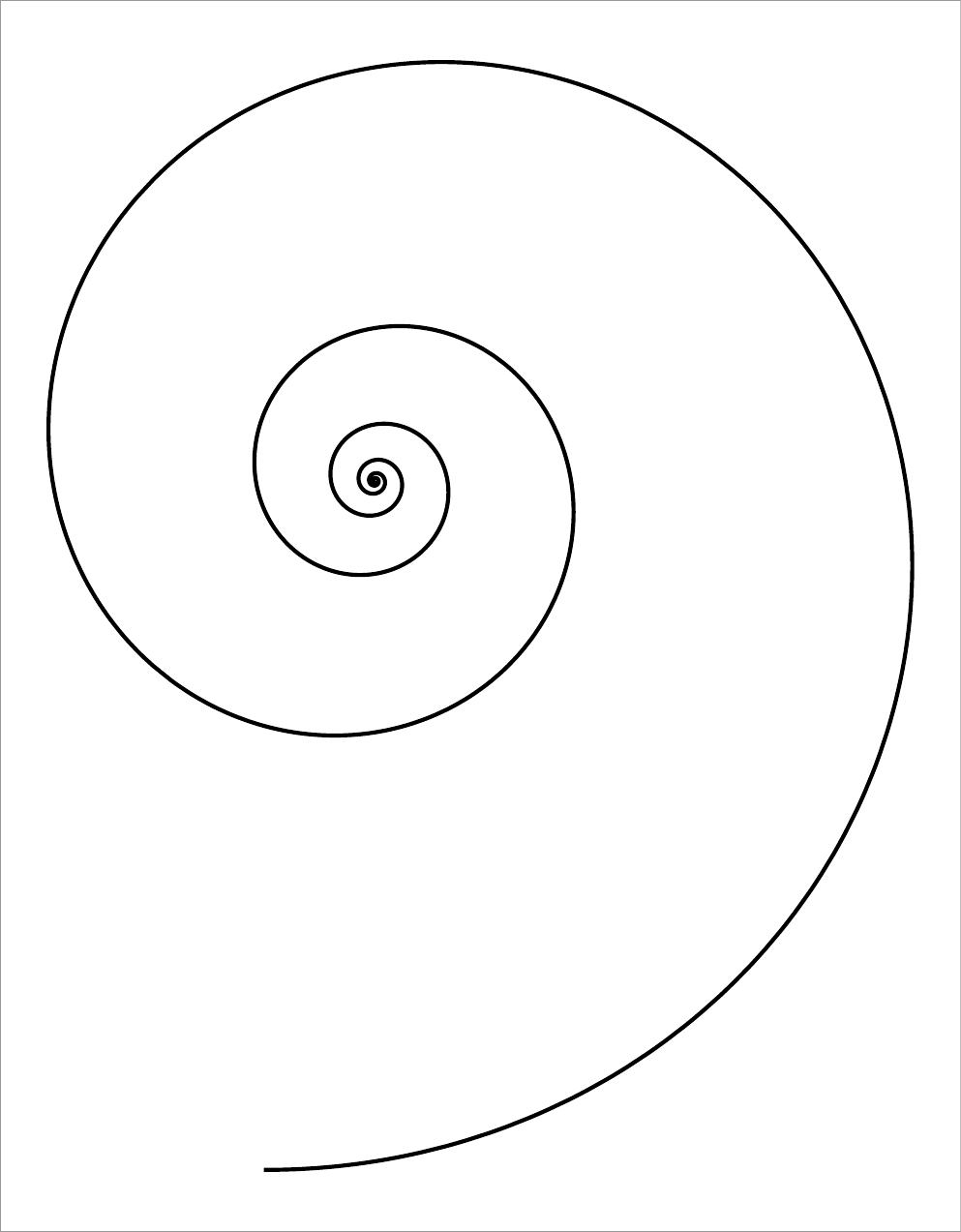}}
			\put(-87.2,70){\color{black}\vector(0,1){15}}
			\put(-88.2,86){$\uvc{z}$}
			\put(-87.2,70){\color{black}\vector(0.03,-0.99){0.4}}
			\put(-87.8,52.3){$\uvc{v}$}
		\end{overpic}}
		\vspace*{4mm}
		\caption{A logarithmic spiral with damped sinusoidal stress is a rigid catenary.  As in Figures \ref{curve90}-\ref{semim90T} and \ref{loop}-\ref{clamp}, with parameters $\phi = -88.5^\circ$, $|\bv| = 0.79$, $T=0$, $q =0.61$, $R_\nu = 0.5$, $m=1$, $\epsilon=0$.  Rigidly translating infinite length curve subtending infinite angle due to the special condition $k = \bar{T} = 0$.}
		\label{spiral}
	\end{figure}

Another curious special case is found at either of two triple points, where the three regions meet.  This occurs on the equator, where $T=0$ and the curve rigidly translates.  In Figure \ref{pseudophase2}, we show three curves taken from the $T=0$ latitude line for a translational angle $\phi = -88.5^\circ$ and Stokes drag anisotropy $R_\nu = 0.5$.  At the triple points, both exponents $\bar{T}$ and $k$ are zero, and the solutions are logarithmic spirals $\partial_s\theta = Ce^{(1-R_\nu) \bar{U} \bar{Q}\theta}$.  The tension is a damped sinusoid, and the compressive regions make the curves unstable.  A portion of spiral subtending $12\pi$, scaled to be 12 units long, is shown in Figure \ref{spiral}.  No cutoff $\epsilon$ is necessary to generate this curve, as all the poles and fixed points have disappeared.   For perfectly downward translation $\phi = -90^\circ$, one triple point spiral becomes a straight line with $\bv  = - q\uvc{z}$, and the other becomes a circle $\partial_s\theta = C$ with $\bv = - \frac{2 q}{1 + R_\nu}\uvc{z}$ and an undamped sinusoidal tension. To our knowledge, the circle is the only periodic solution, excepting the arbitrary shapes (mentioned in Section \ref{fullproblem}) that are possible for isotropic drag $R_\nu = 1$.  It is also a freely sedimenting solution.  
It is perhaps worth pointing out that, unlike elastic curves with bending resistance, and with the exception of $R_\nu = 1$ arbitrary shapes, all of our augmented catenaries are non-inflectional.

\begin{figure}[H]
	\hspace*{0.85cm}
	\scalebox{1}{
		\begin{overpic}[scale=0.4]{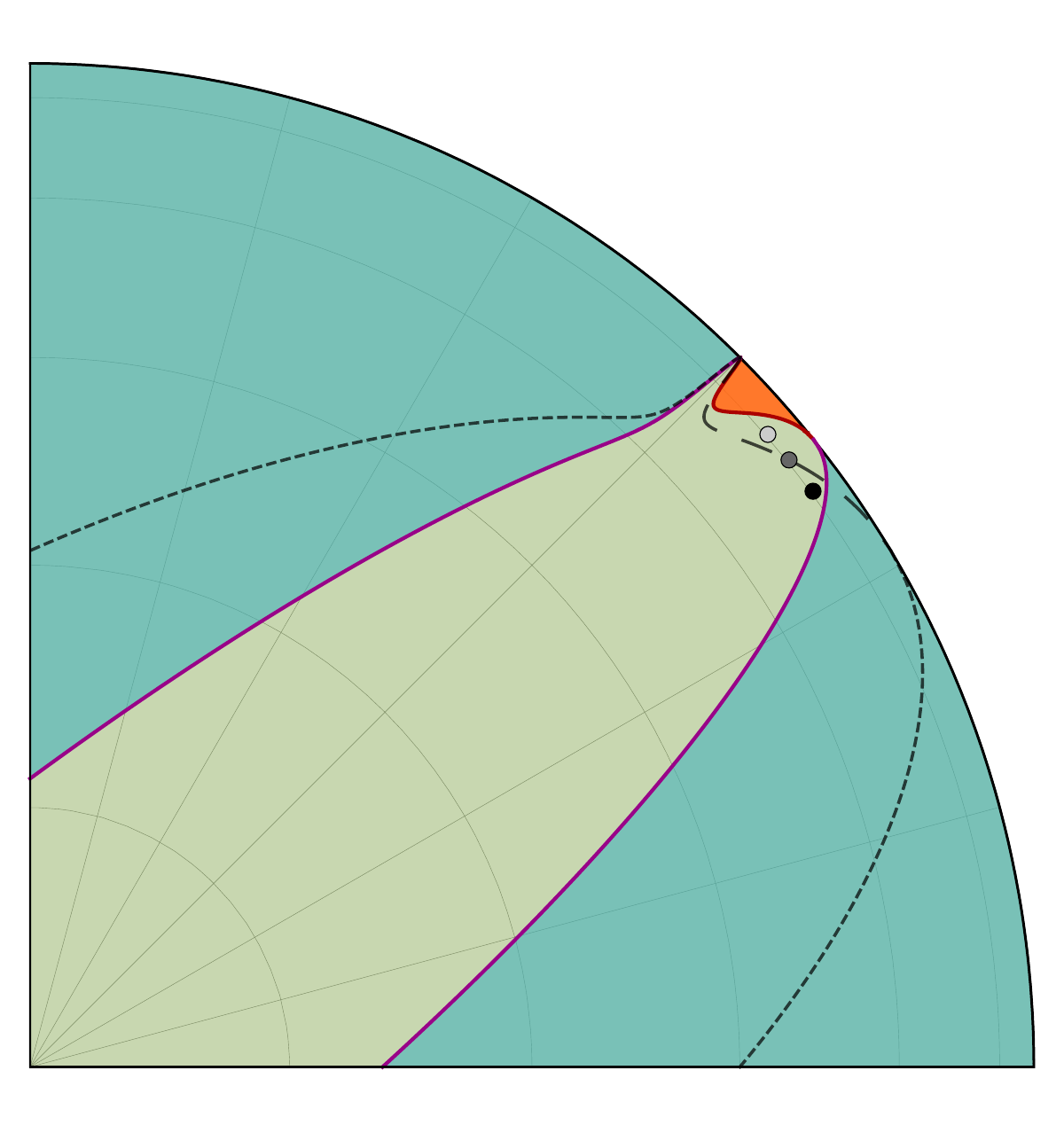}
			
			\put(69,68){\includegraphics[scale=0.1]{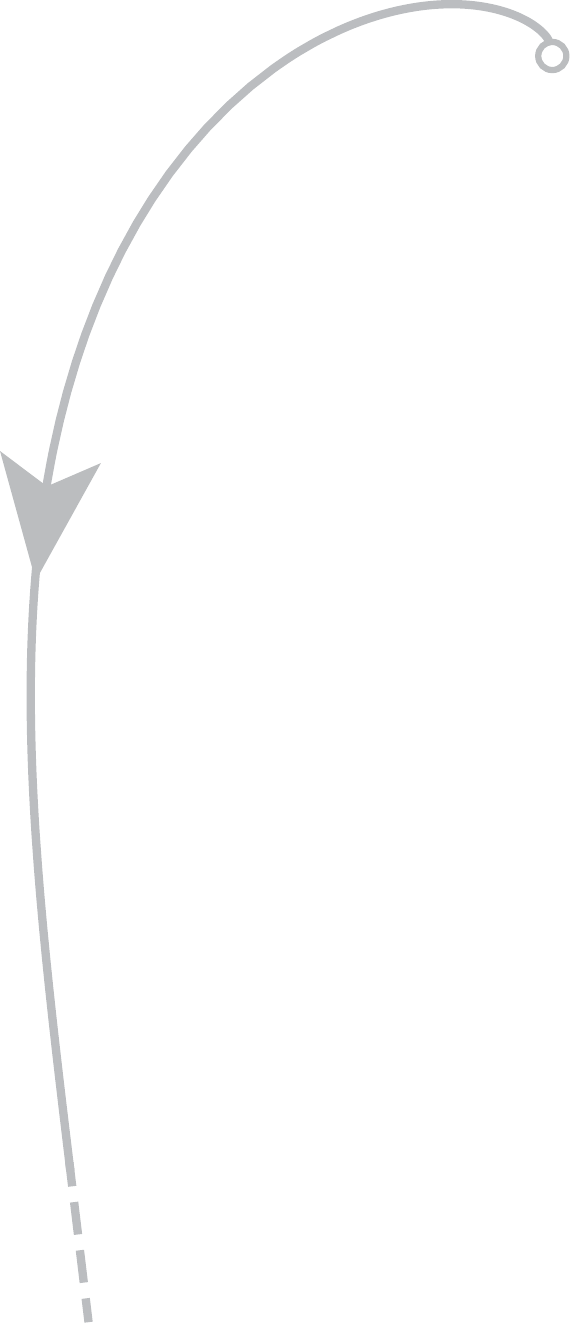}}
			\put(77,61){\includegraphics[scale=0.1]{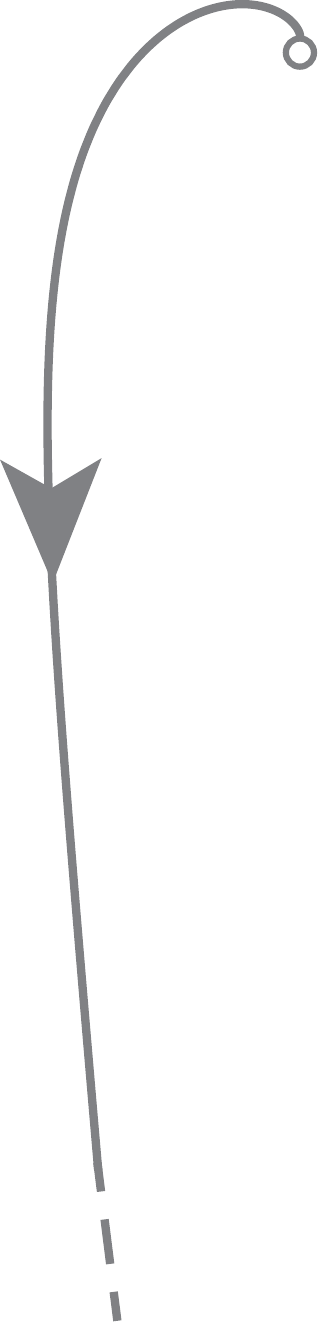}}
			\put(84,52){\includegraphics[scale=0.1]{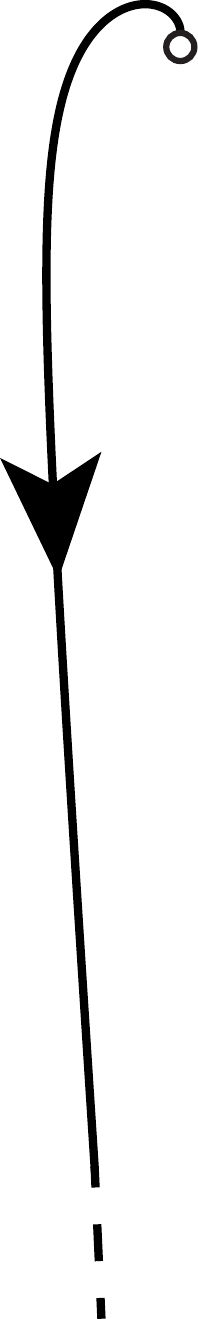}}

			\put(118,0){\includegraphics[scale=0.34]{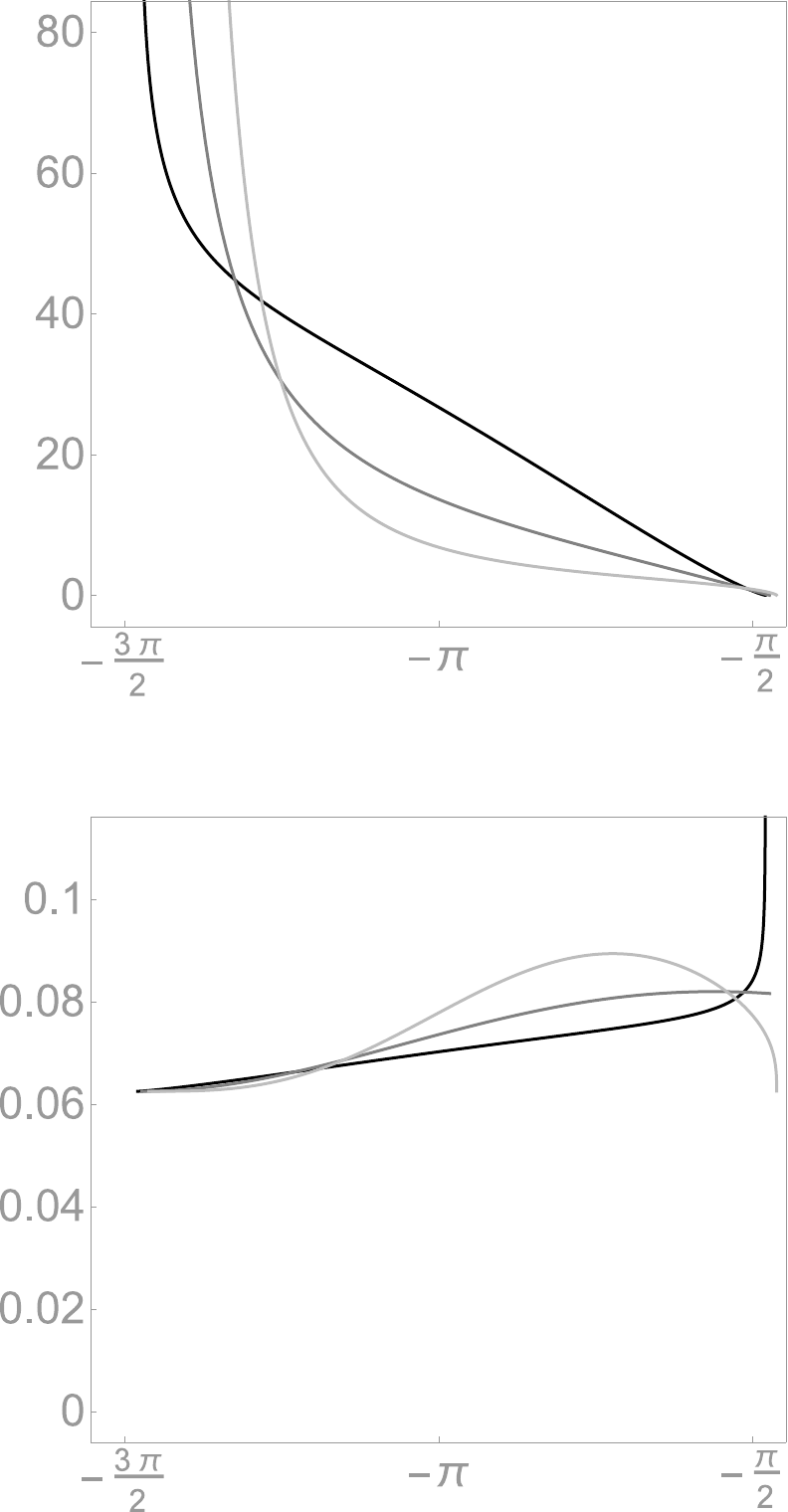}}
			\put(63,91){\color{black}\vector(0,1){12}}
			\put(62,104){$\uvc{z}$}
			\put(63,91){\color{black}\vector(0,-1){12}}
			\put(61.5,75){$\uvc{v}$}

			\put(113,79){$\partial_s\theta$}
			\put(147,51){$\theta$}
			
			\put(116,23){$\sigma$}
			\put(147,-4){$\theta$}
			
			\put(-1.5,93){$q$}
			\put(88,-0.5){$|\bv|$}
			\put(-1,-0.5){$T$}
		\end{overpic}}
		\vspace*{0.2cm}
		\caption{Projection and configurations (left), orbits (upper right), and stresses (lower right) for three single-sided blowup catenaries taken from the dotted locations along the $T = 0.25$ latitude line, with fixed parameters $\phi=-89^\circ$ and $R_\nu = 0.6$.  Darker curves indicate higher velocities.  In crossing the long-dashed boundary line, the stress on the leading side of the body transitions from {\bf{infinite}} to the \textcolor{Lgr}{finite} value $mT^2$, with the special case on the boundary taking on an \textcolor{Mgr}{intermediate finite} value.
The other parameters are $|\bv| =$ (\textcolor{Lgr}{0.74}, \textcolor{Mgr}{0.76}, 0.78), $q =$ (\textcolor{Lgr}{0.63}, \textcolor{Mgr}{0.60}, 0.57), $m=1$, $\epsilon=0.005$.}
		\label{stress1}
	\end{figure}

\begin{figure}[H]
	\hspace*{0.85cm}
	\scalebox{1}{
		\begin{overpic}[scale=0.4]{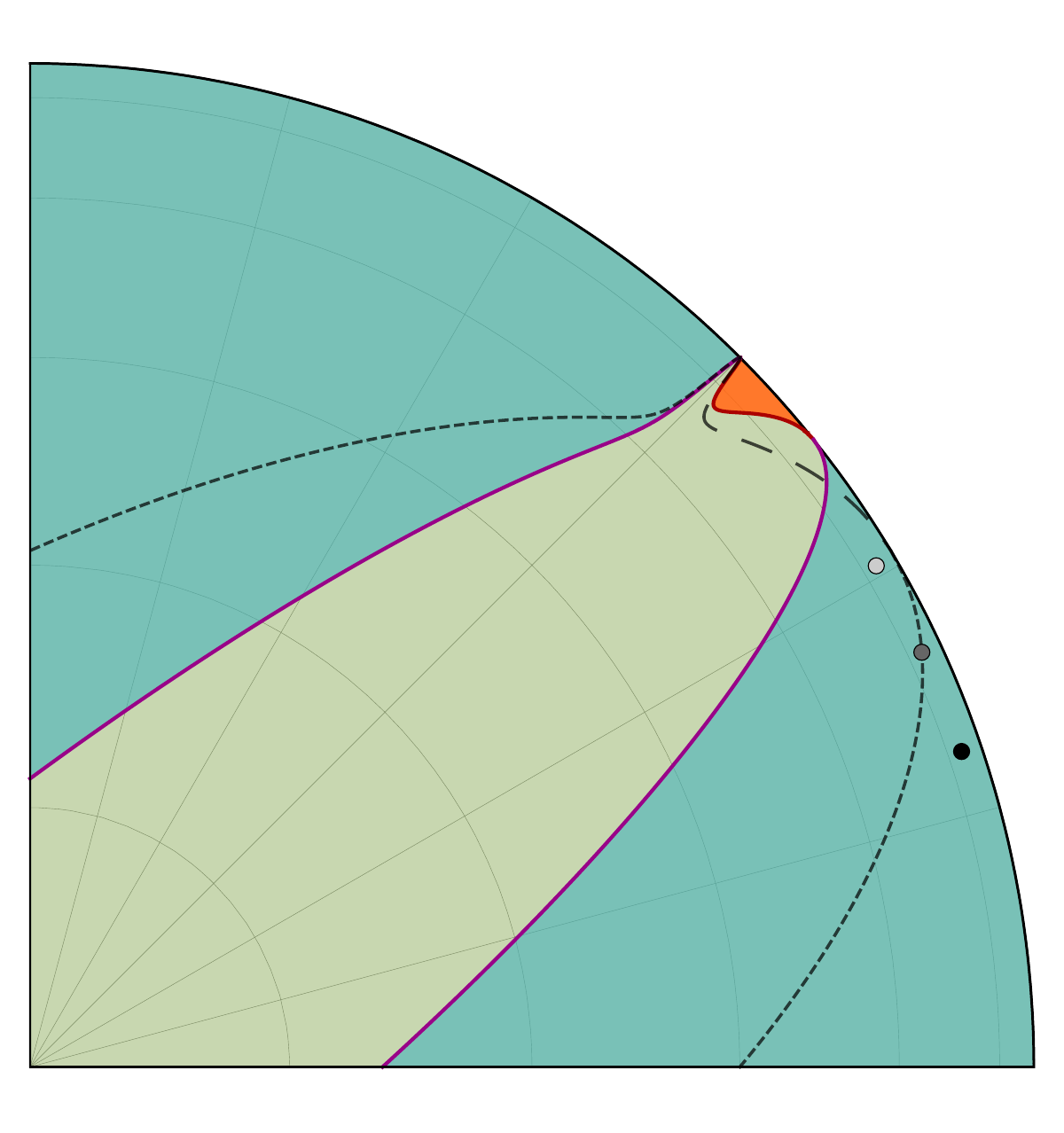}
			
			\put(78,56){\includegraphics[scale=0.1]{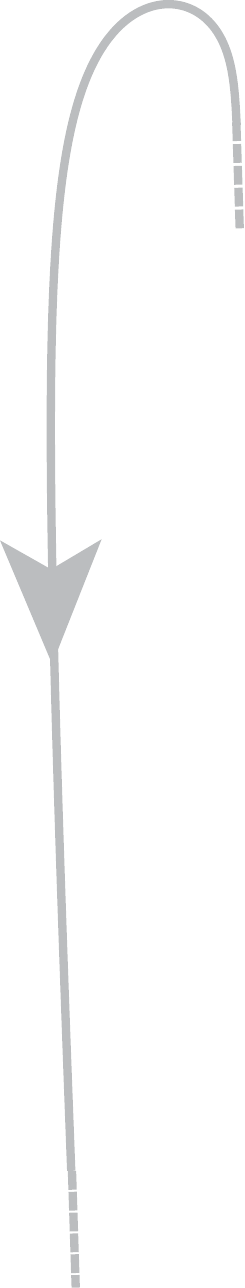}}
			\put(84,44){\includegraphics[scale=0.1]{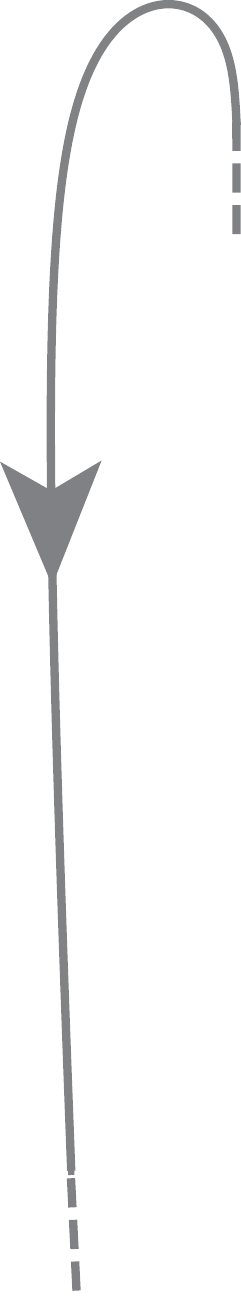}}
			\put(89,33){\includegraphics[scale=0.1]{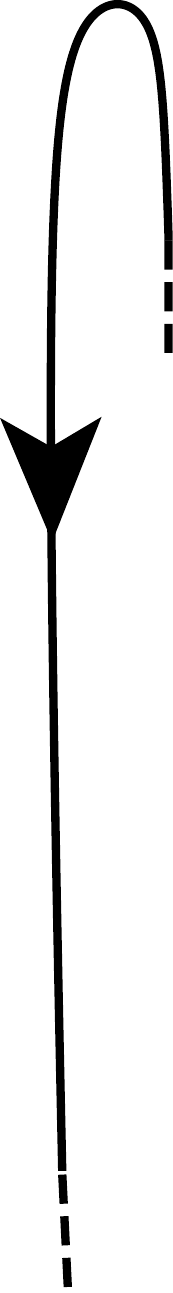}}

			\put(118,0){\includegraphics[scale=0.34]{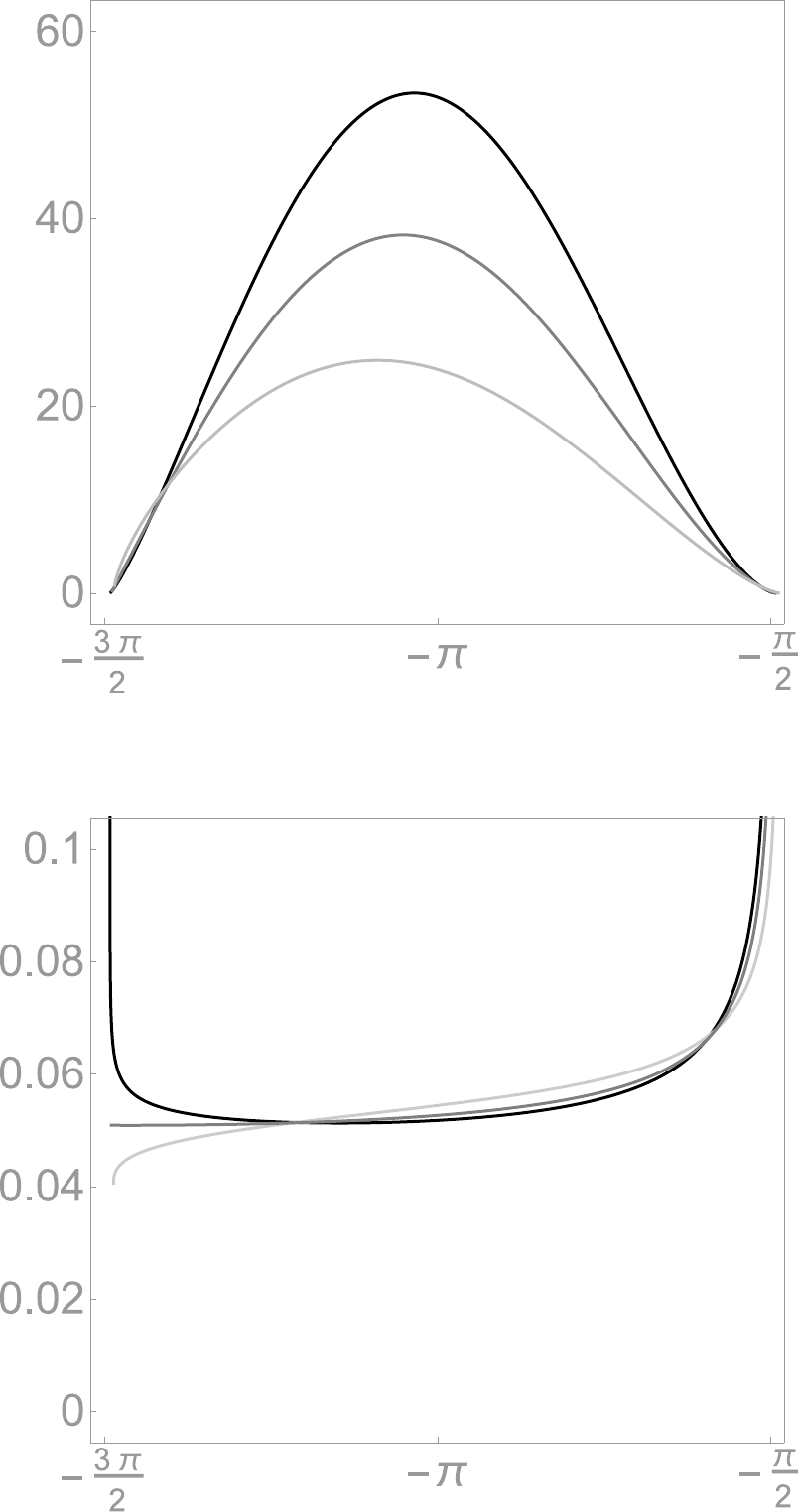}}
			\put(65,90){\color{black}\vector(0,1){12}}
			\put(64,103){$\uvc{z}$}
			\put(65,90){\color{black}\vector(0,-1){12}}
			\put(63.5,74){$\uvc{v}$}

			\put(113,79){$\partial_s\theta$}
			\put(147,51){$\theta$}
			
			\put(116,24){$\sigma$}
			\put(147,-4){$\theta$}
			
			\put(-1.5,93){$q$}
			\put(88,-0.5){$|\bv|$}
			\put(-1,-0.5){$T$}
		\end{overpic}}
		\vspace*{0.2cm}
		\caption{As in prior figure, for three heteroclinic catenaries taken from the dotted locations along the $T = 0.2$ latitude line, with $|\bv| =$ (\textcolor{Lgr}{0.84}, \textcolor{Mgr}{0.88}, 0.93), $q =$ (\textcolor{Lgr}{0.49}, \textcolor{Mgr}{0.41}, 0.30).  
		In crossing the short-dashed boundary line, the stress on the trailing side of the body transitions from {\bf{infinite}} to the \textcolor{Lgr}{finite} value $mT^2$, with the special case on the boundary taking on an \textcolor{Mgr}{intermediate finite} value.
}
		\label{stress2}
	\end{figure}

Examples of stress transitions upon crossing dashed lines are shown in Figures \ref{stress1} and \ref{stress2}.
The solutions are generated with $R_\nu = 0.6$ and a small value of the cutoff $\epsilon=0.005$.
While the configurations and orbits display no qualitative changes, the tension on one side of the curve changes from asymptotically infinite to the finite value $mT^2$.  For the special cases on the boundaries, the tension takes on an intermediate finite value.     In these figures, we encounter the unexpected phenomena of infinitely long curves with finite tension on one side, and semi-infinite curves with finite tension on both sides.  There also exist cases, not shown, of infinitely long curves with finite tension on both sides.

While this behavior is not fully understood, we may gather a clue to its origin by considering rigid collinear translations ($T = \bar{U}=0$) in the part of the heteroclinic region with finite tension.  As is evident from inspection of Figure \ref{planarmap}, these solutions occur when $0<k<1$.  Equation \eqref{keq}, the constraints on twice-rescaled quantities, and the positivity of $\bar{Q}$ then indicate that $R_\nu\bar{W} + \bar{Q} > 0$ while $\bar{W} + \bar{Q} = -1$.  In terms of the once-rescaled quantities, $R_\nu|\bv| < q$ and $|\bv| > q$.  This means that portions of the curve aligned with the direction of translation and body force are moving slower than they would be if allowed to free-fall at their terminal velocity, while portions perpendicular are moving faster than their terminal velocity.  	

\begin{figure}[H]	
	\centering
	\subcaptionbox{$\phi=90^o$}[.31\linewidth][c]{%
		\begin{tikzonimage}[scale=0.25]{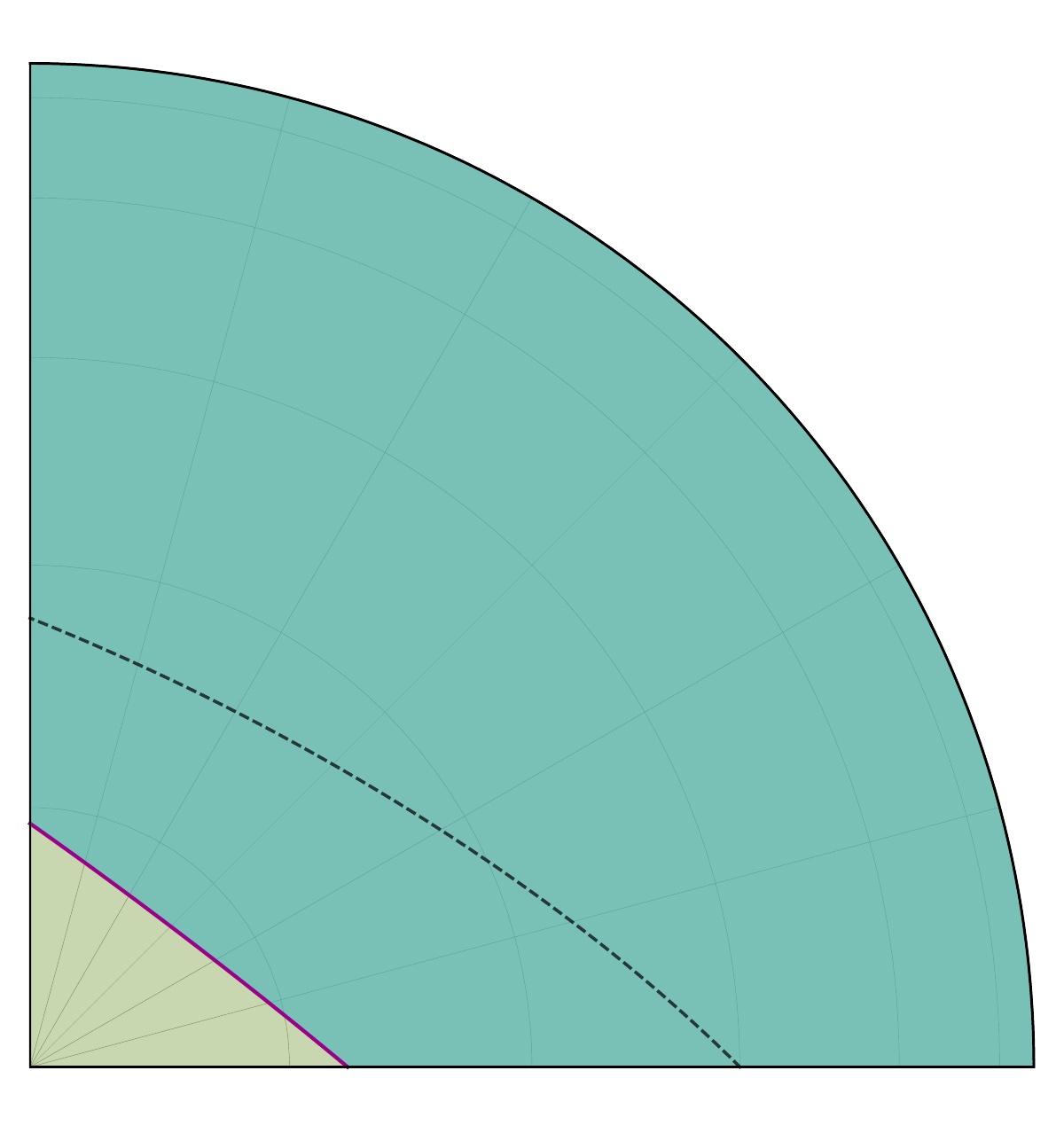} 
			\draw[->,>=stealth] (0.86,0.8,0) -- (0.86,0.93,0) node[above] {$\uvc{z} \hspace{1mm} \uvc{v}$};
			\draw  (0.93,0) node[]{ $|\bv|$};
			\draw  (-0.03,0.93) node[]{q};
			\draw (0.05,0) node[]{T};
		\end{tikzonimage}}\quad
		\subcaptionbox{$\phi=0^o$}[.31\linewidth][c]{%
			\begin{tikzonimage}[scale=0.25] {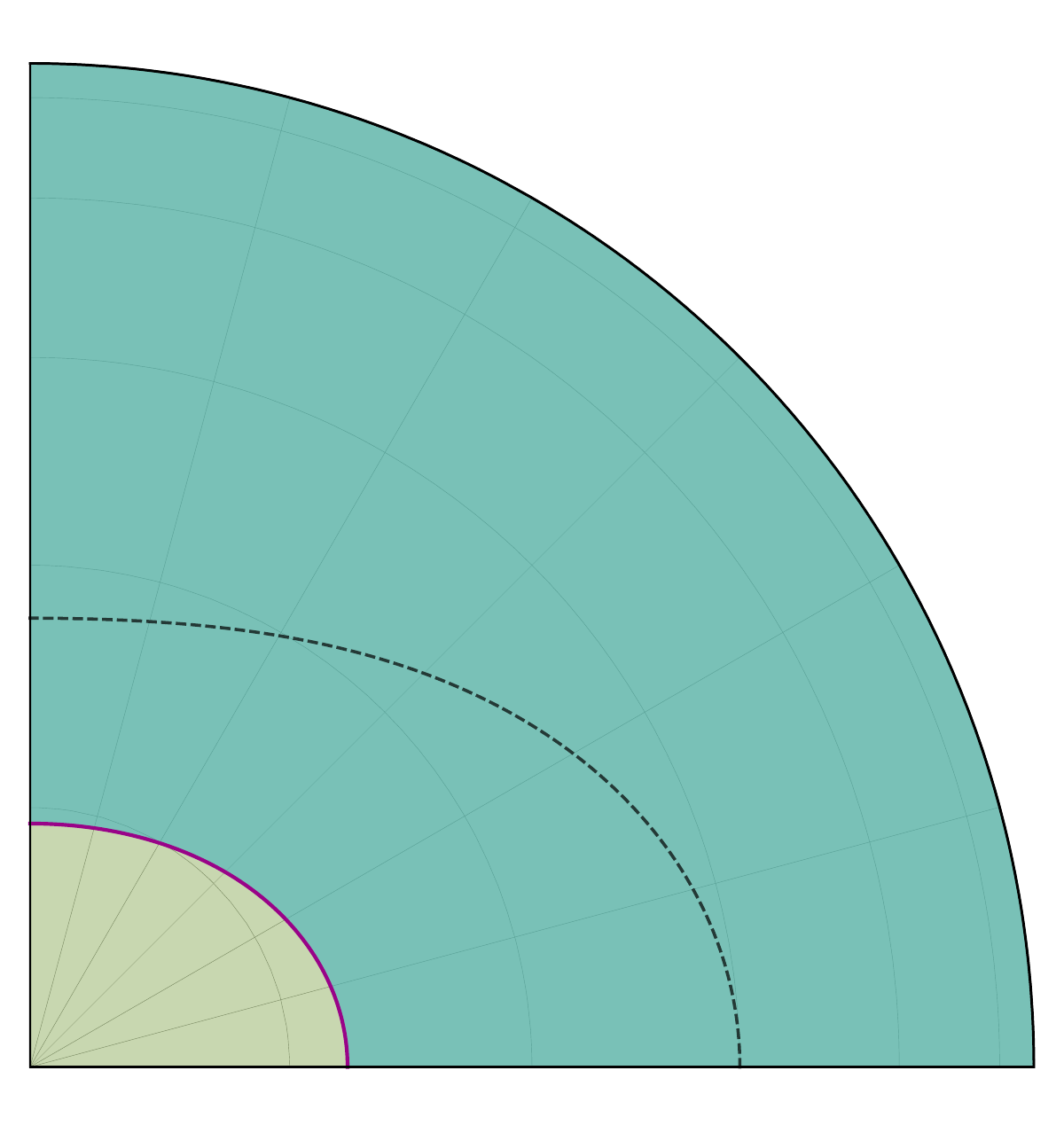} 
				\draw[->,>=stealth] (0.86,0.8,0) -- (0.86,0.93,0) node[above] {$\uvc{z}$};
				\draw[->,>=stealth] (0.86,0.8,0) -- (0.99,0.8,0);
				\draw (1.02,0.81) node[] {$\uvc{v}$};
				\draw  (0.93,0) node[]{ $|\bv|$};
				\draw  (-0.03,0.93) node[]{q};
				\draw (0.05,0) node[]{T};
			\end{tikzonimage}}\quad
			\subcaptionbox{$\phi=-60^o$}[.31\linewidth][c]{%
				\begin{tikzonimage}[scale=0.25]{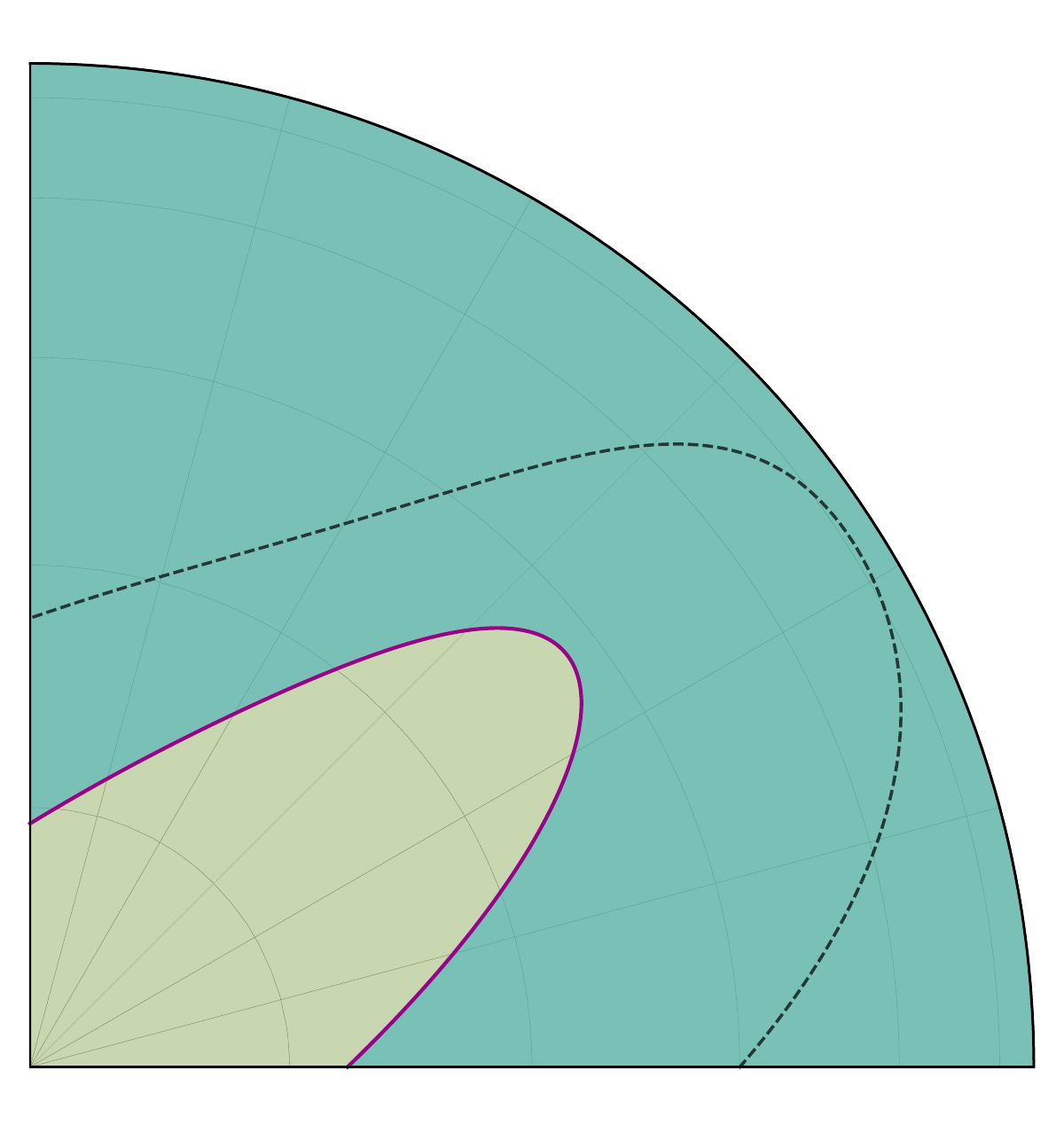} 
					\draw[->,>=stealth] (0.86,0.8,0) -- (0.86,0.93,0) node[above] {$\uvc{z}$};
					\draw[->,>=stealth] (0.86,0.8,0) -- (0.925,0.6875,0); \draw (0.932,0.645) node[] {$\uvc{v}$};
					\draw  (0.93,0) node[]{ $|\bv|$};
					\draw  (-0.03,0.93) node[]{q};
					\draw (0.05,0) node[]{T};
				\end{tikzonimage}}\quad
				
				\bigskip
				
				\subcaptionbox{$\phi=-81.79^o$}[.31\linewidth][c]{%
					\begin{tikzonimage}[scale=0.25] {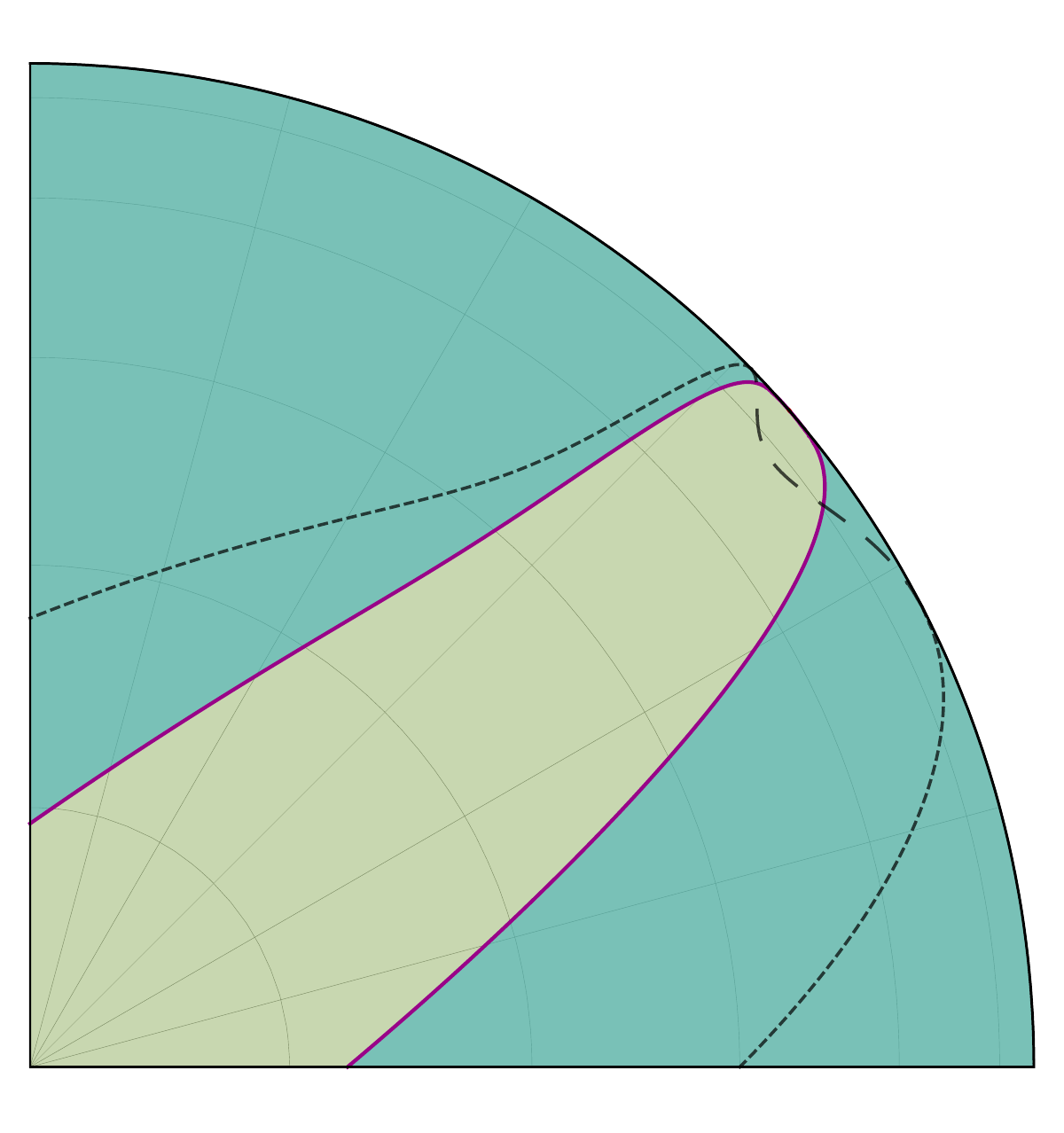} 
						\draw[->,>=stealth] (0.86,0.8,0) -- (0.86,0.93,0) node[above] {$\uvc{z}$};
						\draw[->,>=stealth] (0.86,0.8,0) -- (0.8786,0.6713,0); \draw (0.88,0.63) node[]  {$\uvc{v}$};
						\draw  (0.93,0) node[]{ $|\bv|$};
						\draw  (-0.03,0.93) node[]{q};
						\draw (0.05,0) node[]{T};
					\end{tikzonimage}}\quad
					\subcaptionbox{$\phi=-88^o$}[.31\linewidth][c]{%
						\begin{tikzonimage}[scale=0.25] {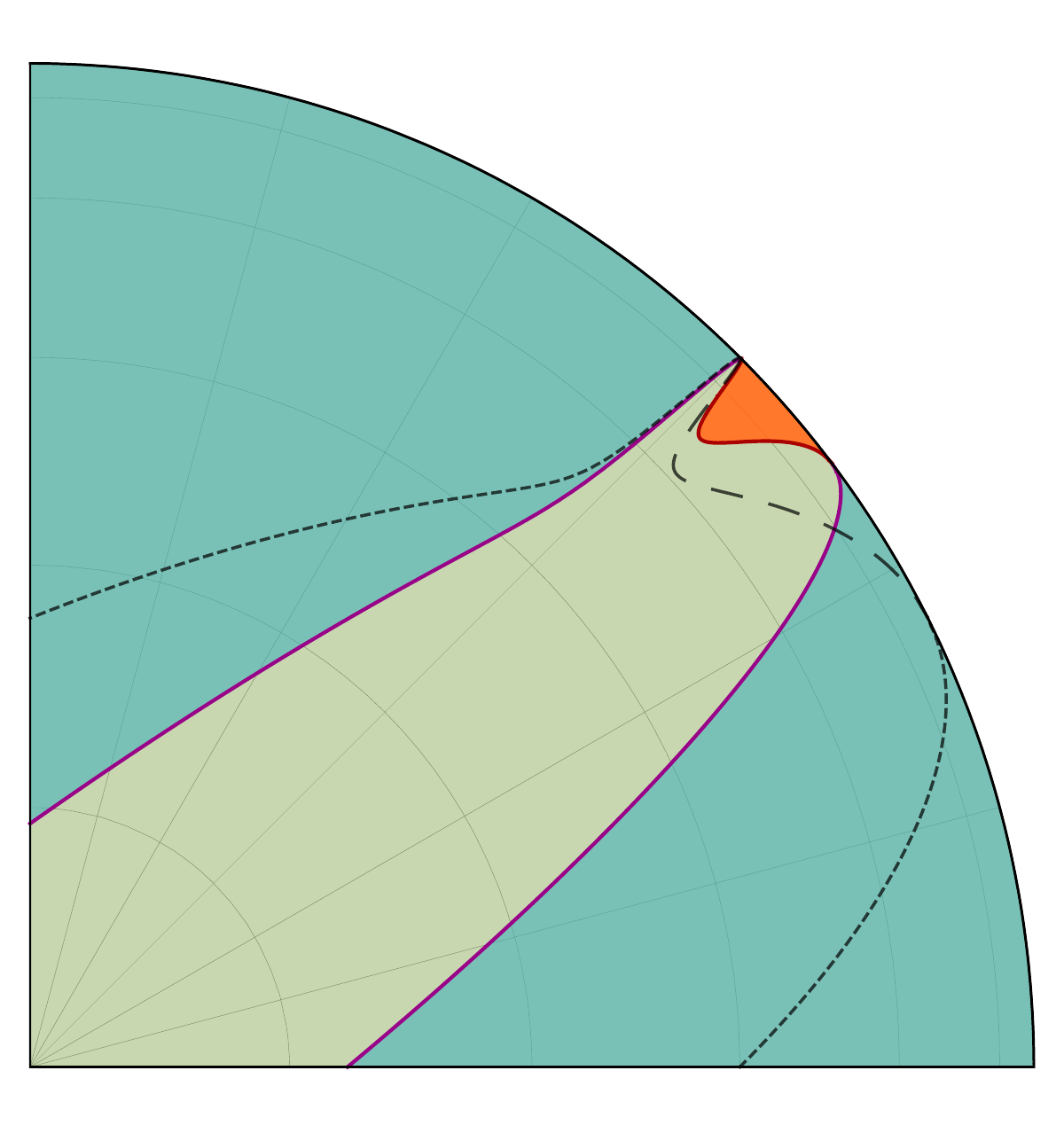}						
							\draw[->,>=stealth] (0.86,0.8,0) -- (0.86,0.93,0) node[above] {$\uvc{z}$};
							\draw[->,>=stealth] (0.86,0.8,0) -- (0.865,0.67,0); \draw (0.87,0.619) node[]  {$\uvc{v}$};
							\draw  (0.93,0) node[]{ $|\bv|$};
							\draw  (-0.03,0.93) node[]{q};
							\draw (0.05,0) node[]{T};
						\end{tikzonimage}}\quad
						\subcaptionbox{$\phi=-90^o$}[.31\linewidth][c]{%
							\begin{tikzonimage}[scale=0.25]{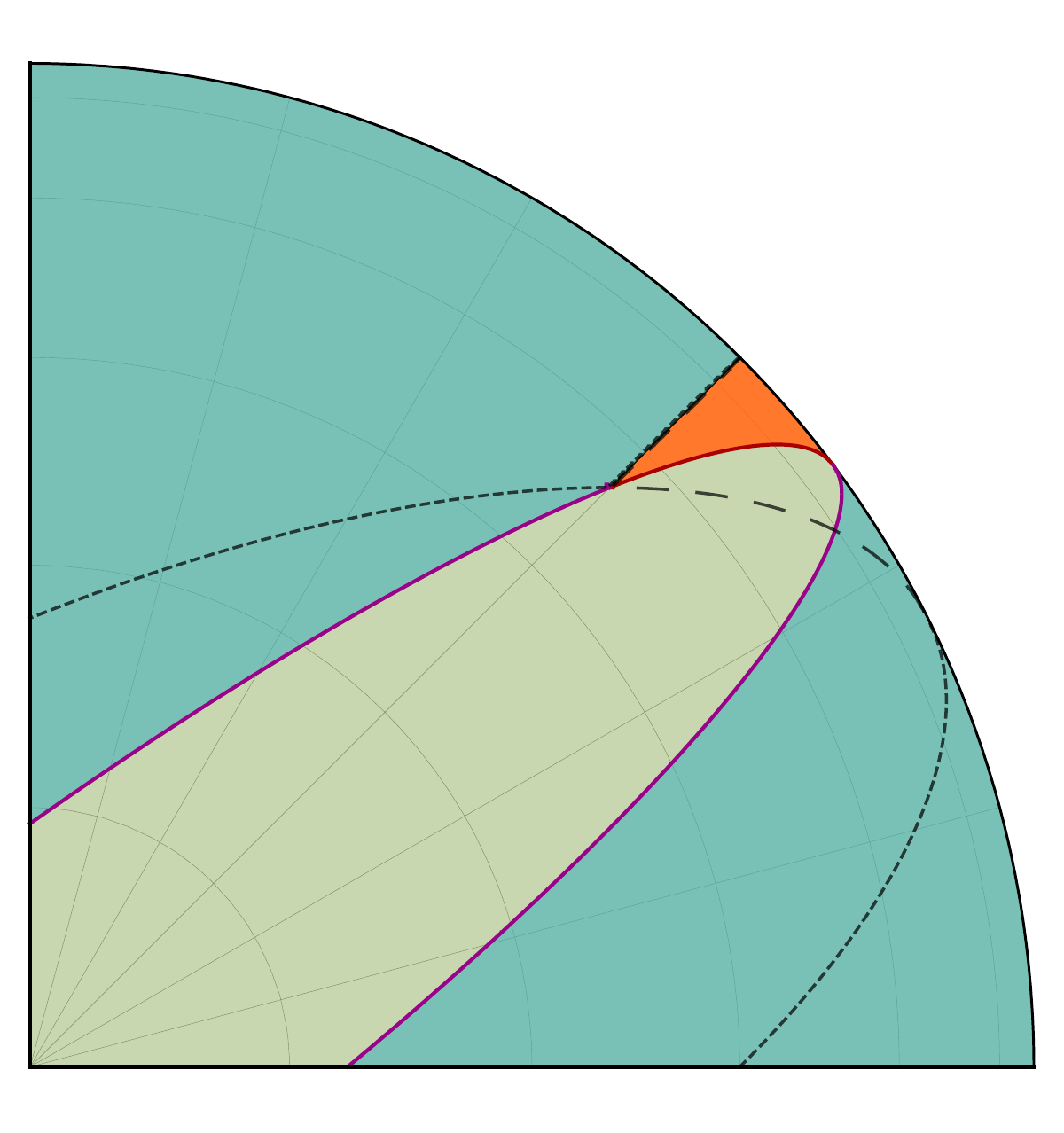}  
								\draw[->,>=stealth] (0.86,0.8,0) -- (0.86,0.93,0) node[above] {$\uvc{z}$};
								\draw[->,>=stealth] (0.86,0.8,0) -- (0.86,0.67,0); \draw (0.868,0.613) node[]  {$\uvc{v}$};
								\draw  (0.93,0) node[]{ $|\bv|$};
								\draw  (-0.03,0.93) node[]{q};
								\draw (0.05,0) node[]{T};
							\end{tikzonimage}}\quad
							\caption{Effect of translational angle $\phi$ on spherical maps with fixed drag anisotropy $R_\nu = 0.5$, shown through orthographic projections onto the $|\bv|-q$ plane.  
							Colored regions, colored lines, and dashed lines have the same meanings as in Figure \ref{sphericalmap}.}
							\label{phivar}
						\end{figure}
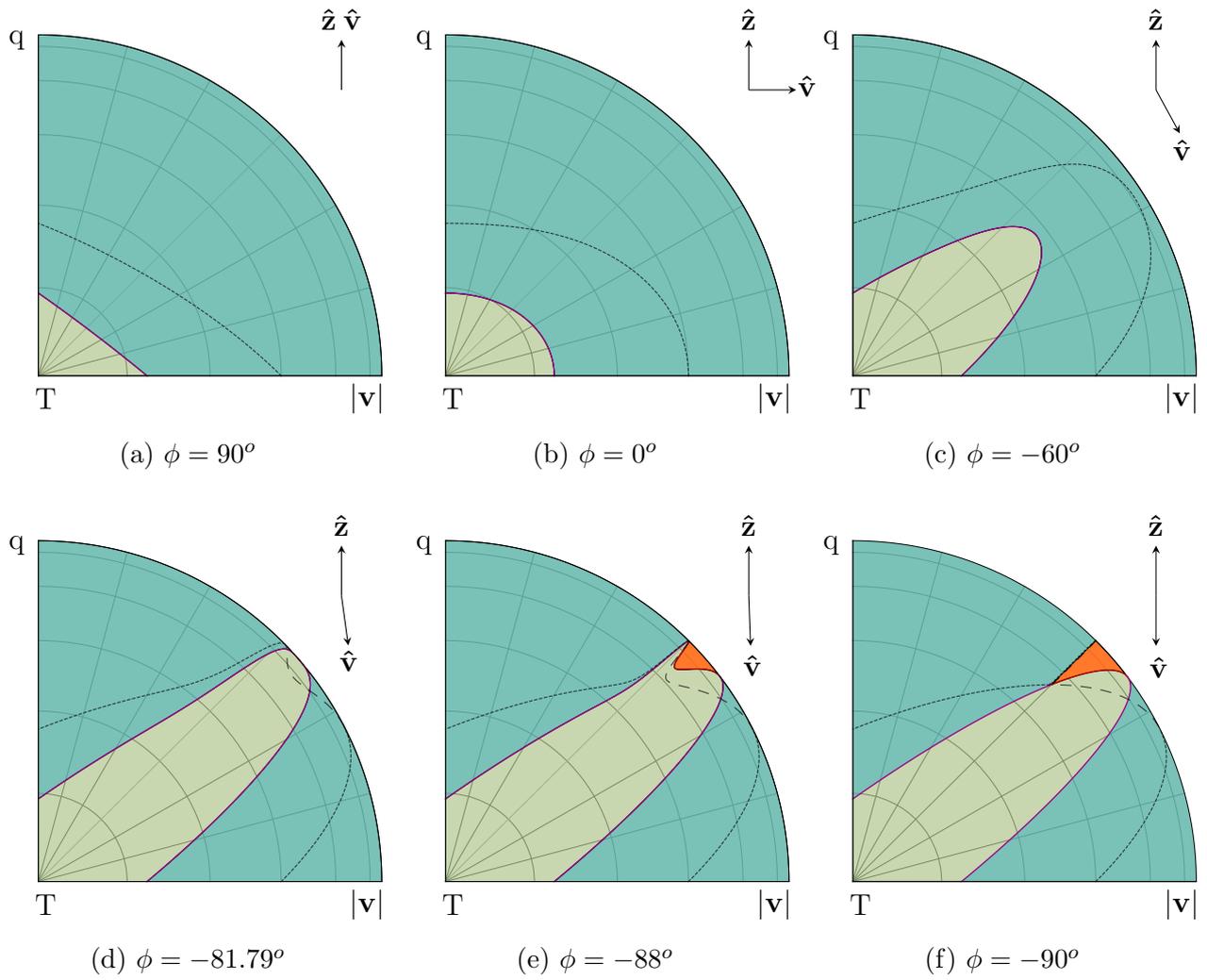

\begin{figure}[H]
							\centering
							\subcaptionbox{$R_\nu$=0}[.31\linewidth][c]{%
								\begin{tikzonimage}[scale=0.25]{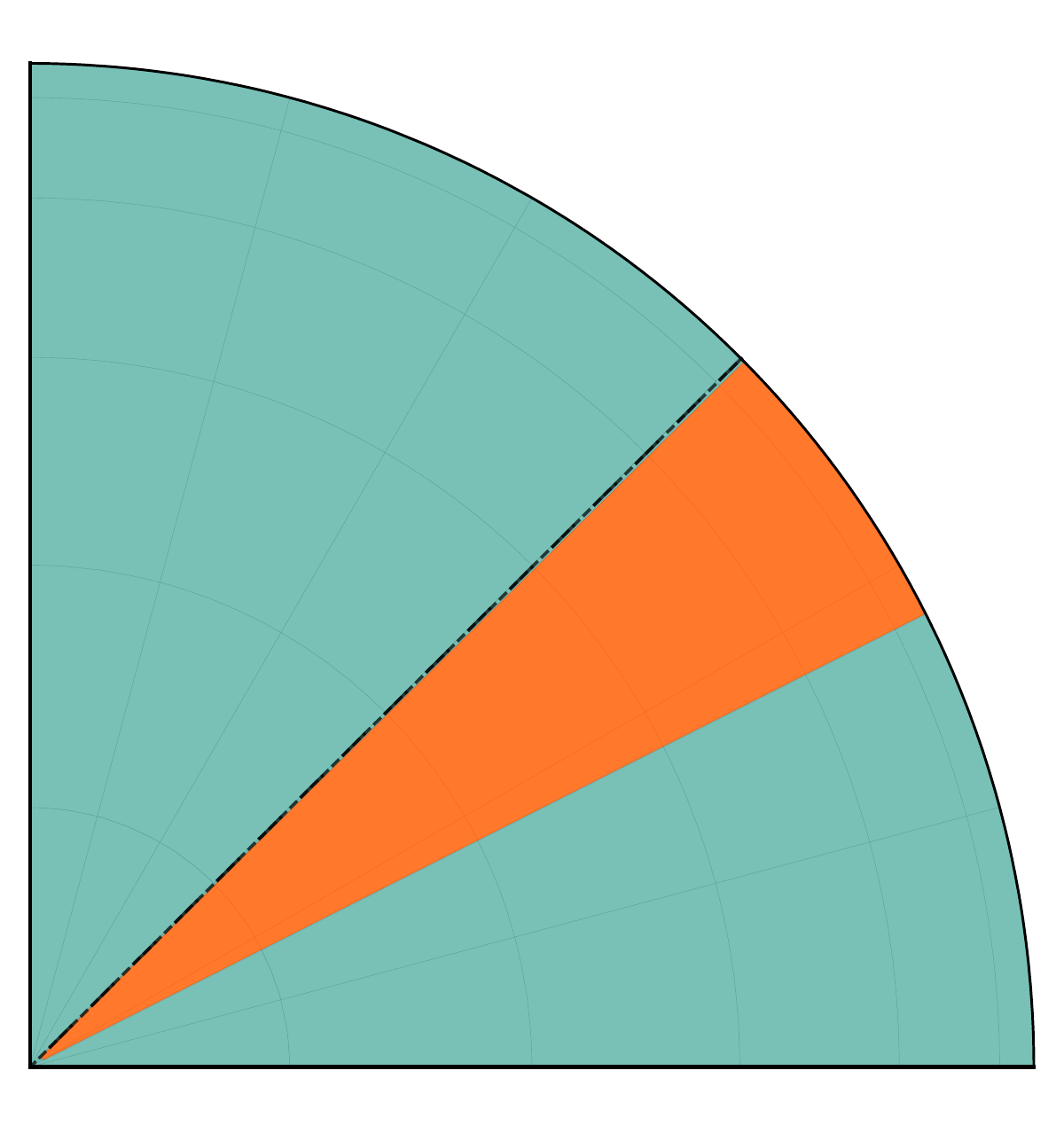} 
									\draw[->,>=stealth] (0.86,0.8,0) -- (0.86,0.93,0) node[above] {$\uvc{z}$};
									\draw[->,>=stealth] (0.86,0.8,0) -- (0.871,0.6704,0); \draw (0.88,0.63) node[]  {$\uvc{v}$};
									
									\draw  (0.93,0) node[]{ $|\bv|$};
									\draw  (-0.03,0.93) node[]{q};
									\draw (0.05,0) node[]{T};					
								\end{tikzonimage}}\quad
								\subcaptionbox{$R_\nu$=0.1}[.31\linewidth][c]{%
									\begin{tikzonimage}[scale=0.25] {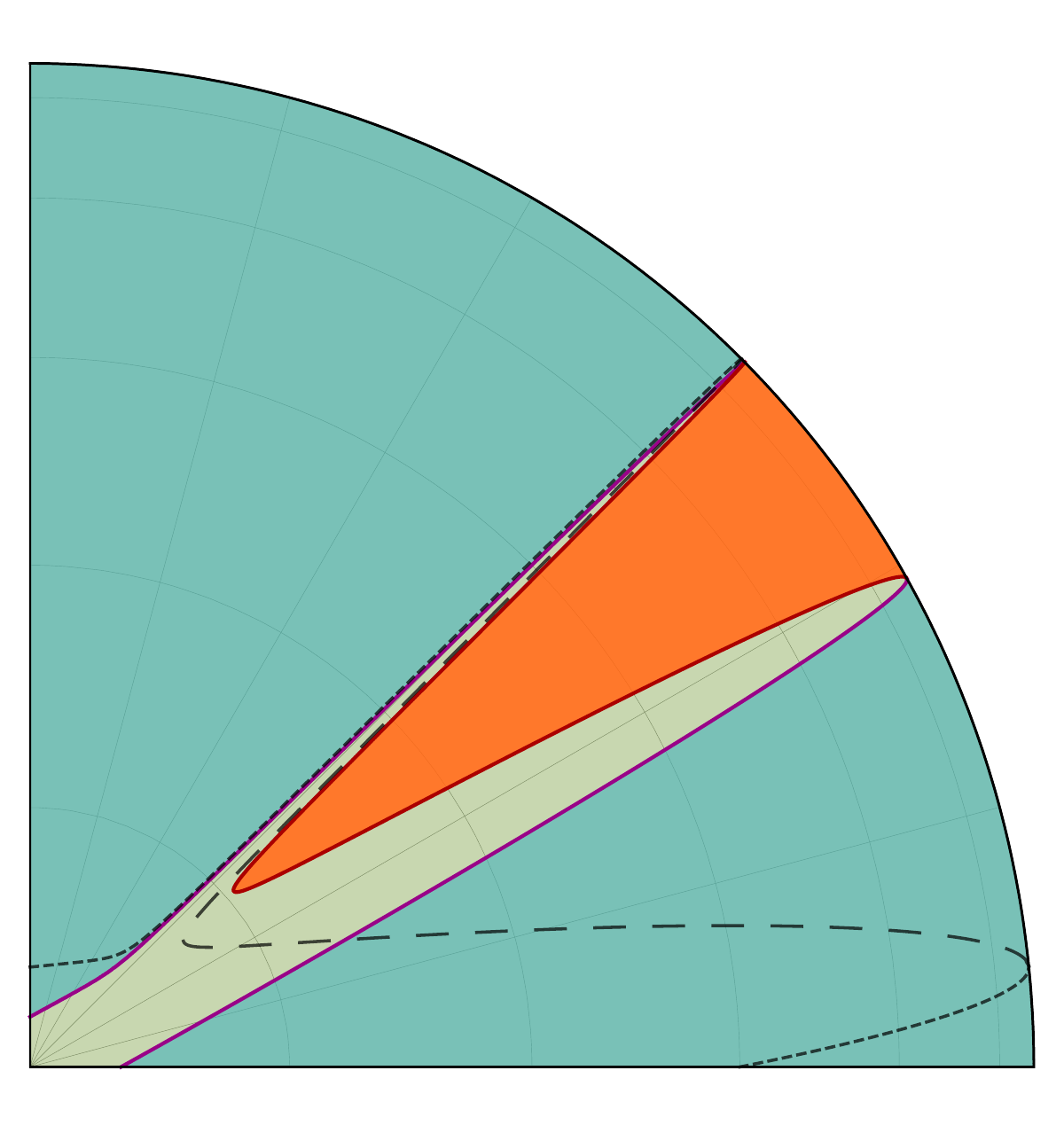}
										\draw  (0.93,0) node[]{ $|\bv|$};
										\draw  (-0.03,0.93) node[]{q};
										\draw (0.05,0) node[]{T};
									\end{tikzonimage}}\quad
									\subcaptionbox{$R_\nu$=0.2}[.31\linewidth][c]{%
										\begin{tikzonimage}[scale=0.25] {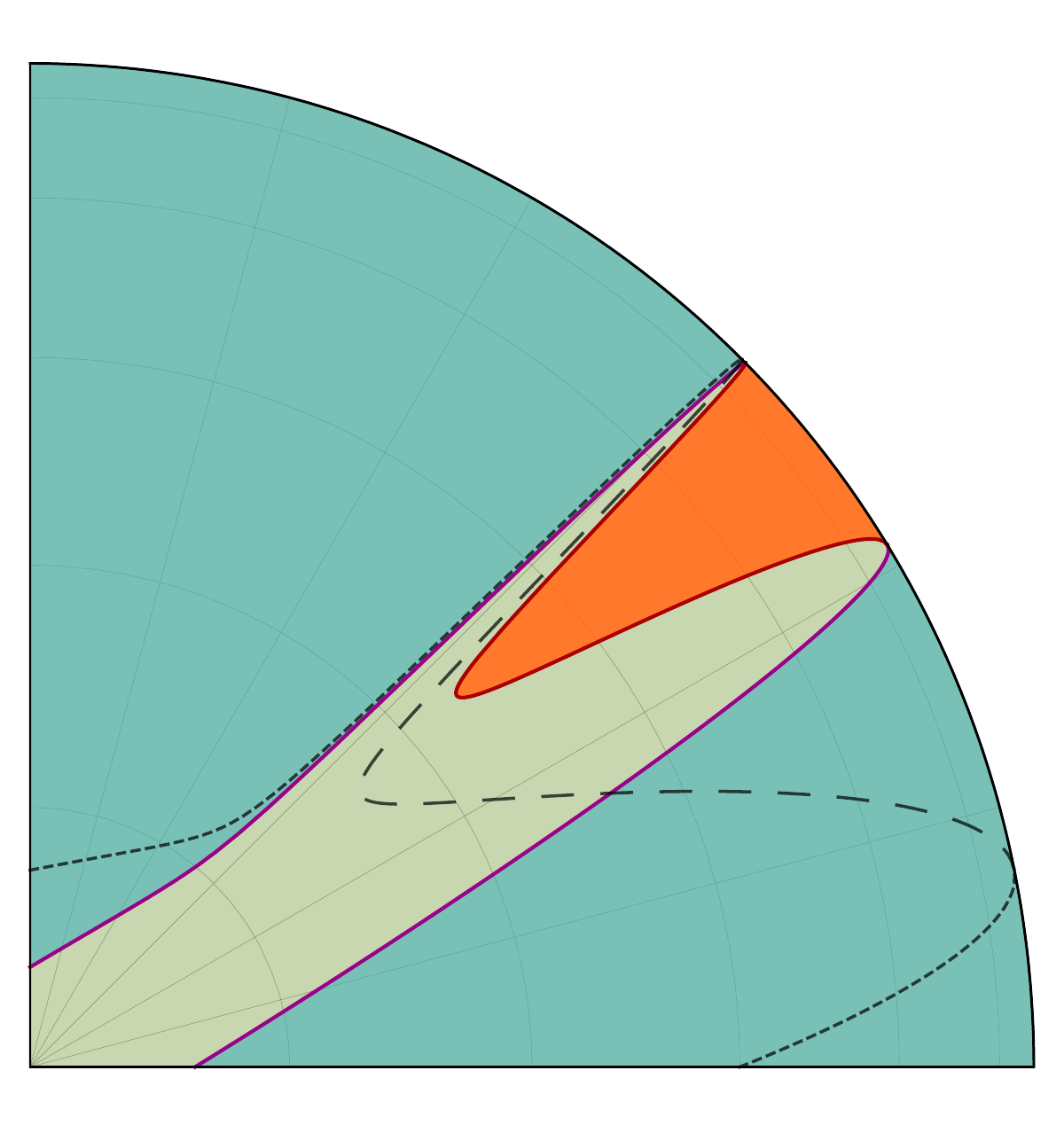}
											\draw  (0.93,0) node[]{ $|\bv|$};
											\draw  (-0.03,0.93) node[]{q};
											\draw (0.05,0) node[]{T};
										\end{tikzonimage}}\quad
										
										\bigskip
										\vspace*{2mm}
										
						\subcaptionbox{$R_\nu$=0.4}[.31\linewidth][c]{%
						\begin{tikzonimage}[scale=0.25] {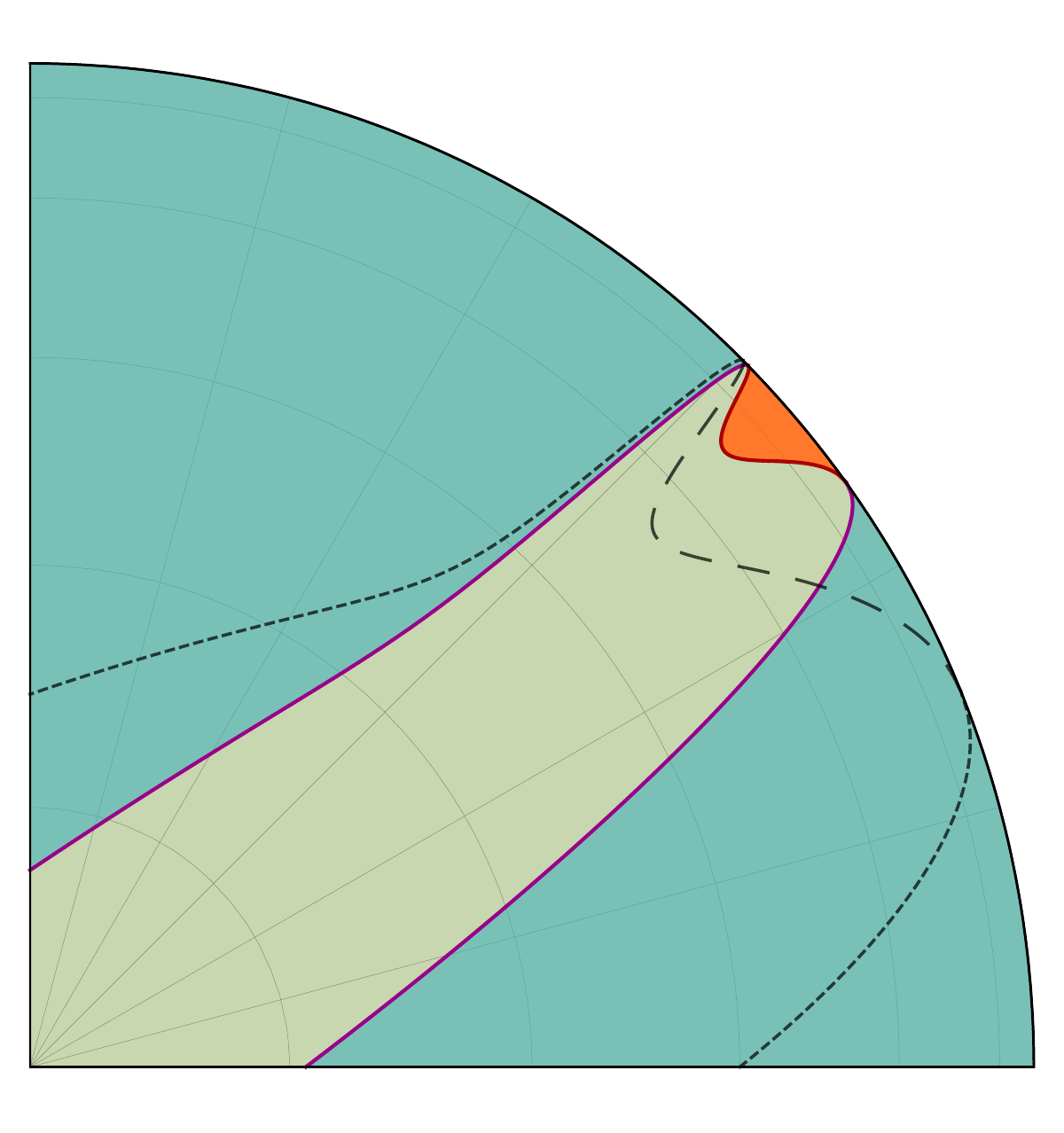}
							\draw  (0.93,0) node[]{ $|\bv|$};
							\draw  (-0.03,0.93) node[]{q};
							\draw (0.05,0) node[]{T};
							\end{tikzonimage}}\quad
							\subcaptionbox{$R_\nu$=0.6}[.31\linewidth][c]{%
						\begin{tikzonimage}[scale=0.25] {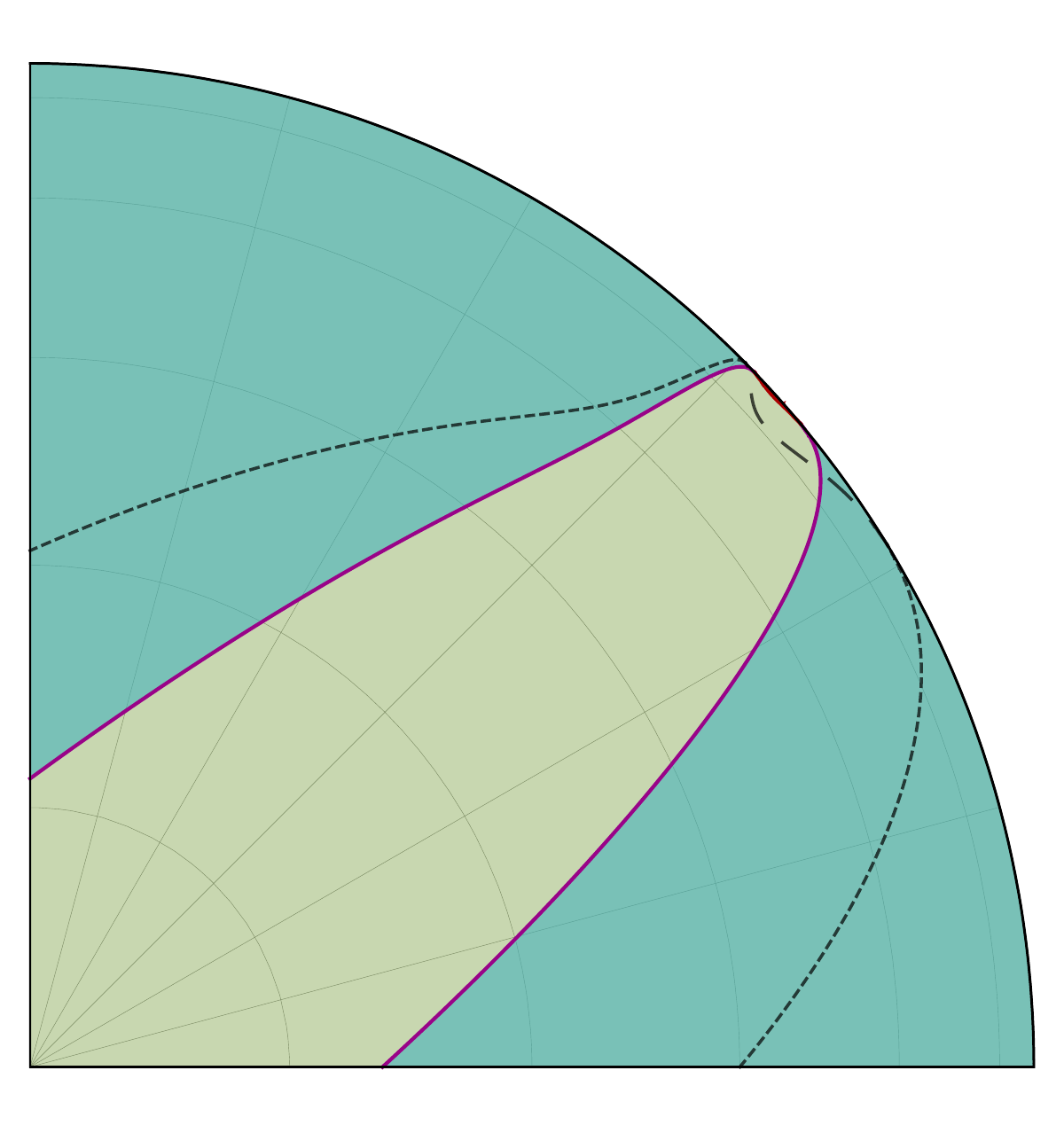}
							\draw  (0.93,0) node[]{ $|\bv|$};
							\draw  (-0.03,0.93) node[]{q};
							\draw (0.05,0) node[]{T};
						\end{tikzonimage}}\quad
						\subcaptionbox{$R_\nu$=1}[.31\linewidth][c]{%
						\begin{tikzonimage}[scale=0.25] {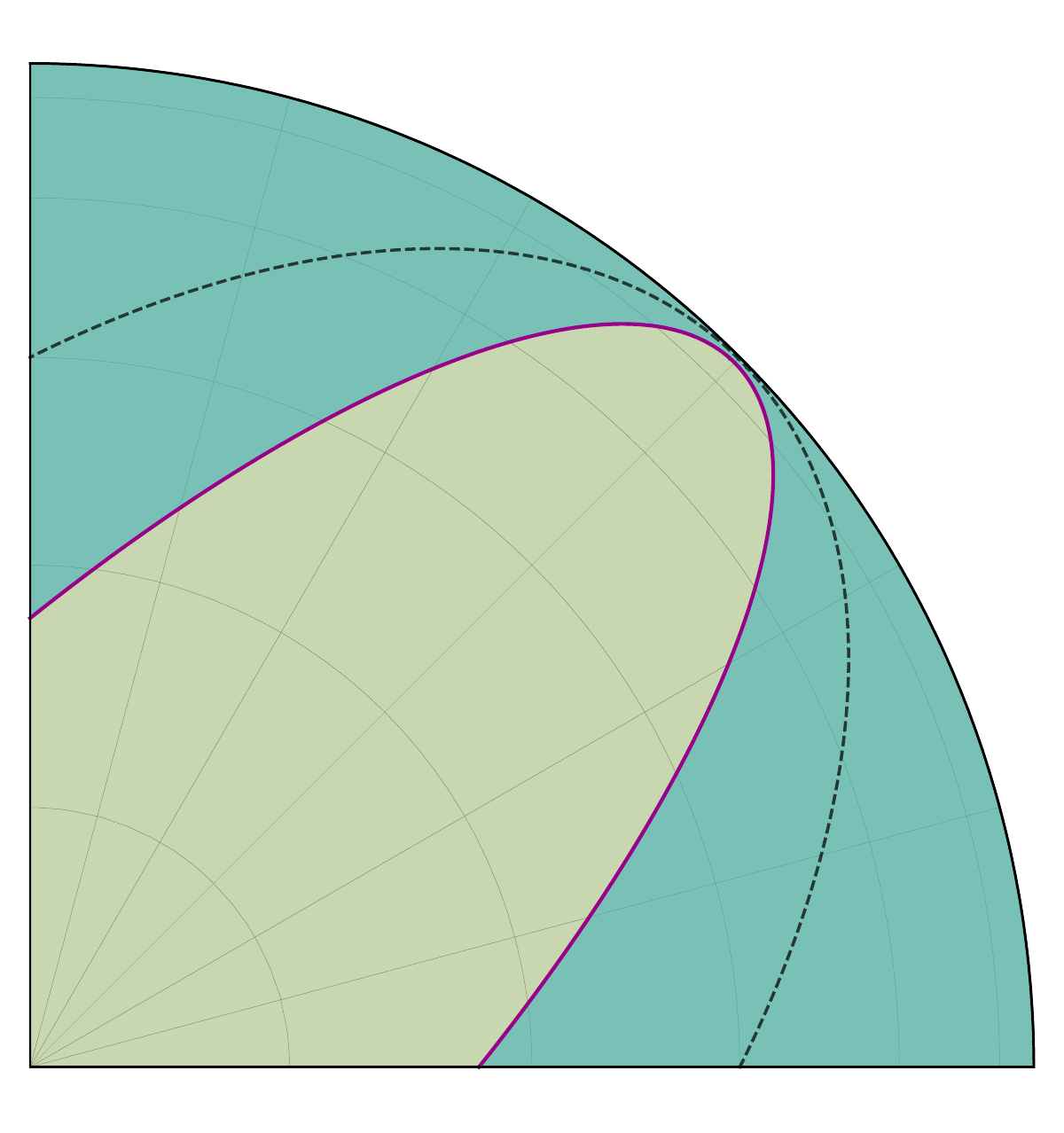}
						\draw  (0.93,0) node[]{ $|\bv|$};
						\draw  (-0.03,0.93) node[]{q};
						\draw (0.05,0) node[]{T};												
						\end{tikzonimage}}\quad
													
	\caption{Effect of drag anisotropy $R_\nu$ on spherical maps with fixed translational angle $\phi = -85^\circ$, shown through orthographic projections onto the $|\bv|-q$ plane.	Colored regions, colored lines, and dashed lines have the same meanings as in Figure \ref{sphericalmap}.  The solid and dashed lines in the bottom right panel do not touch the $T=0$ equator.}
	\label{Rnuvar}
	\end{figure}
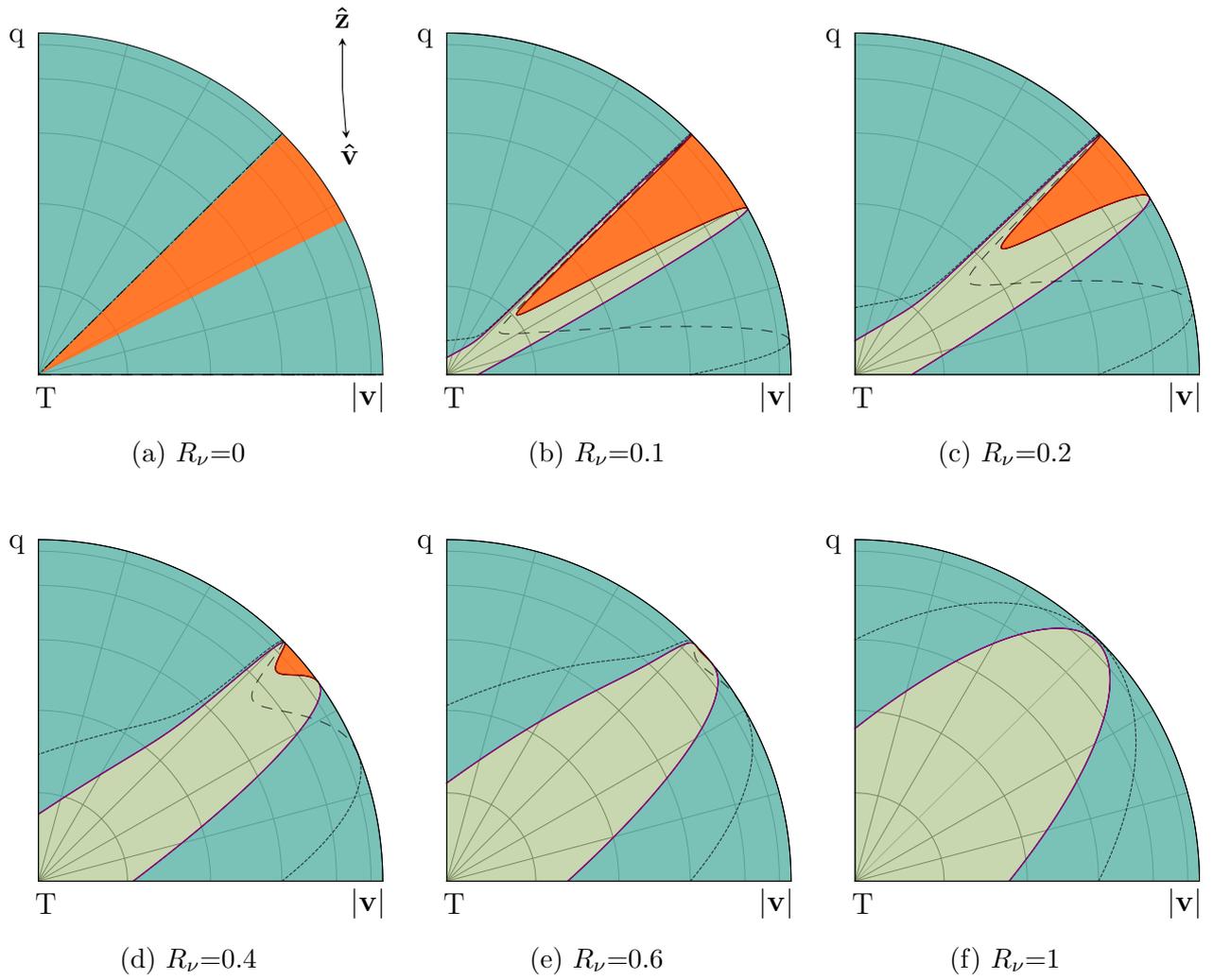

Finally, we show the effects on the spherical map of varying the two remaining relevant parameters.  Figure \ref{phivar} shows maps for six angles of translation at fixed Stokes anisotropy $R_\nu = 0.5$.  The double-sided blowup region is only present for sufficiently aligned translation.  As the direction of translation approaches that of the body force, the dotted stress transition line collides with the equator at a point, which then splits into two spreading points on either side of a newly formed finite-stress region.  Likewise, the green single-sided blowup region collides at a different point, its contacts then spreading around a new double-sided blowup region.
No changes occur on the meridians where $|\bv|$ or $q$ are zero, as the quantity $\phi$ has no meaning there.
Figure \ref{Rnuvar} shows maps for six different drag anisotropies at fixed angle of translation $\phi=-85^\circ$.  The double-sided blowup and counterintuitive infinite-length, finite-stress region expand with increasing anisotropy; the latter occupies half of the sector when $R_\nu = 0$.  At this limit, there is no tangential drag, the axial flow parameter $T$ has no effect on the shape, and the entire sector is radially symmetric.  Both regions disappear as $R_\nu$ approaches unity and the anisotropy becomes sufficiently small.  Note that for the isotropic $R_\nu=1$ map in this figure, the single-sided region and stress transition line do not touch the equator.
The stress transition on the $q=0$ meridian occurs when the axial flow $T$ and translational velocity magnitude $|\bv|$ are equal, independently of the value of $R_\nu$.
At certain limits, the single-sided blowup region becomes a perfect ellipse, the orthographic projection of a region between two small circles on the sphere.
When $\phi = -90^\circ$, the ellipse is given by $[1+(1+R_\nu)^2] |\bv|^2 + 4(1+R_\nu)|\bv|q + 5q^2 =1$,
and when $R_\nu=1$, it is given by $5|\bv|^2 + 8|\bv|q\sin\phi + 5q^2 =1$.

%
%

			
\vspace*{0.25cm}

\newpage

\section{Boundary value problems}\label{bvp}

So far, we have attempted to compactly display a multiparameter family of curves using bulk physical quantities
such as velocities, and geometric variables such as angles.  These quantities are only implicitly related to practical boundary conditions, which usually depend on positions, or to global constraints such as the total length of a cable.  
   In this section, we discuss a few boundary value problems of potential interest.  The first example involves a fully three-dimensional solution, where we recognize that our planar curves and the forces acting on them are embedded in a three-dimensional world.  The last example lives in the limit of vanishing translational velocity, and so requires a new choice of rescaling.
   
As described below in Section \ref{towing}, a three-dimensional problem is first reduced to a two-dimensional problem corresponding to the system described in Section \ref{planarproblem}.  Two-dimensional problems are solved using two-parameter Newton shooting.  In the following examples, the curves have specified endpoint positions and parameters, with the exception of the scaling parameter $C$.  The system described in Section \ref{planarproblem} is integrated over the appropriate arc length $s$ from one end with a pair of given Cartesian coordinates, shooting for the coordinate pair of the other end by adjusting $C$ and the initial angle $\theta$.

\subsection{Towing a cable from two points} \label{towing}

We return to dimensionful quantities.
Consider a fixed length of cable, acted on by a body force $\tbq$, being towed with velocity $\tbv$ by two spools separated by a vector $\bR$.  In general, these are non-orthogonal vectors spanning three dimensions, and our first task is to determine the plane containing the cable.  This plane is independent of any axial flow that might be present.  We will identify it by identifying the uniform binormal vector $\uvc{b}$ of the curve.  For this we use the force balance on the binormals, equation \eqref{bproj}, and the two additional equations $\uvc{b}\cdot\uvc{b} = 1$ and $\uvc{b}\cdot\bR = 0$.  Once the binormal is known, we may define the projections of the translational velocity and body force as $\bv = \tbv - \tbv\cdot\uvc{b}\uvc{b}$ and $-q\uvc{z} = \tbq - \tbq\cdot\uvc{b}\uvc{b}$.  The translational angle $\phi$ is such that $\uvc{v}\cdot\uvc{x} = |\bv|\cos\phi$, with $\uvc{x}$ specified by the relations $\uvc{x}\cdot\uvc{b} = 0$,  $\uvc{x}\cdot\uvc{z} = 0$, and $\uvc{x}\cdot\uvc{x} = 1$.  Nondimensionalizing leads to the quantities appearing in equation \eqref{dimlesseq}.

\begin{figure}[H]		
	\centering
\hspace*{-8.3cm} 	\begin{overpic}[scale=0.39]{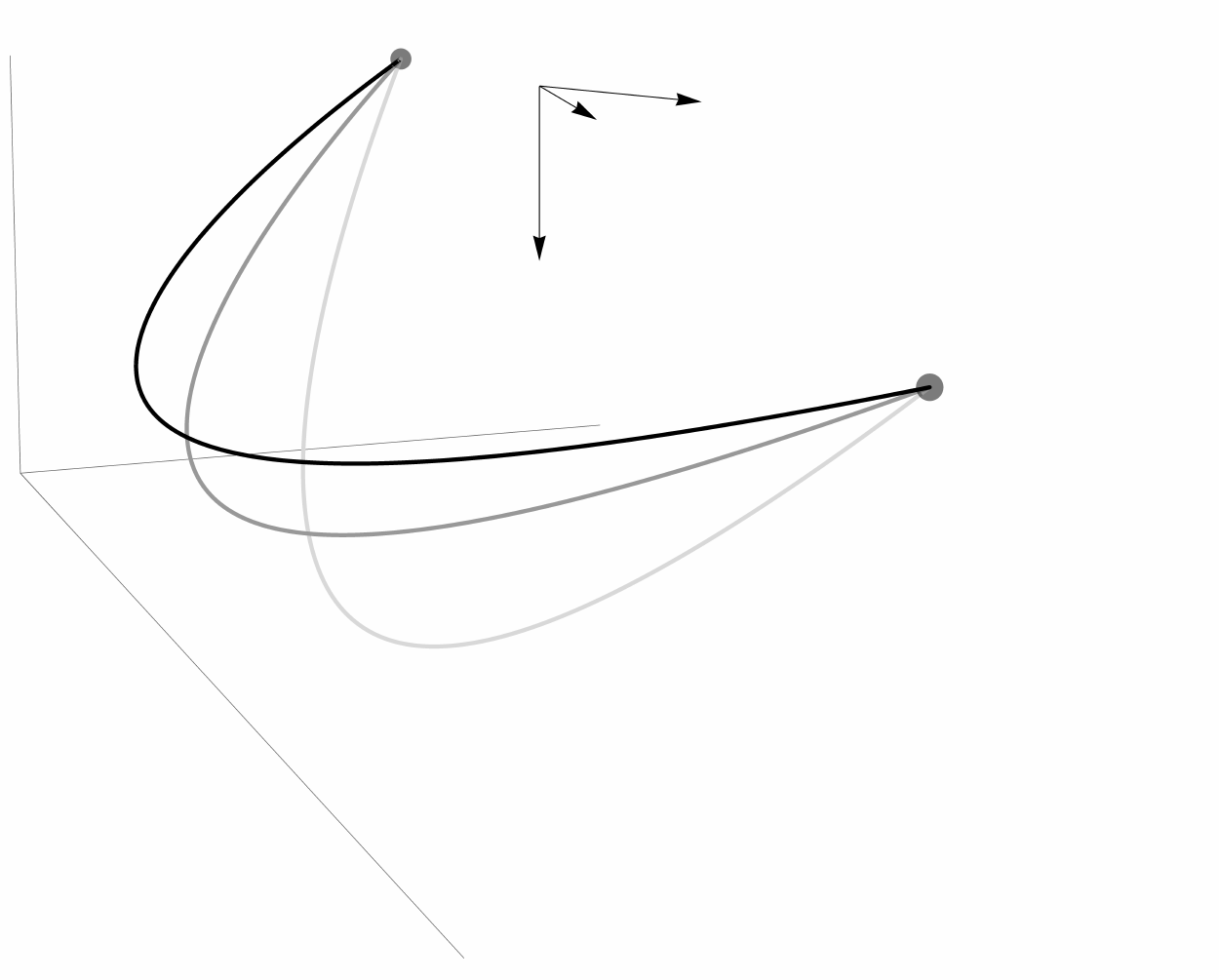}
		\put(43.5,56){\scriptsize {$\hat{\tbq}$}}
		\put(58,70){\scriptsize{$\hat{\tbv}$}}
		\put(49,67){\scriptsize {$\uvc{R}$}}
		\put(90,23){\includegraphics[scale=0.44]{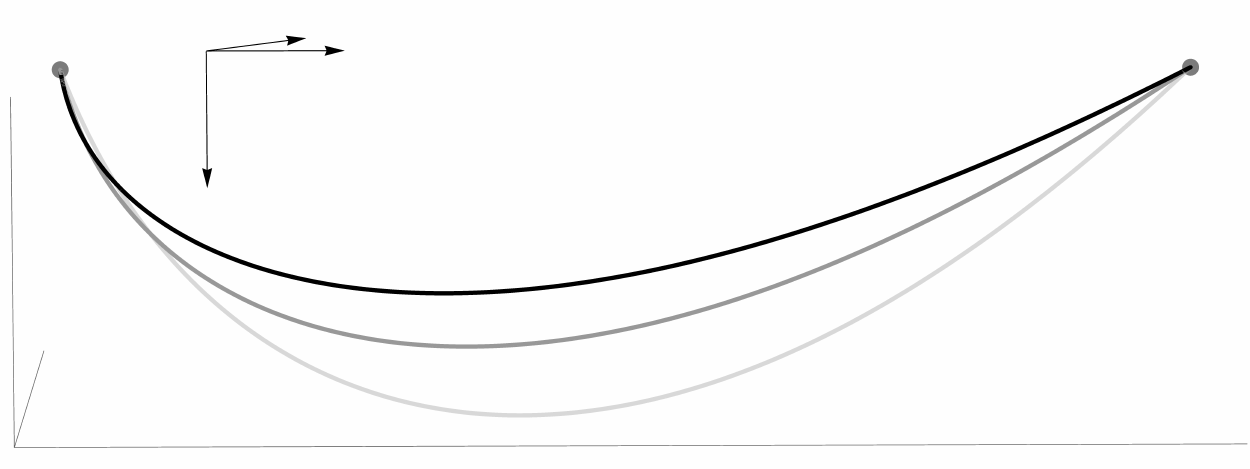}}
		\put(107.5,45){\scriptsize {$\hat{\tbq}$}}
		\put(122,59.5){\scriptsize {$\uvc{R}$}}
		\put(118,62){\scriptsize{$\hat{\tbv}$}}
	\end{overpic}
	\vspace*{-0.25cm}
	\caption{Two views of oblique towing of a unit length cable in a direction perpendicular to the body force. Darker curves indicate higher velocities.  Unit vectors in the directions of the body force $\tbq$, translational velocity $\tbv$, and end-to-end separation $\bR$ are shown. 	The angle between $\tbv$ and $\uvc{R}$ is $45^\circ$, and they are both perpendicular to $\tbq$.	Dimensionful parameters are $\tbv =$ ({\textcolor{Lgr} 4}, {\textcolor{Mgr} 8}, 12), $|\tbq| = 4$, $\nu_\pll = 0.4$, $\nu_\perp = 0.8$, and $|{\bf{R}}|= 0.8$.}
	\label{3dcurves}
\end{figure}

Three curves, representing three states of a single cable, are shown in Figure \ref{3dcurves}.  The curves are rigidly towed in the same direction at different speeds, are identical in length, and their ends are separated by the same vector $\bR$.  For this example, $\bR$ and the velocity $\tbv$ are perpendicular to the body force $\tbq$, but not to each other.  As the towing velocity increases, the curves pivot to become more aligned with the velocity, and also change shape within their planes.

\begin{figure}[H]
	\centering \begin{overpic}[scale=0.7]{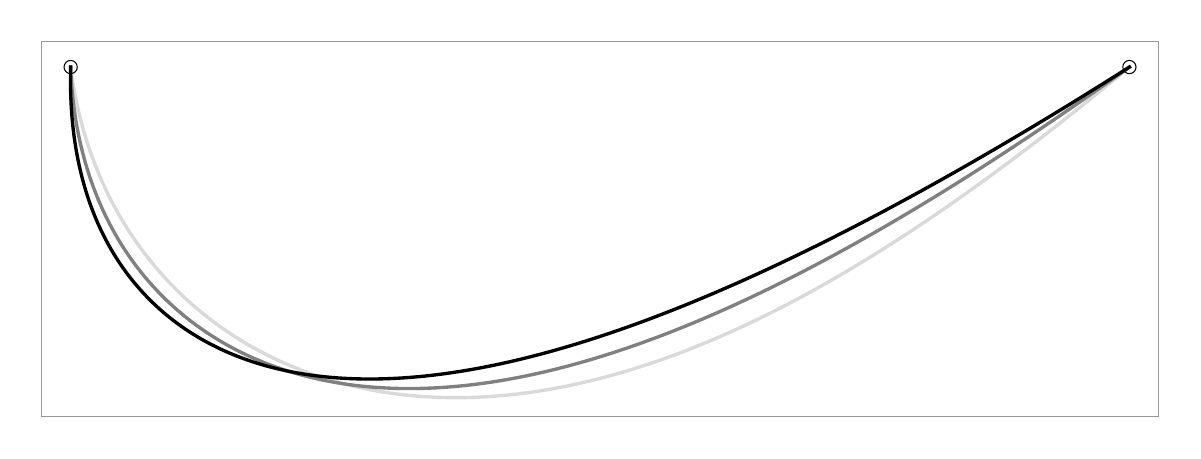}
		\put(40,25){\color{black}\vector(1,1){5}}
		\put(45,30){$\uvc{v}$}
	\end{overpic}
	\vspace*{-0.5cm}
	\caption{Oblique towing of neutrally buoyant cables with different drag anisotropies $R_\nu = $({\textcolor{Lgr} 0}, {\textcolor{Mgr} {0.5}}, 1).  Darker curves indicate lower anisotropy.  The dimensionless length of the cables is unity, and the end-to-end separation magnitude $|{\bf{R}}|= 0.8$.}
	\label{puretow}
\end{figure}

The effect of drag anisotropy is shown in Figure \ref{puretow} for the case of pure rigid towing in the absence of body force.  The problem is inherently two-dimensional for these neutrally buoyant cables.  It is also equivalent to the collinear case discussed in Appendix \ref{collinear}.  The shape equation takes the particularly simple form $\partial_s\theta = C\cos^{(1+R_\nu)}\theta$ for positive $\cos\theta$.  For isotropic drag $R_\nu = 1$ the solutions are equivalent to the classical catenaries; the translational velocity has the same effect as a body force pointing in the opposite direction.  Another case amenable to further integration is that of purely normal drag $R_\nu = 0$, where the arc length may be implicitly defined through $s = \frac{1}{C} \ln \frac{1 + \tan (\theta/2)}{1- \tan (\theta/2)}\,$.

\vspace*{-0.5cm}

\subsection{Reeling a cable between two points}\label{reeling}

Consider an axially flowing catenary reeled between fixed supports.
In this limit, $\nu_\perp$ is meaningless, so we must rescale the quantities in equation \eqref{force1} differently.  Let us use an arbitrary reference length scale $L$ and a velocity scale $\Gamma \equiv \sqrt{T^2 + \left(q/\nu_\pll \right)^2}\,$, so that $s \rightarrow s/L$, $|\bv| \rightarrow |\bv|/\Gamma$, $T \rightarrow T/\Gamma$, $q \rightarrow q/(\nu_\pll\Gamma)$, and $m \rightarrow m\Gamma/(\nu_\pll L)$.  These rescalings lead to the constraint $T^2 + q^2 = 1$ on the axial flow and body force.  Recall from Section \ref{fullproblem} that if the body force $q$ vanishes, then the only equilibrium solution is a straight line.

Following the same steps as in Section \ref{general} leads to the shape and stress equations
\begin{align}
	\partial_s \theta &= C \left[\frac{1 - \tan \frac{\theta}{2}}{1+ \tan \frac{\theta}{2}}\right]^{\frac{T}{q}} \cos^2\theta \, , \label{thetaeqreel} \\
	\sigma &= m T^2  + \frac{q \cos \theta }{\partial_s \theta} \, . \label{sigmaeqreel}
\end{align}
We note in passing that essentially the same derivation can be performed if the drag is any nonlinear function of the axial velocity.  This is relevant to the ``lariat chain'' configuration of a loop of string hanging from a rotating wheel 
\cite{LariatChain, Aitken1878, Gray59}.

Due to the constraint, the dynamical system depends on only two parameters, the scaling constant $C$ and the ratio $T/q$ indicating the relative importance of the axial flow.  The transition from infinite to semi-infinite stress occurs at $T/q = 1$, while that from heteroclinic to single-sided phase portraits occurs at $T/q = 2$. 

\begin{figure}[H]
	\centering
	\begin{tikzonimage}[scale=0.8]{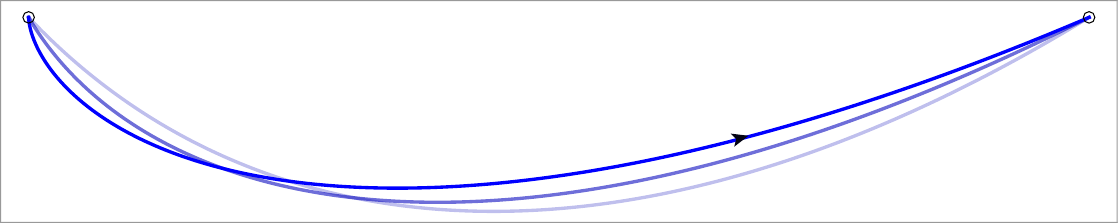}	
	\draw[->,>=stealth] (-0.03,0.6,0) -- (-0.03,0.85,0) node[above] {$\uvc{z}$};
	\end{tikzonimage}
	\caption{Reeling of a unit length cable at different speeds, with end-to-end separation magnitude $|{\bf{R}}|= 0.92$.  Darker curves indicate faster reeling. 	Parameters, rescaled using the tangential drag coefficient $\nu_\pll$, are $T=$ (\textcolor{Lblu}{0.7}, \textcolor{Mblu}{0.88}, \textcolor{Dblu}{0.931}) and $q =$ (\textcolor{Lblu}{0.71}, \textcolor{Mblu}{0.47}, \textcolor{Dblu}{0.365}).}
	\label{reelingcurves}
\end{figure}

\begin{figure}[H]
	\hspace*{0.4cm}
	\scalebox{1}{
		\begin{overpic}[scale=0.5]{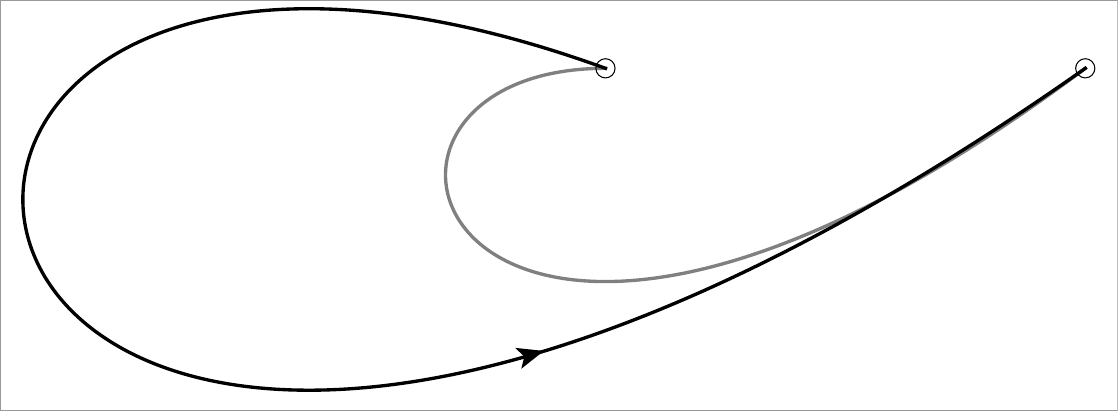}
			\put(107,-13){\includegraphics[scale=0.29]{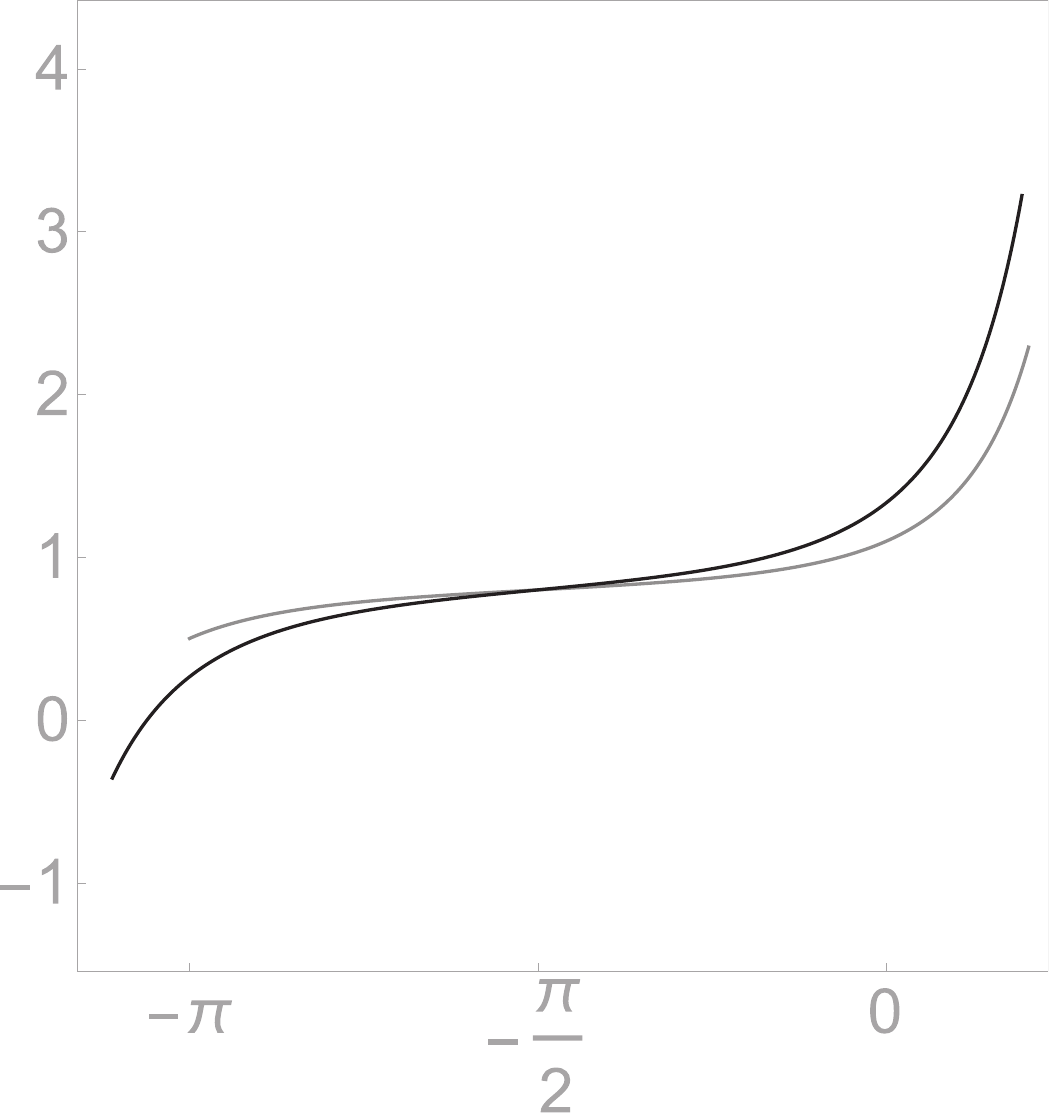}}
			\put(-4,20){\color{black}\vector(0,1){9}}
			\put(-5,30){$\uvc{z}$}
			\put(141,-12){$\theta$}
			\put(106,18){$\sigma$}
		\end{overpic}}
		\vspace*{1cm}
		\caption{Configurations (left) and stresses (right) for curves cut from the $2\pi$-subtending heteroclinic special cases of the reeling problem.  The ratio of axial flow speed to body force is $T/q=2$, the ratios of length $L$ to end-to-end separation magnitude are $L / |{\bf{R}}| =$ ({\textcolor{Mgr} {2}}, 4), and the mass coefficient $m=1$.}
		\label{reelingcurves2pi}
	\end{figure}
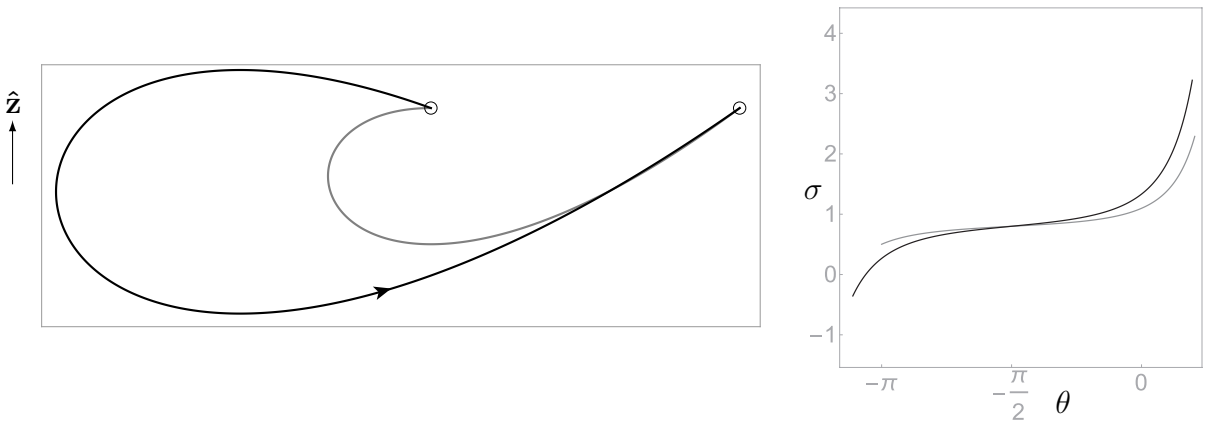

Figure \ref{reelingcurves} shows the effect of increasing axial flow on the curve configurations.  The cable lifts and, somewhat counterintuitively, shifts towards its trailing end.  These qualitative effects can be observed in real ``lariat chains''.  Figure \ref{reelingcurves2pi} shows two curves of different lengths corresponding to the special $2\pi$-subtending heteroclinic case $T/q = 2$ at the purple boundary between heteroclinic and single-sided phase portraits.  Unlike the generic solutions in Figure \ref{reelingcurves}, the curves in Figure \ref{reelingcurves2pi} can subtend an angle greater than $\pi$.  Note that a portion of the longer cable is in compression, so this solution is unstable.  However, increasing the mass coefficient $m$ could bring the entire cable into tension.

\vspace*{0.25cm}

\vspace*{0.75cm}

\section{Discussion}\label{discussion}

Our study has brushed against several established topics in the mechanics of fluid-flexible structure interaction, and raised a few new questions.  We briefly discuss these issues here.

Our analysis allows us to generate a collection of spherical maps, exemplified by Figure \ref{ortho}, and begin to think about what paths through this terrain might correspond to relevant boundary value problems or interesting bifurcations.  Bifurcations that create or destroy solutions likely tell us something about stability, a topic we have not explored in this work.
We did not encounter such phenomena in our incomplete survey of boundary value problems in Section \ref{bvp}, but they might be found somewhere in the five-dimensional parameter space of configurations.  The six-dimensional space of solutions includes bifurcations in the asymptotic behavior of the tension.  It remains to be seen whether such transitions have any significant effect on realistic physical situations when the cables are of finite, indeed moderate, length.

With respect to paths through parameter space, Figure \ref{pseudophase1} is particularly intriguing and raises an interesting question.
Given a family of dynamical systems that depend on several parameters, when can one generate a phase portrait by following a path through parameter space?  The paths we found to generate Figures \ref{pseudophase1} and \ref{pseudophase2} do not change the location of fixed points and poles, but do involve bifurcations that convert one type into the other.  The ends of the path lie inside the parameter space rather than on one of its boundaries.  We have not established that these paths generate true phase portraits.  If they do, it implies that there exists a physically or geometrically relevant parameter not captured by our analysis, and implicitly related to the parameters we chose for representation of our system.

The free sedimentation of a filament is a limiting case of our system.  This problem's history begins with the question of motion and reorientation of DNA or other macromolecules under centrifugation or an electric field \cite{Zimm74, Elvingson92, Allison93, vandenHeuvel08}.  
Many have studied the sedimentation of relatively stiff fibers 
\cite{XuNadim94, SchlagbergerNetz05, Lagomarsino05, Li13} and fiber suspensions \cite{Herzhaft96, TornbergShelley04} at low Reynolds number.  Unlike the generic catenaries, these shapes can bend through angles greater than $\pi$ \cite{Lagomarsino05, Li13}. One simulation demonstrated that at sufficiently low bending stiffness, the system bifurcates from steady sedimentation to periodic motion \cite{Lagomarsino05}.   In the limit of highly flexible thermal chains \cite{Lehtola07, SchlagbergerNetz08}, the behavior appears to be chaotic, resembling the recirculating blob and escaping tail formed by clouds of falling particles at low to moderate Reynolds numbers \cite{Adachi78, NitscheBatchelor97}.
Our findings are consistent with chaotic behavior for an athermal system in the perfectly flexible limit, in that there exists a multiparameter continuum of shapes and terminal velocities corresponding to the double-sided blowup solutions of Section \ref{organization}.  We would expect transient or terminal chaos if some or all of these are unstable.  Note, however, that any bending stiffness would not only regularize the curvature singularities at the ends of the filament, but require vanishing curvature there to satisfy a moment balance boundary condition.  Hence, inferences about even weakly stiff filaments from results in our singular limit may not be well advised.  On the other hand, for an elastic rod--- as opposed to a macromolecule with elastic properties that could be more exotic than those of a simple rod--- at low Reynolds number, nonlocal hydrodynamic forces will dominate over bending.  These next-order interactions are likely responsible for much of the nonlinear phenomena observed in the literature.  They are also likely to destabilize many, if not all, of the sedimentation solutions found in the present study.
All this aside, the nature of the transition from steady to periodic to presumably aperiodic sedimentation with decreasing bending stiffness is a topic worthy of further investigation.

Extending the towing results of Section \ref{towing} to steady rotations and the signed quadratic drag forces reflective of high Reynolds number would aid in understanding towed cable and drogue dynamics in the air or ocean.  In the context of single-end towing problems, much attention has been paid to linear and circular maneuvers, as well as combinations of these basic elements to form hairpin turns \cite{Pode51, Eames68, SchramReyle68, Crist70, HuffmanGenin71, ChooCasarella72, PaulSoler72, IversMudie73, RussellAnderson77, DeZoysa78, Leonard79, Sanders82, AblowSchechter83, Matuk83, Wingham83, Chapman84, LeGuerch87, Huang94, Clifton95, Murray96, KishoreGanapathy96, ZhuRahn98, Grosenbaugh07}.  Equilibrium shapes and tensions are of interest, as are small vibrations superimposed on these background solutions.
Pode captured the salient features of the planar rigid towing problem in terms of four quadratures \cite{Pode51}.
Building on this work, Eames \cite{Eames68} provided analytical expressions for the Cartesian position vector of a cable towing a heavy drogue and translating perpendicularly to the body force, for several different drag assumptions including isotropic linear drag.  He also discusses other interesting boundary value problems.
While the non-analyticity of signed quadratic drag is inconvenient, it is a surmountable difficulty, and it appears upon preliminary glance that these equilibria can be reduced to either first- or second-order dynamical systems.  Such representations would be of lower order than the systems of equations in some of the preceding references, and of considerable advantage, in the search for equilibria, over the full-blown numerical schemes in the others. 

In the limiting case of reeling discussed in Section \ref{reeling}, any nonlinear relationship between velocity and drag force can be accommodated in a shape equation much like
\eqref{thetaeqreel}.  Human-scale experiments using chains or ropes in a ``lariat chain'' configuration \cite{LariatChain}, for which we expect the drag to be quadratic in form, should afford a test of the theory.  Another theoretical study of this configuration may be found in an early paper by Gregory \cite{Gregory49}, whose results are consistent with our shape bifurcation from heteroclinic to single-end blowup when drag forces become sufficiently large.
It should also be possible to experimentally search for configurations corresponding to transitions between solution types.  At a particular speed, the cable should be able to subtend an angle greater than $\pi$, though for a given length the stability, and therefore the observability, of this solution will depend on the weight of the material.  The existence of such a solution only under such restrictive conditions on speed is surprising, and this result will remain suspect until it is observed experimentally.

The problem of laying and recovery of cable is a hybrid of towing and reeling problems, with a history that begins with the work of Thomson \cite{Thomson1857both} and Airy \cite{Airy1858} in the mid-19th century, and continues into the modern day \cite{Zajac57, Pedersen75, LeonardKarnoski90, VazPatel00}.
Thomson neglected drag forces, while Airy unrealistically assumed the forces to be linear and isotropic, corresponding to our limit $R_\nu = 1$.  
The mid-20th century work of Zajac \cite{Zajac57} is particularly worthy of note in the light of the present study in that, with particular assumptions on the configurations and boundary conditions, he considered the two-dimensional translation problem with axial flow and quadratic drag as a pair of coupled first order ordinary differential equations akin to our equations \eqref{dimlesstproj} and \eqref{dimlessnproj}.  

Extension of this work to non-uniform high Reynolds number flows, and the non-planar shapes they can induce, is relevant to sonar, cable laying, mooring lines, and newer applications such as tethered wind energy devices \cite{Gilbert78, LeonardKarnoski90, VazPatel00, AirborneWindEnergy}.  It may also have some biological relevance.
The deformation of plant life in response to non-uniform winds and waves, euphemized as ``reconfiguration'' by Vogel \cite{Vogel94}, is a topic attracting much recent activity in the form of experimental and modeling studies of living and inanimate blade-like planar structures \cite{DennyGaylord02, Abdelrhman07, deLangre08, DijkstraUittenbogaard10, LuharNepf11, deLangre12, Alben04, Gosselin10, BaroisdeLangre13, HenriquezBarrero-Gil14}.  Possible complications here include bending resistance, and non-uniform buoyancy due to variation in morphology or health along the blade \cite{Abdelrhman07, LuharNepf11}.  In some of these, as well as other, high-Reynolds number systems, it may be reasonable and analytically helpful to neglect tangential drag forces.
 
From a more fundamental point of view, the question of rotating, axially flowing equilibria subjected to linear drag forces would provide a simple framework to study the interaction between gyroscopic and dissipative mechanisms for breaking the axial flow symmetry of a one-dimensional continuum living in three dimensions.

\section{Summary}

We have presented and organized a six-parameter family of solutions, containing a five-parameter family of shapes, that extend the classical problem of the catenary to include linear drag forces resulting from translation and axial motion.  The shapes are described by a first-order dynamical system.  Limiting cases include sedimentation, towing, and reeling of cables.
The effects of variation of parameters on solutions were displayed, first in a comprehensive but abstract way, and then with reference to a few specific problems with physically realistic boundary conditions and an integral constraint on length.  These solutions may serve as a starting point for more advanced studies of flexible structure equilibria in fluid flows.

\section*{Acknowledgments}

We thank G Judd for bringing our attention to reference \cite{Gregory49}, and E de Langre for a discussion of preliminary results.

\begin{appendices}

\section{Symmetries of the shape equation}\label{zeroesandpoles}

We wish to show that the two functions
\begin{align*}
	f(\theta) &= \frac{ 1- \bar{U} - \left(\bar{W} + \bar{Q}\right)\tan\frac{\theta}{2} }{ 1+  \bar{U} + \left(\bar{W} + \bar{Q}\right)\tan\frac{\theta}{2} }\quad \mathrm{and} \\
	g(\theta) &= -\bar{U}\sin\theta + \left(\bar{W} + \bar{Q} \right)\cos\theta 
\end{align*}
have either a pole or a zero at the same values of $\theta$.  The function $g(\theta)$ has zeroes at angles $\arctan\frac{\bar{W}+\bar{Q}}{\bar{U}}$, spaced $\pi$ apart.  The denominator or numerator in $f(\theta)$ will vanish at angles $2\arctan\frac{1\pm\bar{U}}{\bar{W}+\bar{Q}}$.  We use the identities $\arctan x = 2\arctan\frac{x}{1+\sqrt{1+x^2}}$ and $\arctan x + \arctan\frac{1}{x} = \frac{\pi}{2}$, and the constraint $\bar{U}^2 + \left(\bar{W}+\bar{Q}\right)^2 = 1$, which can also be written as $\frac{\bar{W}+\bar{Q}}{1+\bar{U}} = \frac{1-\bar{U}}{\bar{W}+\bar{Q}}$, 
 to conclude that $\arctan\frac{\bar{W}+\bar{Q}}{\bar{U}} = 2\arctan\frac{\bar{W}+\bar{Q}}{1+\bar{U}} = 2\arctan\frac{1-\bar{U}}{\bar{W}+\bar{Q}} = \pi - 2\arctan\frac{1+\bar{U}}{\bar{W}+\bar{Q}}\,$.
 
 We also wish to justify the rescaling of 
 \begin{equation*}
 	\partial_s\theta = Ce^{\left(1-R_\nu\right)\bar{U}\bar{Q} \theta} \left| f(\theta) \right|^{\bar{T}} \left| g(\theta) \right|^k
\end{equation*}
after a shift of $\pi$ and change in sign of $\bar{T}$,
\begin{align*}\label{scaleax}
	\partial_s \theta \left(\theta_0 - \pi + \theta, -\bar{T}\right) &= e^{-(1-R_\nu)\bar{U}\bar{Q}\pi} \left(\frac{1+\bar{U}}{1-\bar{U}} \right)^{\bar{T}}  \partial_s \theta \left(\theta_0 + \theta , \bar{T}\right) \, , \\
	\mathrm{where}\quad \tan\tfrac{\theta_0}{2} &= \tfrac{\bar{W}+\bar{Q}}{1+\bar{U}} = \tfrac{1-\bar{U}}{\bar{W}+\bar{Q}} \, .
\end{align*}
The function $\left| g(\theta) \right|$ is $\pi$-periodic, the behavior of the exponential under the transformation is simple, and the only place where $\bar{T}$ appears is in the exponent of the function $\left| f(\theta) \right|$.  We need only demonstrate what happens to this function.  We use the identities $\tan\left(x+y\right) = \frac{\tan x+\tan y}{1-\tan x\tan y}$ and $\tan\left(x\pm\tfrac{\pi}{2}\right) = -\mathrm{cot}\,x$ to find that $f(\theta_0 + \theta) = \frac{\bar{U}-1}{\bar{W}+\bar{Q}}\tan\frac{\theta}{2}$ and $f(\theta_0 -\pi + \theta) =  \frac{\bar{W}+\bar{Q}}{1+\bar{U}}\cot\frac{\theta}{2}$, and therefore
\begin{equation}
	\left[f(\theta_0 -\pi + \theta)\right]^{-\bar{T}} =  \left(\frac{1+\bar{U}}{\bar{U}-1} \right)^{\bar{T}} \left[f(\theta_0 + \theta)\right]^{\bar{T}} .
\end{equation}
Finally, the constraint on $\bar{U}$, $\bar{W}$, and $\bar{Q}$ implies that $\bar{U} \le 1$, so $\left| \bar{U} - 1 \right|$ = $1 - \bar{U}$.

\begin{figure}[H]
	\centering
	\begin{tikzonimage}[scale=0.5]{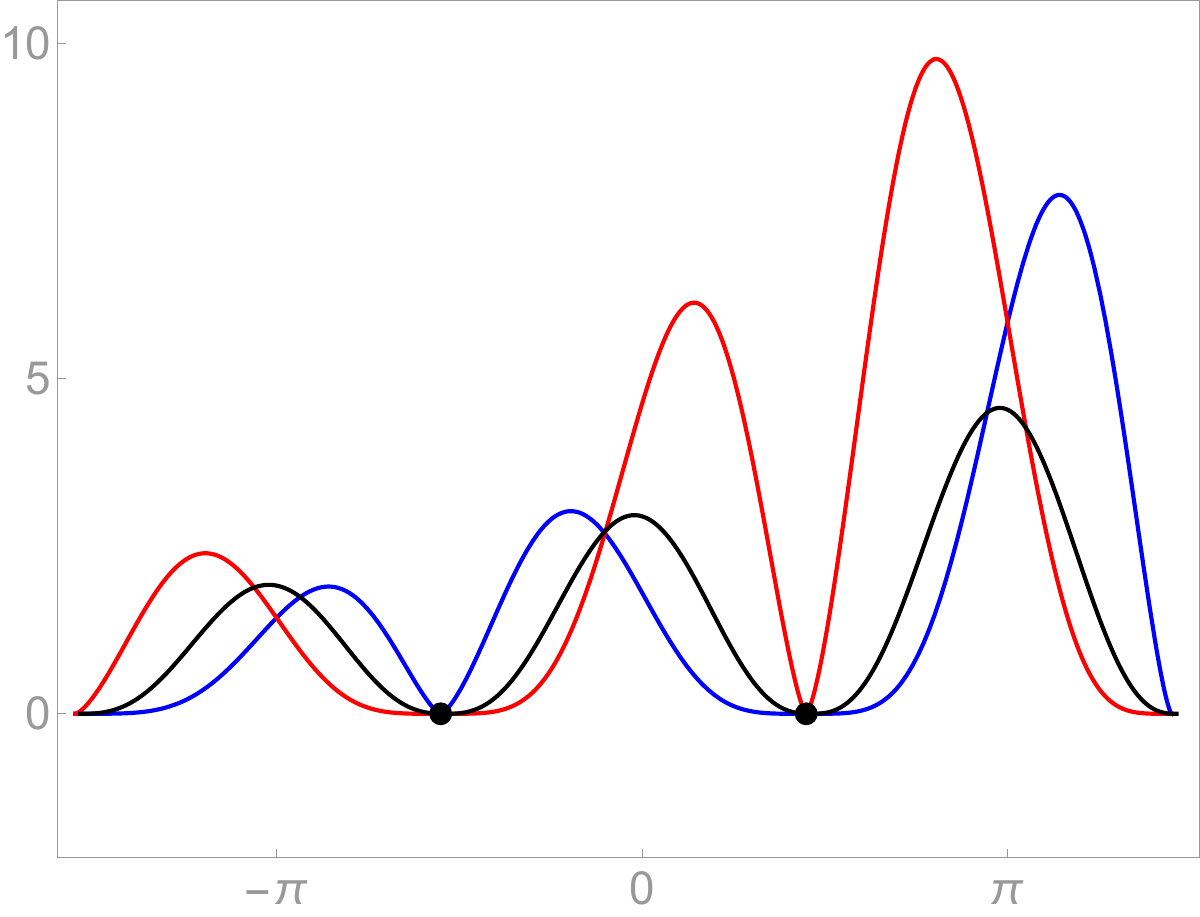} 
		\draw  (-0.03,0.58) node[]{$\partial_s \theta$};
		\draw  (0.55,0.-0.04) node[]{$\theta$};	
		\draw  (0.68,0.17) node[]{$\theta_0$};		
		\draw  (0.38,0.17) node[]{$\theta_0-\pi$};	
	\end{tikzonimage}
	\vspace*{-0.25cm}
	\caption{The effect of $T$ on the function $\partial_s\theta$.  Two roots are labeled with black dots.  The parameters are $\phi = -85^\circ$, $|\bv| = 0.5$, $T= $(\textcolor{red}{-0.4}, 0, \textcolor{blue}{0.4}), $q = $ (\textcolor{red}{0.77}, 0.87, \textcolor{blue}{0.77}), $C = 3$.  The transformation \eqref{scaleax} and additional prefactor scaling relates a blue orbit to a red orbit in an adjacent lobe.  The corresponding reflected red orbit (not shown) below the $\theta$ axis generates the same physical configuration as the original blue orbit.}
	\label{symmetry}
\end{figure}
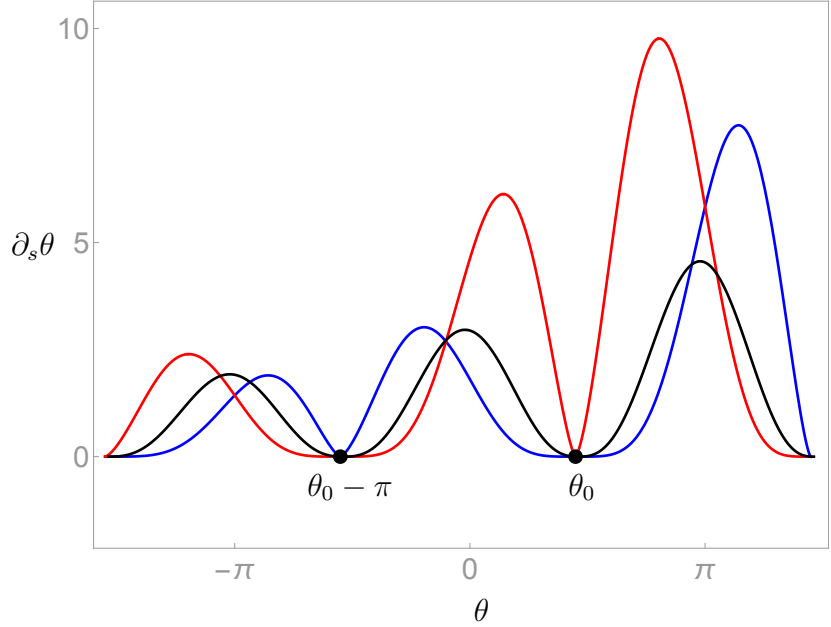

The transformations and scaling are displayed graphically in Figure \ref{symmetry} for an exemplary choice of parameters.

\section{Alternate derivation and representation of \\ collinear systems}\label{collinear}

Here we consider rigid translations collinear with the body force, including as limiting cases the classical catenaries and neutrally buoyant towed cables.  This section makes use of the relations \eqref{framerotation} that describe derivatives of the frame vectors.

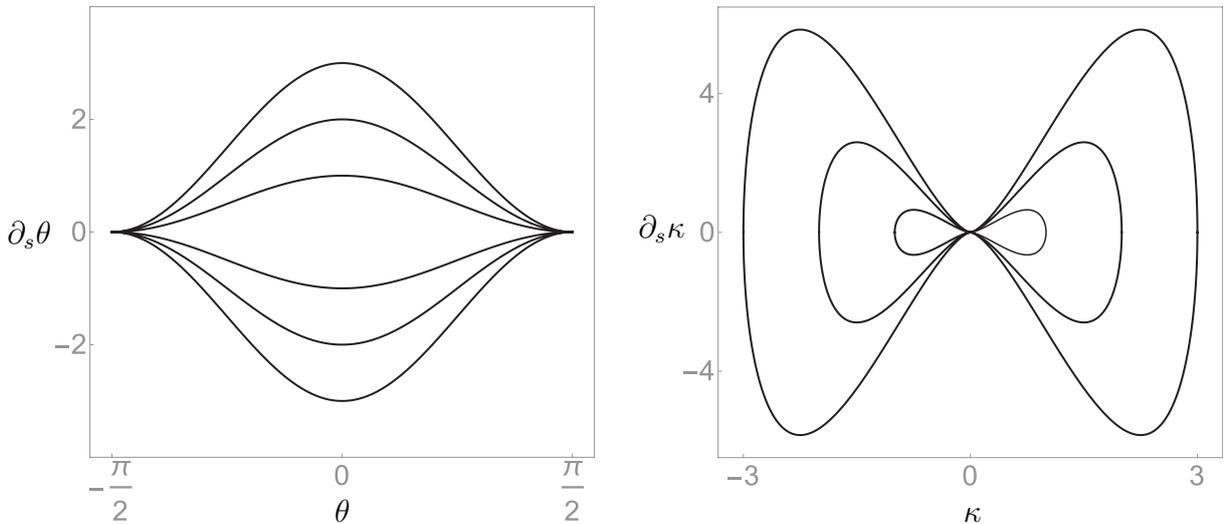
\begin{figure}[H]
	\centering
	\hspace{0.25cm}
	\begin{overpic}[scale=0.35]{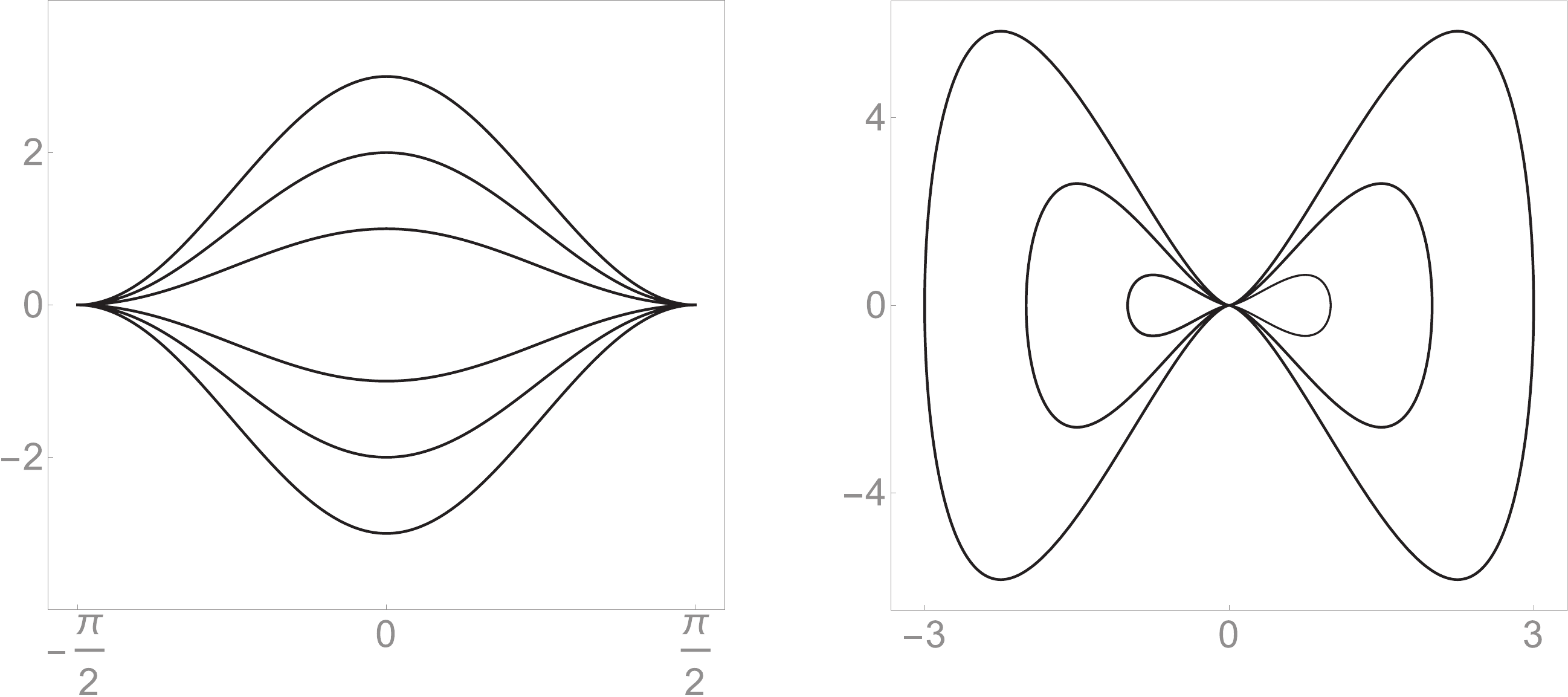}
	\put(-4,24){$\partial_s \theta$}
	\put(50,24.5){$\partial_s \kappa$}	
	\put(24,0){$\theta$}
	\put(77.8,0){$\kappa$}
	\end{overpic}
	\vspace*{-0.1cm}
\caption{Two phase space representations of the classical catenaries, using an angle or curvature variable.  On the left, $\partial_s\theta = C\cos^2\theta$, on the right, $\left( \partial_s\kappa \right)^2 = 4 \kappa^3 \left(C_\nu - \kappa \right)$, using $C_\nu = C = \pm(1,2,3)$.}
\label{catenaryphaseportraits}
\end{figure}

For rigid translations, $T = 0$, and the projections (\ref{tproj}-\ref{bproj}) of the dimensionful three-dimensional problem onto the frame vectors are modified accordingly,
\begin{align}
\nu_\pll\tbv\cdot \uvc{t} &= \partial_s \sigma + \tbq\cdot \uvc{t} \label{tprojrigid} \, , \\ 
\nu_\perp \tbv \cdot \uvc{n} &= \sigma \kappa + \tbq \cdot \uvc{n} \label{nprojrigid} \, , \\ 
\nu_\perp \tbv \cdot \uvc{b} &= \tbq \cdot \uvc{b} \label{bprojrigid} \, .
\end{align}
We may note immediately that combining \eqref{nprojrigid} and an $s$-derivative of \eqref{bprojrigid} leads to the equation $\sigma\kappa\tau = 0$, implying that any solutions with nontrivial stress $\sigma$ and curvature $\kappa$ must be planar ($\tau = 0$).  Therefore, we restrict consideration to the planar problem obtained by projecting the rigid form of the dimensionless equation \eqref{dimlesseq} onto $\uvc{t}$ and $\uvc{n}$.  We also let $\bv = v\uvc{z}$, collinear with the body force, so that
\begin{align}
\left(R_{\nu}v+q\right)\uvc{z}\cdot\uvc{t} &= \partial_s \sigma \label{dimlesstprojrigidcoll} \, , \\ 
\left(v+q\right)\uvc{z}\cdot\uvc{n} &= \sigma \kappa \label{dimlessnprojrigidcoll} \, .
\end{align}
Take a derivative of \eqref{dimlesstprojrigidcoll} and use \eqref{dimlessnprojrigidcoll}, and take a derivative of \eqref{dimlessnprojrigidcoll} and use \eqref{dimlesstprojrigidcoll}, to obtain two equations,
\begin{align}
	\partial_s^2\sigma - A_\nu\sigma\kappa^2 &= 0\, , \label{tmaster} \\
	\left(\tfrac{1+A_\nu}{A_\nu}\right)\partial_s\sigma\kappa + \sigma\partial_s\kappa &= 0 \, , \label{nmaster}
\end{align}
depending on a single parameter $A_\nu \equiv \tfrac{R_\nu v + q}{v + q}$.  
This parameter is equal to $R_\nu$ for neutrally buoyant towed cables ($q=0$), and equal to unity for the classical catenaries ($v=0$) and for the equivalent case of isotropic drag ($R_\nu=1$).
Divide \eqref{nmaster} by $\sigma\kappa$ to get, for positive $\sigma$ and $\kappa$,
\begin{align}
	\left(\tfrac{1+A_\nu}{A_\nu}\right)\partial_s\ln\sigma + \partial_s\ln\kappa = 0 \, , \nonumber \\
	\kappa\sigma^{\tfrac{1+A_\nu}{A_\nu}} = B \label{kappasigma} \, , 
\end{align}
$B$ constant, then insert \eqref{kappasigma} into \eqref{tmaster} to get
\begin{equation}
	\partial_s^2\sigma - A_\nu B^2\sigma^{-\left(\tfrac{2+A_\nu}{A_\nu}\right)} = 0 \, ,
\end{equation}
and finally multiply by $\partial_s\sigma$ and integrate to get
\begin{equation}
	\left(\partial_s\sigma\right)^2 + A_\nu^2B^2\sigma^{-\tfrac{2}{A_\nu}} = C_\nu \label{sigmaquad} \, .
\end{equation}
$C_\nu$ constant.
If the stress $\sigma$ is obtained from equation \eqref{sigmaquad}, the Ces{\`{a}}ro equation of the curve follows as 
\begin{equation}
	\kappa = B\sigma^{-\left(\tfrac{1+A_\nu}{A_\nu}\right)} \, .
\end{equation}
For many values of $A_\nu$, the analytical solution of equation \eqref{sigmaquad} is implicit and complicated.  However, when $A_\nu=1$, corresponding to the classical catenaries or isotropic drag, the stress is a square root of a quadratic polynomial in $s$, and the curvature $\kappa$ is the reciprocal of that polynomial.  This is a well known representation of the catenaries.  For general values of $A_\nu$, the curve shape is more easily obtained by inserting \eqref{kappasigma} in \eqref{sigmaquad} and multiplying through to eliminate denominators,
\begin{equation}\label{kappaode}
	\left( \partial_s\kappa \right)^2 +\left(A_\nu+1\right)^2 \kappa^{\tfrac{2(2A_\nu+1)}{A_\nu+1}} \left( \kappa^{\tfrac{2}{A_\nu+1}} - C_\nu\right) = 0 \, .
\end{equation}
The dynamical system \eqref{kappaode} can be derived from the rigid collinear form of \eqref{thetaeq}, $\partial_s\theta = C\cos^k\theta$ for positive $\cos\theta$.  We note that $k = 1 + A_\nu$ and identify $C_\nu = C^{\tfrac{2}{1+A_\nu}}$.  In Figure \ref{catenaryphaseportraits} we display two representations of the classical catenaries, for which $C_\nu = C$, as heteroclinic orbits in $\theta$-$\partial_s\theta$ space and as homoclinic orbits in $\kappa$-$\partial_s\kappa$ space.

\vspace{0.1cm}

\end{appendices}

\bibliographystyle{unsrt}

\end{document}